\documentclass[prb,aps,amssymb,twocolumn,showpacs,floatfix]{revtex4}
 \usepackage[dvips]{graphicx}
 \renewcommand{\theequation}{\arabic{section}.\arabic{equation}}
\def\be{\begin{equation}}
\def\ee{\end{equation}}
\def\bea{\begin{eqnarray}}
\def\eea{\end{eqnarray}}
\def\br{{\bf r}} 
\def\bk{{\bf k}} 
\def\bq{{\bf q}} 
\def\bn{{\bf n}} 
\def\bp{{\bf p}} 

\begin{document}

\title{Interaction-induced magnetoresistance in a two-dimensional electron gas}

\author{I.~V.~Gornyi$^{1,a}$}
\author{A.~D.~Mirlin$^{1,2,b}$} 
\affiliation{$^{1}$Institut f\"ur Nanotechnologie,
Forschungszentrum Karlsruhe, 76021 Karlsruhe, Germany \\
$^{2}$Institut f\"ur Theorie der kondensierten Materie, Universit\"at
Karlsruhe, 76128 Karlsruhe, Germany}

\date{\today} 

\begin{abstract} 
We study the
interaction-induced quantum correction $\delta\sigma_{\alpha\beta}$ to 
the conductivity tensor of electrons in two dimensions 
for arbitrary $T\tau$ (where 
$T$ is the temperature and $\tau$ the transport scattering time), 
magnetic field, and type of disorder.  
A general theory is developed, allowing us to
express $\delta\sigma_{\alpha\beta}$ 
in terms of classical propagators (``ballistic diffusons'').  
The formalism is used to 
calculate the interaction contribution to the longitudinal and the Hall resistivities
in a transverse magnetic field in the whole 
range of temperature from the diffusive   
($T\tau\ll 1$) to the ballistic ($T\tau\gtrsim 1$) regime, both in
smooth disorder and in the presence of short-range
scatterers. Further, we apply the formalism to anisotropic systems
and demonstrate that the interaction
induces novel quantum oscillations in the resistivity of
lateral superlattices.  
\end{abstract} 

\pacs{
72.10.-d, 
73.23.Ad, 
71.10.-w, 
73.43.Qt  
}

\maketitle

\section{Introduction}
\setcounter{equation}{0}
\label{I}

The magnetoresistance (MR) in a transverse field $B$ is one of the 
most frequently studied characteristics of the two-dimensional (2D) 
electron gas 
 \cite{AA,Been}. 
Within the 
Drude-Boltzmann theory, 
the longitudinal resistivity of an isotropic degenerate system is 
$B$--independent,
\be
\rho_{xx}(B)=\rho_0=(e^2\nu v_F^2 \tau)^{-1}, 
\label{drude}
\ee
where $\nu$ is the density of states per spin direction, 
$v_F$ the Fermi velocity, and $\tau$ the transport scattering time. 
Deviations from the constant $\rho_{xx}(B)$ are customarily called a
positive or negative MR, depending on the sign of the deviation. 
There are several distinct sources of a non-trivial MR, 
which reflect the rich physics of the magnetotransport in 2D systems. 

First of all, it has been recognized recently that even within the  
quasiclassical theory memory effects may lead to strong MR  
\cite{Hauge,Baskin98,dmitriev01,FSDP,PMR,Antidots,polyakov01}. 
The essence of such effects is that a
particle ``keeps memory'' about 
the presence (or absence) 
of a scatterer 
in a spatial region which it has already visited. As a
result, if the particle returns back, the new scattering event is
correlated with the original one, yielding a correction to the
resistivity (\ref{drude}). Since the magnetic field enhances the
return probability, the correction turns out to be $B$-dependent.
As a prominent example, memory effects
in magnetotransport of composite fermions subject to an effective 
smooth random magnetic field explain a positive 
MR around half-filling of the lowest Landau level \cite{PMR}.
Another type of memory effects taking place in systems with rare strong
scatterers is responsible for 
a negative MR in disordered antidot arrays \cite{Hauge,Baskin98,dmitriev01,
Antidots,polyakov01}. 
However, such effects turn out to be of a relatively minor 
importance for the low--field 
quasiclassical magnetotransport in semiconductor heterostructures with  
typical experimental parameters, while at higher $B$ they are obscured by  
the development of the Shubnikov-de Haas oscillation (SdHO). 

Second, the negative MR induced by the suppression of the quantum interference 
by the magnetic field is a famous manifestation of weak
localization~\cite{AA}. While the weak-localization correction to
conductivity is also related to the return probability, it has
(contrary to the quasiclassical memory effects) an
intrinsically quantum character, since it is governed by quantum
interference of time-reversed paths. As a result it is suppressed
already by a classically negligible magnetic field, which changes
relative phases of the two paths. Consequently,
the corresponding correction to  $\rho_{xx}$ in high-mobility structures
is very small and restricted to the range of very weak magnetic fields. 

Finally, another quantum correction to MR is induced by the
electron--electron interaction. While this effect is similar to those
discussed above in its connection with the return probability
(see Sec.~\ref{IV} below), it is distinctly different in several crucial aspects.  
In contrast to the memory effects, this contribution is of quantum
nature and is therefore strongly $T$-dependent at low temperatures.
On the other hand, contrary to the weak localization, the interaction
correction to conductivity is not destroyed by a strong magnetic
field. As a result, it induces an appreciable MR in the range of
classically strong magnetic fields. This effect will be the subject of
the present paper.  
 
It was discovered by Altshuler and Aronov \cite{AA} 
that the Coulomb 
interaction enhanced by the diffusive motion of electrons 
gives rise to a quantum correction to conductivity, 
which has in 2D the form 
(we set $k_B=\hbar=1$) 
\be 
\delta\sigma_{xx} \simeq {e^2\over 2\pi^2} \left(1-{3\over 2} {\cal F}\right) 
\ln T\tau, \; \qquad  T\tau \ll 1. 
\label{AAdiff} 
\ee 
The first term in the factor $(1-{3\over 2} {\cal F})$ originates from
the exchange contribution, and the second one from the Hartree
contribution. In the weak-interaction regime, $\kappa \ll k_F$,
where $\kappa=4\pi e^2\nu$ is the inverse screening length, 
the Hartree contribution is small, ${\cal F}\sim
(\kappa/k_F)\ln(k_F/\kappa)\ll 1$. The conductivity correction
(\ref{AAdiff}) is then dominated by the exchange term and is
negative. The condition $T\tau\ll 1$ under which Eq.~(\ref{AAdiff}) 
is derived \cite{AA} implies that electrons move diffusively 
on the time scale $1/T$ and is termed the ``diffusive regime''. 
Subsequent works \cite{SenGir,girvin82} showed that Eq.~(\ref{AAdiff}) 
remains valid in a strong magnetic field, leading (in combination with 
$\delta\sigma_{xy}=0$) to a parabolic 
interaction--induced quantum MR, 
\be 
{\delta\rho_{xx}(B)\over \rho_0} \simeq \left(1-{3\over 2} {\cal F}\right)   
{{(\omega_c\tau)^2-1}\over \pi k_F l} \ln T\tau,  \; \qquad  T\tau \ll 1, 
\label{MRAA} 
\ee 
where $\omega_c=eB/mc$ is the cyclotron frequency and $l=v_F\tau$ the  
transport mean free path. Indeed, a $T$--dependent negative MR was 
observed in experiments \cite{PTH83,Choi,Sav,Minkov02,Minkov03} 
and attributed to the interaction effect.  
However, the majority of experiments~\cite{PTH83,Choi,Sav} cannot be
directly compared with 
the theory \cite{AA,SenGir,girvin82} since they were performed at higher
temperatures, $T\tau \gtrsim 1$.
(In high-mobility GaAs heterostructures conventionally
used in MR experiments, $1/\tau$ is typically $\sim 100\  {\rm mK}$
and becomes even smaller with improving quality of samples.) 
In order to explain the experimentally observed $T$-dependent 
negative MR in this temperature range the authors of Refs.~\onlinecite{PTH83,Choi} 
conjectured various {\it ad hoc} extensions of Eq.~(\ref{MRAA}) 
to higher $T$. 
Specifically, Ref.~\onlinecite{PTH83} conjectures that the logarithmic behavior  
(\ref{MRAA}) with $\tau$ replaced by the quantum time
$\tau_s$ is valid up to 
$T \sim 1/\tau_s$, while Ref.~\onlinecite{Choi} proposes to replace
$\ln T\tau$ by  $-\pi^2/2T\tau$. 
These proposals, however, were not supported by theoretical 
calculations.
There is thus a clear need for a theory of the MR 
in the ballistic regime, $T \gtrsim 1/\tau$. 
 
In fact, the effect of interaction 
on the conductivity at $T\gtrsim 1/\tau$ has been already considered
in the literature 
\cite{GD86,ZNA-sigmaxx,ZNA-rhoxy,ZNA-MRpar,ZNA-deph,reizer,fluct-supercond}. 
Gold and Dolgopolov \cite{GD86} analyzed the
correction to conductivity arising from the $T$-dependent 
screening of the impurity potential. They
obtained a linear-in-$T$ correction $\delta\sigma\sim e^2T\tau$.
In the last few years, this effect attracted a 
great deal of interest in a context
of low-density 2D systems showing a seemingly  
metallic behavior~\cite{AKS01,Alt01}, 
$d\rho/dT>0$. 
Recently, Zala, Narozhny, and Aleiner 
\cite{ZNA-sigmaxx,ZNA-rhoxy,ZNA-MRpar} developed 
a systematic theory of the interaction corrections 
valid for arbitrary $T\tau$. They showed that the
temperature-dependent screening of Ref.~\onlinecite{GD86} has in fact a
common physical origin with the Altshuler-Aronov effect but that the
calculation of Ref.~\onlinecite{GD86} took only the Hartree term into account and
missed the exchange contribution.
In the ballistic range of temperatures, the theory of 
Refs.~\onlinecite{ZNA-sigmaxx,ZNA-rhoxy,ZNA-MRpar}, 
predicts, in addition to the linear-in-$T$ correction to  
conductivity $\sigma_{xx}$, a $1/T$ correction to the Hall coefficient~\cite{ZNA-rhoxy}
$\rho_{xy}/B$ at $B \to 0$, and describes the MR in a {\it parallel} field~\cite{ZNA-MRpar}.

The consideration of Ref.~\onlinecite{ZNA-sigmaxx,ZNA-rhoxy,ZNA-MRpar} 
is restricted, however, 
to {\it classically weak} transverse fields,  
$\omega_c\tau \ll 1$, and to the {\it white-noise} disorder. 
The latter assumption is believed to be justified for Si-based 
and some (those with a very large spacer) GaAs structures, and the results of 
Refs.~\onlinecite{ZNA-sigmaxx,ZNA-rhoxy,ZNA-MRpar} have been 
by and large confirmed by most recent 
experiments~\cite{Cole02,Shashkin02,proskuryakov02,kvon02,vitkalov02,pudalov02,noh02}
on such systems. On the other hand, the random potential in  
typical GaAs heterostructures 
is due to remote donors and has a long--range  
character. Thus, the impurity scattering is predominantly of a small--angle  
nature and is characterized by two relaxation times, the transport 
time $\tau$ 
and the single-particle (quantum) time $\tau_s$ governing damping of SdHO, 
with $\tau \gg \tau_s$. Therefore, a description of the MR in such
systems requires a more general theory valid also in the range of
strong magnetic fields and for smooth disorder.
[A related problem of the tunneling density of states
in this situation was studied in Ref.~\onlinecite{RAG}.]

In this paper,
we develop a general theory of the interaction--induced corrections
to the conductivity tensor of 2D  
electrons valid for arbitrary temperatures, transverse 
magnetic fields, and range of random potential.   
We further apply it to the problem of magnetotransport 
in a smooth disorder at $\omega_c\tau \gg 1$.
In the ballistic limit, $T\tau \gg 1$
(where the character of disorder is crucially important),
we show that while the correction to $\rho_{xx}$
is exponentially suppressed for $\omega_c\ll T$, 
a MR arises at stronger $B$ 
where it scales as $B^2T^{-1/2}$. 
We also study the temperature-dependent correction to the Hall
resistivity and show that it scales as $BT^{1/2}$
in the ballistic regime and for strong $B$.
We further investigate a ``mixed-disorder''
model, with both short-range and long-range impurities
present. We find that a sufficient concentration of
short-range scatterers strongly enhances the MR
in the ballistic regime.

The outline of the paper is as follows. In Sec.~\ref{II} we present our
formalism and derive a general formula for the conductivity
correction. We further demonstrate (Sec.~\ref{IIc}) that in the
corresponding limiting cases our theory reproduces all previously known
results for the interaction correction. In Sec.~\ref{III} we apply our
formalism to the problem of interaction-induced MR in strong magnetic
fields and smooth disorder. Section~\ref{IV} is devoted to a physical
interpretation of our results in terms of a classical return
probability. In Sections~\ref{V} and \ref{VI} we present several
further applications of our theory. Specifically, we analyze the
interaction effects in systems with short-range scatterers 
and in magnetotransport in modulated systems (lateral superlattices). 
A summary of our results, a comparison with experiment,
and a discussion of
possible further developments is presented in Sec.~\ref{VII}.
Some of the results of the paper
have been published in a brief form in the Letter~\cite{prl}.

\section{General formalism}
\label{II}
\setcounter{equation}{0}

\subsection{Smooth disorder}
\label{IIa}

We consider a 2D electron gas (charge $-e$, mass $m$, density $n_e$)
subject to a transverse magnetic field $B$ and to a random potential
$u({\bf r})$ characterized by a correlation function
\be
\label{corr-function}
\langle u({\bf r}) u({\bf r'})\rangle = w(|{\bf r}-{\bf r'}|)
\ee
with a spatial range $d$. The total ($\tau_s^{-1}$) and the transport
($\tau^{-1}$) scattering rates induced by the random potential are
given by
\bea
{1\over \tau_s} &=& 2\pi\nu \int_0^{2\pi}{d\phi\over 2\pi} 
W(\phi), \label{taus}\\
{1\over \tau} &=& 2\pi\nu \int_0^{2\pi}{d\phi\over 2\pi} 
W(\phi) (1-\cos\phi), \label{tautr}
\eea
where $W(\phi)=\tilde{w}(2 k_F\sin{\phi\over 2})$ is the
scattering cross-section. 
We begin by considering the case of smooth disorder, 
$k_F d\gg 1$, when $\tau/\tau_s\sim (k_F d)^2\gg 1$; generalization
onto systems with arbitrary $\tau/\tau_s$ will be presented in
Sec.~\ref{IIb}. We assume that the magnetic field is not too strong,
$\omega_c\tau_s\ll 1$, so that the Landau quantization is destroyed by
disorder. Note that this assumption is not in conflict with a
condition of classically strong magnetic fields ($\omega_c\tau\gg 1$),
which is a range of our main interest in the present paper. 

\begin{figure}
  \medskip
\centerline{\includegraphics[width=8cm]{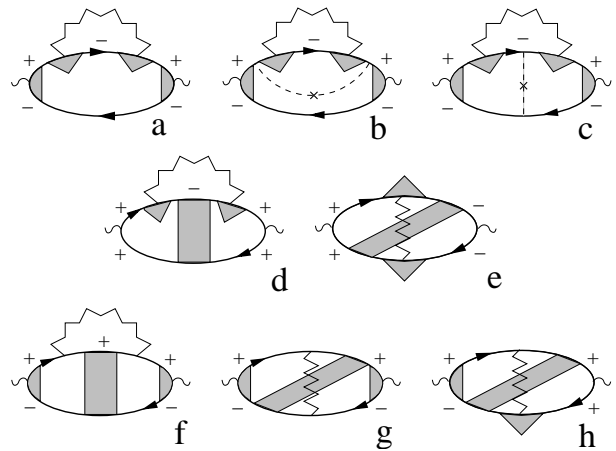}}
\caption{ Exchange diagrams for the interaction  
correction to $\sigma_{\alpha\beta}$. 
The wavy (dashed) lines denote the interaction
(impurity scattering),
the shaded blocks are impurity ladders,
and the $+/-$ symbols denote the signs 
of the Matsubara frequencies. The diagrams obtained by a flip  
and/or by an exchange $+ \leftrightarrow -$ should also 
be included. ``Inelastic'' part of the diagrams {\it f}, {\it g} is
canceled by a contribution of
the Coulomb-drag type, see Appendix~\ref{A1}. }
\label{fig1} 
\end{figure}

We consider two types of the electron-electron interaction
potential $U_0({\bf r})$: (i) point-like interaction, $U_0({\bf r})=V_0$,
and (ii) Coulomb interaction, $U_0({\bf r})=e^2/r$.
In order to find  the interaction-induced correction 
$\delta\sigma_{\alpha\beta}$ to the conductivity tensor, 
we make use of the ``ballistic'' 
generalization of the diffuson diagram technique of 
Ref.~\onlinecite{AA}. We consider the exchange contribution first and will discuss  
the Hartree term later on. Within the Matsubara formalism, 
the conductivity is expressed via the Kubo formula through the
current-current correlation function,
\begin{eqnarray}
&& \sigma_{\alpha\beta}(i\Omega_k)={n_e e^2\over m
\Omega_k}\delta_{\alpha\beta} \nonumber \\
&& - {1\over \Omega_k}\int_0^{1/T} \!\! d\tau\int d^2r
\langle {{\cal T}_\tau \hat j_{\alpha}({\bf r},\tau )}{\hat
j_{\beta}(0,0)} \rangle  e^{ i \Omega_k \tau },\nonumber \\
\label{kubo}
\end{eqnarray}
where $\Omega_k=2\pi k T$ is the bosonic Matsubara frequency. Diagrams
for the leading-order interaction correction are shown in Fig.~\ref{fig1}
and can be generated in the following way. First, there are two essentially
different ways to insert an interaction line into the bubble formed by
two electronic Green's function. Second, one puts signs of electronic
Matsubara frequencies in all possible ways. On the third step, one
connects lines with opposite signs of frequencies
$\epsilon_n>0,\ \epsilon_m<0$ by impurity--line ladders (which are not
allowed to cross each other). Finally, in the case of the diagram {\it a},
where four electronic lines form a ``box'', one should include two
additional diagrams, {\it b} and {\it c}, with an extra impurity line 
(``Hikami box'').

The impurity--line ladders are denoted by shaded blocks in Fig.~\ref{fig1}; 
we term them ``ballistic diffusons''. Formally,
the ballistic diffuson is defined as an impurity average (denoted below
as $\langle \dots \rangle_{\rm imp}$) of a product of a retarded and
advanced Green's functions,
\bea
&&{\cal D}(i\epsilon_m,i\epsilon_n;\br_1,\br_2;\br_3,\br_4)\nonumber \\
&&=\theta(-\epsilon_m\epsilon_n)\langle G(\br_1,\br_2;i\epsilon_m) 
G(\br_3,\br_4;i\epsilon_n) \rangle_{\rm imp}.\nonumber \\
\label{balldiffrr}
\eea
Following the standard route of the quasiclassical formalism~\cite{semi,AL,RAG-dos-corr}, 
we perform the Wigner transformation,
\bea
&&{\cal D}(i\epsilon_m-i\epsilon_n;{\bf R}_1,\bp_1;{\bf R}_2,\bp_2)=
\int d\br  d\br' \nonumber \\ 
\displaystyle
&&\times e^{-i[\bp_1-(e/c)\bf A(\bf R_1)]\br}
e^{-i[\bp_2-(e/c)\bf A(\bf R_2)]\br'} \nonumber \\ 
&&\times
{\cal D}(i\epsilon_m,i\epsilon_n;{\bf R}_1,\br; 
{\bf R}_2,\br'),
\label{balldiff}
\eea
where ${\bf R}_1=(\br_4+\br_1)/2$, ${\bf R}_2=(\br_2+\br_3)/2$,
$\br=\br_4-\br_1$, and $\br'=\br_2-\br_3$.
Note that the factors depending on the vector potential make the
ballistic diffuson (\ref{balldiff}) gauge-invariant.  
Finally, we integrate out the absolute values of momenta $p_{1,2}$
and get the final form of the ballistic diffuson
\bea
&&{\cal D}(i\omega_l;{\bf R}_1,\bn_1;{\bf R}_2,\bn_2) \nonumber \\
&&={1\over 2\pi\nu}
\int {p_1 dp_1\over 2\pi}\int {p_2 dp_2\over 2\pi}
{\cal D}(i\omega_l;{\bf R}_1,\bp_1;{\bf R}_2,\bp_2),
\nonumber \\
\label{calBRn}
\eea
which  describes 
the quasiclassical propagation of an electron in the phase space 
from the point ${\bf R}_2,\bn_2$ to ${\bf R}_1,\bn_1$. 
Here $\bn$ is the unit vector characterizing the direction of velocity 
on the Fermi surface. 
The ballistic diffuson satisfies the quasiclassical Liouville-Boltzmann
equation
\bea 
\left[|\omega_l| + i v_F q \cos(\phi-\phi_q)+
\omega_c{\partial\over\partial\phi} 
\right. 
&+&\left.{\hat C}\right] 
{\cal D}(i\omega_l,\bq;\phi,\phi') 
\nonumber \\ 
&=& 
2\pi\delta(\phi-\phi'), 
\label{LB} 
\eea 
where $\phi\  (\phi_q)$ is the polar angle of $\bn \ (\bq)$ and
${\hat C}$ is the collision integral, determined by the
scattering cross-section
$W({\bf n},{\bf n'}).$
For the case of a smooth disorder, the collision integral is given by 
\be
{\hat C}=-{1\over \tau}{\partial^2\over \partial \phi^2}.
\label{Csmooth}
\ee
In contrast to the diffusive regime, where ${\cal D}$ has a universal and 
simple structure ${\cal D}(i\omega_l,\bq)=1/(Dq^2+|\omega_l|)$ 
determined by the diffusion constant $D$ only, its form in the  
ballistic regime is much more complicated. 
We are able, however, to get a general expression for  
$\delta\sigma_{\alpha\beta}$ in terms of the ballistic propagator 
${\cal D}(\omega_l,\bq;\bn,\bn')$.

The temperature range 
of main interest in the present paper is restricted by $T\tau_s \ll 1$, 
since at higher $T$ the MR will be small in the whole range of 
the quasiclassical transport  $\omega_c\tau_s \ll 1$ (see below). 
In this case the ladders are dominated by contributions with many 
($\gg 1$) impurity lines. We will assume this situation when
evaluating diagrams in the present subsection. A general case of
arbitrary $T\tau_s$ and $\tau_s/\tau$ will be addressed in Sec.~\ref{IIb}.

\begin{figure}
  \medskip
\includegraphics[width=8cm]{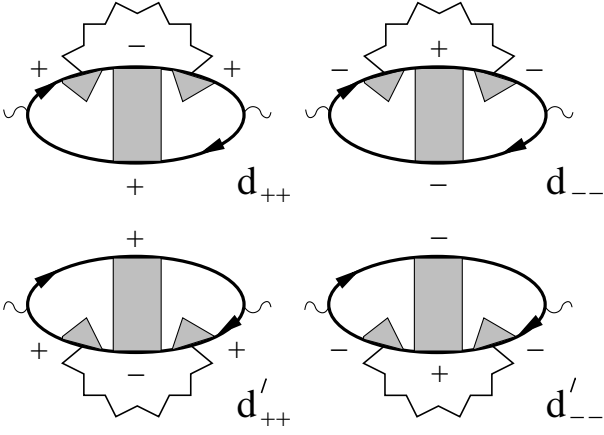}
\caption{  Diagrams  obtained by a flip  
and/or by an exchange $+ \leftrightarrow -$ from the diagram d.} 
\label{figd} 
\end{figure} 

\begin{figure}
  \medskip
\includegraphics[width=8cm]{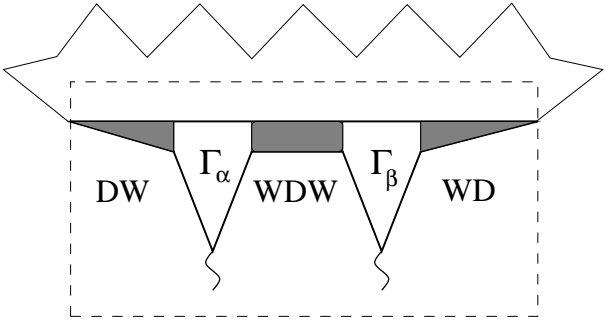}
\caption{  Diagram $d$ drawn in a different way in order to
visualize the structure of Eq.~(\ref{Bdef}). 
The dashed frame encloses 
the electronic part ${\cal B}_{\alpha\beta}^{d}$.} 
\label{kroko} 
\end{figure}

We start with the diagrams {\it d} and {\it e} that give rise
to the logarithmic correction in the diffusive regime \cite{AA}.
Let us fix the sign of the external frequency, $\Omega_k > 0.$ 
Each of the diagrams {\it d} and {\it e} generates four diagrams 
by a flip with respect to the horizontal line or
by exchange $+ \leftrightarrow -$, see Fig.~\ref{figd}. 
Consider first the diagram $d_{++}$. 
There are two triangular boxes containing each
a current vertex and three
electron Green's functions (Fig.~\ref{kroko}).
In the quasiclassical regime $\omega_c\tau_s \ll 1$
one may neglect the effect of magnetic field on the Green's functions
(keeping $\omega_c$ in the ballistic propagators only).
Furthermore, using $T\tau_s\ll 1$, 
we neglect the difference in momenta and frequencies in
the Green's functions,
since typical values of frequencies $i\Omega_k$, $i\omega_l$
and momenta ${\bf q}$ carried by the ballistic diffusons 
are set by the temperature.
Each triangle then reads
\bea
\Gamma_\alpha(\bn)&=&{e\over m}\int {p dp\over 2\pi}  
{p \ n_\alpha\over (-\xi_p+i/2\tau_s)^2
(-\xi_p-i/2\tau_s)}\nonumber \\
 &\simeq&  i  2\pi  \nu \tau_s^2 e v_F n_\alpha,
\eea
where $\xi_p=p^2/2m-\mu$.
Combining the triangles with the three ballistic propagators 
separated by the impurity lines (see Fig.~\ref{kroko}),
we obtain the following expression for the 
electronic part of the diagram $d_{++}$,
\bea
&&(2\pi\nu)^3\int \prod_{i=1}^{6}{d\phi_i\over 2\pi}
{\cal D}(i\omega_l,\bq;\phi,\phi_1)W(\phi_1-\phi_2)
\Gamma_\alpha(\phi_2) \nonumber \\
&&\times W(\phi_2-\phi_3){\cal D}(i\omega_l-i\Omega_k,\bq;\phi_3,\phi_4) 
W(\phi_4-\phi_5)\nonumber \\
&&\times \Gamma_\beta(\phi_5)W(\phi_5-\phi_6)
{\cal D}(i\omega_l,\bq;\phi_6,\phi')\nonumber \\
&&\equiv {4\pi \sigma_0\over \tau}
{\cal B}^{d}_{\alpha\beta}(i\omega_l,-i\Omega_k,\bq;\phi,\phi')
\label{Bdef}. 
\eea
In what follows we will use for brevity 
a short-hand notation 
$$(2\pi\nu)^3
{\cal D} W \Gamma_\alpha W {\cal D} W \Gamma_\beta W
{\cal D}$$ 
for  the l.h.s. of (\ref{Bdef}) and analogous notations for
other structures of this type. 
Making use of the small-angle nature of scattering in a smooth random
potential,  we can replace the $W(\phi_i-\phi_j)$ factors  in (\ref{Bdef})
by $(\nu\tau_s)^{-1}\delta(\phi_i-\phi_j)$, yielding
\bea
&&
{\cal B}^{d}_{\alpha\beta}(q,i\omega_l,i\Omega_k;\phi,\phi')
\nonumber \\
&&\simeq -{\cal D}(i\omega_l,\bq) 
n_\alpha {\cal D}(i\omega_l+i\Omega_k,\bq) 
n_\beta {\cal D}(i\omega_l,\bq). \nonumber
\eea
In the exchange term (calculated in the present subsection)
this structure is further integrated over the angles $\phi$ and $\phi'$,
\be
B^{d}_{\alpha\beta}(i\omega_l,i\Omega_k,\bq)=
\langle {\cal B}^{d}_{\alpha\beta}(i\omega_l,i\Omega_k,\bq;\phi,\phi') 
\rangle.
\label{15}
\ee
The angular brackets $\langle \dots \rangle$ 
denote averaging over velocity directions, e.g.               
$$\langle n_x{\cal D}n_x\rangle\equiv \int {d\phi_1\over 2\pi} {d\phi_2\over 2\pi}
\cos\phi_1{\cal D}(\omega_l,\bq;\phi_1,\phi_2)\cos\phi_2.$$ 
The fermionic frequencies
obey the inequalities $\epsilon_m>0$, $\epsilon_m-\omega_l<0$, 
and $\epsilon_m-\Omega_k>0$, which implies
$\omega_l>\epsilon_m>\Omega_k$, so that 
the summation over $\epsilon_m$ gives the factor 
$(\omega_l-\Omega_k)/2\pi T$.

The diagram $d_{--}$ has the same structure
(both triangles have opposite signs, thus the total sign remains unchanged), 
but the frequency summation is restricted 
by $\epsilon_m<0$, $\epsilon_m-\omega_l>0$, and $\epsilon_m-\Omega_k<0$,
yielding the factor $-\omega_l/2\pi T$ in the conductivity correction.
The diagrams $d'_{++}$ and $d'_{--}$ obtained from $d_{++}$ and $d_{--}$ 
by a flip (or, equivalently, by reversing all arrows) 
double the result.
Combining the four contributions and changing sign of the summation
variable, $\omega_l \to -\omega_l$
in $d_{--}$ and $d'_{--}$ terms, we have
\bea
&&\delta\sigma_{\alpha\beta}^{d}(i\Omega_k)
=-{8\pi\sigma_0\over \tau}{T^2\over \Omega_k} \int{d^2 q\over (2\pi)^2}
\nonumber \\
&&\times 
\left[\ 
\sum_{\omega_l>\Omega_k}{\omega_l-\Omega_k \over 2\pi T} 
U(i\omega_l,\bq) 
B^{d}_{\alpha\beta}(-i\omega_l,i\Omega_k,\bq)
\right. \nonumber \\
&&+
\left. \sum_{\omega_l>0}{\omega_l\over 2\pi T}
U(i\omega_l,\bq) 
B^{d}_{\alpha\beta}(i\omega_l,i\Omega_k,\bq)\ 
\right],
\label{sigmad}
\eea
where $U(i\omega,\bq)$ is the interaction potential equal to 
a constant $V_0$ for point-like interaction and to 
\be 
U(i\omega_l,\bq)= 
\frac{1}{2\nu}\ \frac{\kappa}{q+\kappa[1-|\omega_l|\langle 
{\cal D}(\bq,i\omega_l) \rangle]} 
\label{screen} 
\ee 
for screened Coulomb interaction. 
In (\ref{sigmad}) we used the fact that 
$U({\bf q},-i\omega_l)=U({\bf q},i\omega_l)$
and ${\cal D}({\bf q},-i\omega_l)={\cal D}({\bf q},i\omega_l)$.
Equation (\ref{screen}) is a statement of the random-phase
approximation (RPA), with the polarization operator given by
\be
\label{polarization}
\Pi(i\omega_l,\bq)=2\nu[1-|\omega_l|\langle 
{\cal D}(i\omega_l,\bq) \rangle].
\ee
The first term (unity) in square brackets in (\ref{polarization}) 
comes from the $++$ and
$--$ contributions to the polarization bubble, while the second term 
is generated by the $+-$ contribution (ballistic diffuson).

The diagrams {\it e} are evaluated in a similar way.
In all four diagrams of this type one of the electron triangles
is the same as in diagrams {\it d} while another one has an opposite
sign.
The structures arising after integrating out fast momenta in
electron bubbles coincide with those of {\it d}-type
($B_{\alpha\beta}^d$). Summation over the fermionic frequency
$\epsilon_m$ is constrained by the
condition $\omega_l>\epsilon_m>\Omega_k$ for all the $e$-type diagrams.
The correction due to the diagrams {\it e} therefore reads
\bea
&&\delta\sigma_{\alpha\beta}^{e}(i\Omega_k)
={8\pi\sigma_0\over \tau}{T^2\over \Omega_k} \int\!\!{d^2 q\over (2\pi)^2}
\sum_{\omega_l>\Omega_k}\!\!\!{\omega_l-\Omega_k \over 2\pi T} \:
U(i\omega_l,\bq) \nonumber \\
&&\times\left[ B^{d}_{\alpha\beta}(-i\omega_l,i\Omega_k,\bq)
+
B^{d}_{\alpha\beta}(i\omega_l,i\Omega_k,\bq)\right].
\label{sigmae}
\eea
We see that the first term in square brackets in (\ref{sigmae})
cancels the first term in (\ref{sigmad}).
Thus, the sum of the contributions of diagrams {\it d} and {\it e}
takes the form  
\bea
\delta\sigma_{\alpha\beta}^{d+e}(i\Omega_k)
&=&-{4\sigma_0\over \tau}{T\over\Omega_k} 
\left[
\sum_{\omega_l=0}^{\Omega_k}\omega_l
\Phi^{d}_{\alpha\beta}(i\omega_l,i\Omega_k)\right.
\nonumber \\
&+&
\left. \sum_{\omega_l>\Omega_k}\Omega_k
\Phi^{d}_{\alpha\beta}(i\omega_l,i\Omega_k)
\right],
\label{sigmae+d}
\eea
where we introduced a notation
\be
\Phi^{\mu}_{\alpha\beta}(i\omega_l,i\Omega_k)=
\int{d^2 q\over (2\pi)^2}U(i\omega_l,\bq)
B^{\mu}_{\alpha\beta}(i\omega_l,i\Omega_k,\bq),
\label{F-def}
\ee
with the index $\mu$ labeling the diagram.

Similarly, we obtain for the diagram {\it h}
\bea
\delta\sigma_{\alpha\beta}^{h}(i\Omega_k)
&=&-{4\sigma_0\over \tau}{T\over \Omega_k} 
\left[
\sum_{\omega_l=0}^{\Omega_k}\omega_l
\Phi^{h}_{\alpha\beta}(i\omega_l,i\Omega_k)\right.
\nonumber \\
&+&
\left. \sum_{\omega_l>\Omega_k}\Omega_k
\Phi^{h}_{\alpha\beta}(i\omega_l,i\Omega_k)
\right],
\label{sigmah}
\eea
with
\be
B^{h}_{\alpha\beta}(i\omega_l,i\Omega_k,\bq)=-2 
T_{\alpha\gamma} \langle n_{\gamma} 
{\cal D}(i\omega_l+i\Omega_k,\bq) n_{\beta} {\cal D}(i\omega_l,\bq) \rangle. 
\nonumber\\
\label{bh}
\ee
The tensor $T_{\alpha\gamma}$ appearing in (\ref{bh})
describes the renormalization of a current vertex connecting
two electronic lines with opposite signs of frequencies, 
\bea
\label{t_alphabeta}
T_{\alpha\beta}&=&2\left.\langle n_\alpha 
{\cal D}n_\beta\rangle\right|_{q=0,\omega\to 0}=
{\sigma_{\alpha\beta}\over e^2v_F^2\nu} \nonumber\\
&=& {\tau\over 1+\omega_c^2\tau^2} \left(
\begin{array}{cc} 1       &  -\omega_c\tau \\
             \omega_c\tau & 1
\end{array}
\right).
\eea

We turn now to diagrams {\it f} and {\it g}.
The expressions for the corresponding contributions read
\begin{eqnarray}
&&\delta\sigma_{\alpha\beta}^{f}(i\Omega_k)
=-{4\sigma_0\over \tau}{T\over\Omega_k} 
\left[\sum_{\omega_l \geq 0}\Omega_k
\Phi^{f}_{\alpha\beta}(i\omega_l,i\Omega_k) \right.
\nonumber  \\
&&+
\left. 
\sum_{-\Omega_k<\omega_l<0} \!\!(\Omega_k+\omega_l)
\Phi^{f}_{\alpha\beta}(-i\omega_l,i\Omega_k)
\right],\label{sigmaf} \\
&&\delta\sigma_{\alpha\beta}^{g}(i\Omega_k)
={4\sigma_0\over \tau}{T\over \Omega_k} 
\left[\sum_{\omega_l=0}^{\Omega_k}(\Omega_k-\omega_l)
\Phi^{f}_{\alpha\beta}(i\omega_l,i\Omega_k)
\right.
\nonumber \\
&&+ 
\left. \sum_{-\Omega_k<\omega_l<0} \!\!(\Omega_k+\omega_l)
\Phi^{f}_{\alpha\beta}(-i\omega_l,i\Omega_k)
\right], \label{sigmag}
\end{eqnarray}
with
\be
B^{f}_{\alpha\beta}(i\omega_l,i\Omega_k,\bq)=
T_{\alpha\gamma} \langle n_{\gamma} 
{\cal D}(i\omega_l+i\Omega_k,\bq) n_{\delta}\rangle
T_{\delta\beta}. \nonumber \\
\label{bf}
\ee
The sum of the contributions {\it f} and {\it g}
is therefore given by
\bea
\delta\sigma_{\alpha\beta}^{f+g}(i\Omega_k)
&=&-{4\sigma_0\over \tau}{T\over\Omega_k} 
\left[
\sum_{\omega_l=0}^{\Omega_k}\omega_l
\Phi^{f}_{\alpha\beta}(i\omega_l,i\Omega_k)\right.
\nonumber \\
&+&
\left. \sum_{\omega_l>\Omega_k}\Omega_k
\Phi^{f}_{\alpha\beta}(i\omega_l,i\Omega_k)
\right].
\label{sigmaf+g}
\eea
We see that when the diagrams ${\it f}$ and ${\it g}$ are combined, 
the same Matsubara structure as for other diagrams
[Eqs.~(\ref{sigmae+d}), (\ref{sigmah})] arises. 
In other words, 
the role of the diagrams {\it g} is to cancel the extra contribution
of diagrams {\it f}, which has a different Matsubara structure. 

A word of caution is in order here. In our calculation we have set the
value of velocity coming from current vertices to be equal $v_F$, thus
neglecting a particle-hole asymmetry. If one goes beyond this
approximation and takes into account the momentum-dependence of
velocity (violating the particle-hole symmetry), the above cancellation
ceases to be exact and an additional term with a different Matsubara
structure arises in  $\delta\sigma_{\alpha\beta}^{f+g}$. 
After the analytical continuation is performed,
the corresponding correction to the conductivity has a form
\bea
\delta\sigma^{\rm inel}_{\alpha\beta}&=&
-{2\sigma_0\over \tau}\int_{-\infty}^\infty 
{d\omega \over 2\pi} {\omega\over 2T\sinh^{2}(\omega/2T)}\nonumber \\
&\times&\!\!\int {d^2q\over (2\pi)^2}\ \delta B^{f}_{\alpha\beta}(\omega,q)
\ {\rm Im}\, U(\omega,q),
\label{sigma-inel}
\eea
characteristic for effects governed by inelastic scattering.
This contribution is determined by real inelastic scattering processes with
an energy transfer $\omega\lesssim T$  
and behaves (in zero magnetic field) as $e^2(T\tau)^2$. This implies that the corresponding
resistivity correction, $\delta\rho\sim (T/e E_F)^2$
is independent of disorder. However,  
such a correction should not exist because of total momentum
conservation.
Indeed, an explicit calculation (see Appendix~\ref{A1}) shows that 
this term is canceled by the Aslamazov-Larkin-type diagrams analogous
to those describing the Coulomb drag.

Finally, we consider the diagrams {\it a,b}, and {\it c}.
Already taken separately, each of them has the expected Matsubara
structure (contrary to the diagrams {\it d,e} and {\it f,g}, which
should be combined to get this structure). However, another
peculiarity should be taken into account. The diagrams {\it a,b}, and
{\it c} form together the Hikami box, so that their sum is smaller
by a factor $\sim \tau_s/\tau$ than separate terms. Therefore, some care is
required: subleading terms of order $\tau_s/\tau$ should be retained
when contributions of individual diagrams are calculated. The result reads
\begin{eqnarray}
\delta\sigma_{\alpha\beta}^{a+b+c}(i\Omega_k)
&=&-{4\sigma_0\over \tau}{T\over \Omega_k} 
\left[\sum_{\omega_l=0}^{\Omega_k}\omega_l
\Phi^{a+b+c}_{\alpha\beta}(i\omega_l,i\Omega_k)\right.
\nonumber \\
&+&
\left. \sum_{\omega_l>\Omega_k}\Omega_k
\Phi^{a+b+c}_{\alpha\beta}(i\omega_l,i\Omega_k) \right].
\label{sigmaabc} 
\end{eqnarray}
Here the contributions of individual diagrams {\it a, b,} and {\it c} 
have the form 
\bea
\label{ba}
&&B^{a}_{\alpha\beta}(i\omega_l,i\Omega_k,\bq)\nonumber \\
&&=
{1\over 2}T_{\alpha\gamma}
\left[{1\over \tau_s}\delta_{\gamma\delta}+({\tilde T}^{-1})_{\gamma\delta}
\right] T_{\delta\beta} 
\langle {\cal D}(i\omega_l,\bq)
 {\cal D}(i\omega_l,\bq)\rangle 
\nonumber \\
&& +{1\over 2}T_{\alpha\gamma}
T_{\gamma\beta}\langle {\cal D}(i\omega_l,\bq)\rangle,
\eea
where the matrix ${\tilde T}_{\alpha\beta}$ has the same form 
as $T_{\alpha\beta}$ with a replacement $\tau \to \tau_s$,
\be
\label{bb}
B^{b}_{\alpha\beta}(i\omega_l,i\Omega_k,\bq)=
-{1\over 2\tau_s}T_{\alpha\gamma}
T_{\gamma\beta} \langle {\cal D}(i\omega_l,\bq) 
{\cal D}(q,i\omega_l)\rangle , 
\ee
and
\bea
&&B^{c}_{\alpha\beta}(i\omega_l,i\Omega_k,\bq)=\nonumber \\
&&-{1\over 2}\left({1\over \tau_s}-{1\over \tau}\right)
T_{\alpha\gamma}
T_{\gamma\beta} \langle{\cal D}(i\omega_l,\bq) 
{\cal D}(i\omega_l,\bq)\rangle. \nonumber \\
\label{bc}
\eea
We see that although each of the expressions (\ref{ba}), (\ref{bb}),
and (\ref{bc})  depends on $\tau_s$, the
single-particle time disappears from the total contribution
of the Hikami-box,
\bea
B^{a+b+c}_{\alpha\beta}(i\omega_l,i\Omega_k,\bq)&=&
{1\over 2}T_{\alpha\beta}
\langle {\cal D}(i\omega_l,\bq)
 {\cal D}(i\omega_l,\bq)\rangle
\nonumber \\
&+&{1\over 2}T_{\alpha\gamma}
\langle {\cal D}(i\omega_l,\bq)\rangle T_{\gamma\beta}.
\label{babc}
\eea

The total correction to the conductivity tensor is obtained by
collecting the contributions (\ref{sigmae+d}), (\ref{sigmah}),
(\ref{sigmaf+g}), and (\ref{sigmaabc}). 
Carrying out the analytical continuation to real 
frequencies, we get
\bea
\delta\sigma_{\alpha\beta}(\Omega)&=& 
{\sigma_0\over i\pi\tau\Omega}\int_{-\infty}^{\infty} d\omega\,
\omega\coth{\omega\over2T} \nonumber \\
&\times&
[\Phi_{\alpha\beta}(\omega+\Omega,\Omega)-\Phi_{\alpha\beta}(\omega,\Omega)],
\label{e34}
\eea
where
\bea
\Phi_{\alpha\beta}(\omega,\Omega)&=&\Phi^{a+b+c}_{\alpha\beta}(\omega,\Omega)
+\Phi^{d}_{\alpha\beta}(\omega,\Omega)\nonumber \\
&+&\Phi^{f}_{\alpha\beta}(\omega,\Omega)+\Phi^{h}_{\alpha\beta}(\omega,\Omega).
\label{F-total}
\eea
We are interested in the case of zero external
frequency, $\Omega\to 0$, when Eq.~(\ref{e34})
can be rewritten as
\bea
\delta\sigma_{\alpha\beta}&=&-{\sigma_0\over i\pi\tau}
\int_{-\infty}^{\infty}d\omega \nonumber \\
&\times&\Phi_{\alpha\beta}(\omega,0) 
{\partial\over\partial\omega}\left[\omega\coth{\omega\over
2T}\right].
\label{anal-co}
\eea
Recalling the definition (\ref{F-def}) of $\Phi^\mu$, 
we finally arrive at
the following result
\begin{eqnarray} 
\delta\sigma_{\alpha\beta}&=&-2e^2 v_F^2\nu \int_{-\infty}^\infty 
\frac{d\omega}{2\pi} 
\frac{\partial}{\partial \omega} 
\left[\omega\  \coth{\omega\over{2T}}\right]\nonumber \\ 
&\times&\int \frac{d^2{\bf q}}{(2\pi)^2}\  {\rm Im} 
\left[\ U(\omega,{\bf q})\  B_{\alpha\beta}(\omega,{\bf q})\ \right], 
\nonumber \\
 \label{sigma} 
\end{eqnarray} 
where the tensor $B_{\alpha\beta}(\omega,{\bf q})$ 
is given by 
\begin{eqnarray} 
B_{\alpha\beta}(\omega,{\bf q})&=& 
{T_{\alpha\beta} \over 2}\langle {\cal D}{\cal D}\rangle+ 
T_{\alpha\gamma}\!\left({\delta_{\gamma\delta}\over 2}\langle {\cal D}\rangle  
- \langle n_{\gamma} {\cal D} n_{\delta} \rangle \right)\! T_{\delta\beta} 
\nonumber \\ 
&-&2 T_{\alpha\gamma} \langle n_{\gamma} {\cal D} n_{\beta} {\cal D} \rangle  
- \langle {\cal D}  n_{\alpha} {\cal D} n_{\beta} {\cal D} \rangle.
\label{Bwq} 
\end{eqnarray} 
The first term in (\ref{Bwq}) originates 
from the diagrams {\it a,b,c}, 
the second term from {\it a,f,g}, 
the third term from {\it h}, 
and the last one -- from {\it d} and {\it e}. 
We remind the reader that this result has been obtained under the
assumption $\tau_s\ll \tau,\:T^{-1}$; generalization to arbitrary 
$\tau_s/\tau$ and $\tau_s T$ will be considered in Sec.~\ref{IIb}. It
will be shown there that the conductivity correction retains the form 
(\ref{sigma}) in the general case but the expression (\ref{Bwq}) for
$B_{\alpha\beta}(\omega,\bq)$ is slightly modified.

\subsection{General case}
\label{IIb}

In the previous subsection we have derived the formula for 
the correction to the conductivity tensor for the case of a smooth
disorder (with $\tau_s\ll\tau$) 
assuming $\tau_s \ll T^{-1}$. Since characteristic momenta
$q$ and frequencies $\omega$ are set by the temperature, this
assumption implies $ql_s\ll 1$ and $\omega\tau_s \ll 1$. This allowed
us to simplify the calculation by neglecting the $q$ and $\omega$
dependence of Green's functions connecting  ballistic diffusons
and by considering only the ladders with many impurity lines.
Furthermore, we have used the small-angle nature of scattering 
when calculating the Hikami box contribution (\ref{babc}). 
We are now going to discuss the general case of arbitrary
$\tau_s/\tau$ and $T\tau_s$. 

It turns out that
the expressions (\ref{sigmae+d}), (\ref{sigmah}),
and (\ref{sigmaf+g}) for the contribution of the diagrams 
$d-h$ derived in the
case of a smooth disorder remain valid
in the general situation. The simplest way to show this is to  
use the following technical trick (cf. Refs.~\onlinecite{bmm01,gm02}).
One can add to the system an auxiliary weak smooth random potential with
a long transport scattering time $\tilde \tau \gg \tau$
but short single-particle ${\tilde\tau}_s\ll \tau_s$, 
such that $T{\tilde \tau}_s\ll 1$. 
This potential will not affect the quasiclassical dynamics and thus
should not change the result. On the other hand, it allows us (in view
of the condition  $T{\tilde \tau}_s\ll 1$)
to perform the gradient and frequency expansion in Green's functions
as was done in Sec.~\ref{IIa}. Adding such an auxiliary 
disorder amounts to a re-distribution
between quantum and quasiclassical degrees of freedom: all the information
about the real disorder is now contained in the ballistic propagators.  
It can be verified by a direct calculation (without using the 
additional disorder) that the above procedure yields the correct
result. 

It remains to consider the Hikami-box contribution (\ref{sigmaabc}).
When calculating it in Sec.~\ref{IIa}, we used the small-angle nature
of scattering implying that 
a single scattering line inserted
between two ballistic propagators approximately
preserves the direction of velocity,
$\langle{\cal D}W{\cal D}\rangle \to 
(2\pi\nu)^{-1}\langle{\cal D}{\cal D}\rangle/\tau_s$
and $\langle{\cal D}n_\alpha W n_\beta {\cal D}\rangle \to 
(2\pi\nu)^{-1}\langle{\cal D}{\cal D}\rangle(1/\tau_s-1/\tau)
\delta_{\alpha\beta}$. 
In the more general situation, when the 
scattering is at least partly of the large--angle character,
this is no longer valid and 
Eq.~(\ref{babc}) acquires a slightly more complicated form,  
\bea 
\label{hikami} 
&& B^{a+b+c}_{\alpha\beta}(\bq,i\omega_l,i\Omega_k) = \nonumber \\
&&
\pi\nu T_{\alpha\gamma}[ 
\langle {\cal D} {\cal S}_{\gamma\delta} {\cal D}\rangle  -  
2\langle {\cal D}n_{\gamma} W n_{\delta}{\cal D}\rangle]  
T_{\delta\beta}\nonumber \\
 &&+{1\over 2}T_{\alpha\gamma}
\langle {\cal D}\rangle T_{\gamma\beta}, 
\eea  
where ${\cal S}_{xx}={\cal S}_{yy}=W({\bf n},{\bf n'}), 
\quad {\cal S}_{xy}=-{\cal S}_{yx}=\omega_c/2\pi\nu$.

Summarizing the consideration in this subsection, 
in the general situation the interaction correction retains
the form (\ref{sigma}) with the tensor $B_{\alpha\beta}(\omega,{\bf q})$ 
given by 
\bea
\label{Bwq_general}
&&B_{\alpha\beta}(\omega,{\bf q}) =  T_{\alpha\gamma} \pi\nu[ 
\langle {\cal D} {\cal S}_{\gamma\delta} {\cal D}\rangle  -  
2\langle {\cal D}n_{\gamma} W n_{\delta}{\cal D}\rangle] T_{\delta\beta} 
\nonumber \\
&&+ 
T_{\alpha\gamma}\left({\delta_{\gamma\delta}\over 2}\langle {\cal D}\rangle  
- \langle n_{\gamma} {\cal D} n_{\delta} \rangle \right) T_{\delta\beta} 
\nonumber \\ 
&&-2 T_{\alpha\gamma} \langle n_{\gamma} {\cal D} n_{\beta} {\cal D} \rangle  
- \langle {\cal D}  n_{\alpha} {\cal D} n_{\beta} {\cal D} \rangle.
\eea
The correction $\delta\rho_{\alpha\beta}$ to the resistivity tensor is
then immediately obtained by using
$\delta\hat{\rho}=-\hat{\rho}\,\delta\hat{\sigma}\,\hat{\rho}$.  
This yields 
\begin{eqnarray} 
\delta\rho_{\alpha\beta}&=& {2\over e^2 v_F^2\nu }\int_{-\infty}^\infty 
\frac{d\omega}{2\pi} 
\frac{\partial}{\partial \omega} 
\left[\omega\  {\rm coth}{\omega\over{2T}}\right]\nonumber \\ 
&\times&\int \frac{d^2{\bf q}}{(2\pi)^2}\  {\rm Im} 
\left[\ U(\omega,{\bf q})\  B^{(\rho)}_{\alpha\beta}(\omega,{\bf q})\ \right], 
\nonumber \\
 \label{rho} 
\end{eqnarray} 
where the tensor $B^{(\rho)}_{\alpha\beta}$ is related to
$B_{\alpha\beta}$, Eq.~(\ref{Bwq_general}), via  
\be
 B^{(\rho)}_{\alpha\beta} = (T^{-1})_{\alpha\gamma} 
B_{\gamma\delta} (T^{-1})_{\delta\beta}.
\label{Brho}
\ee
Explicitly, corrections to the components of the resistivity tensor are 
expressed through $\delta\sigma_{xx}=\delta\sigma_{yy}$ and 
$\delta\sigma_{xy}=-\delta\sigma_{yx}$ as follows
\bea
\label{rhoxx-sigma}
\delta\rho_{xx}&=&\rho_0^2
[(\omega_c^2\tau^2-1)\delta\sigma_{xx}+2\omega_c\tau \delta\sigma_{xy}],\\
\delta\rho_{xy}&=&\rho_0^2
[(\omega_c^2\tau^2-1)\delta\sigma_{xy}-2\omega_c\tau \delta\sigma_{xx}].
\label{rhoxy-sigma}
\eea

Note that the results (\ref{Bwq}),\ (\ref{Bwq_general}) for 
$B_{\alpha\beta}(\omega,{\bf q})$  satisfy the requirement 
\be
B_{\alpha\beta}(\omega,0)=0,
\label{B(q=0)}
\ee
as follows from 
$$\left.\langle n_{\alpha} {\cal D} n_{\beta} \rangle\right|_{q=0}=
\sigma_{\alpha \beta}(\omega)/2e^2\nu v_F^2$$ 
and 
$\left.(2\pi)^{-1}\!\int d\phi\ {\cal D}(\phi,\phi')\right|_{q=0}=i/\omega.$
The condition (\ref{B(q=0)}) implies that spatially homogeneous fluctuations
in the potential do not change the conductivity, see 
Refs.~\onlinecite{ZNA-sigmaxx},\onlinecite{Ka-And} for discussion.

\subsection{Limiting cases}
\label{IIc}

Having obtained the general formula, we
will now demonstrate that it reproduces, in the appropriate limits, 
the previously known results for the interaction correction.
Specifically, in Sec.~\ref{IIc1} we will consider the diffusive limit 
$T\tau\ll 1$ studied in Refs.~\onlinecite{AA,SenGir,girvin82}, 
while Sec.~\ref{IIc2} is
devoted to the $B\to 0$ case with a white-noise disorder addressed in
Refs.~\onlinecite{ZNA-sigmaxx,ZNA-rhoxy}. 
In Sec.~\ref{IIc3} we will analyze how the linear-in-$T$
asymptotics of $\delta\sigma(B=0)$ in the ballistic regime obtained in
Ref.~\onlinecite{ZNA-sigmaxx} for a white-noise disorder depends on the character of the
random potential.

\subsubsection{Diffusive limit}
\label{IIc1}

We begin by considering the diffusive limit $T\tau\ll 1$ in which  
we reproduce (for arbitrary $B$ and disorder range)  
the logarithmic correction (\ref{AAdiff}), (\ref{MRAA}) 
determined by the diagrams {\it a-e}. 
Let us briefly outline the corresponding calculation. 
The propagator for $ql, \ \omega\tau \ll 1$ can be decomposed 
as ${\cal D}={\cal D}^{\rm s}+{\cal D}^{\rm reg}$,
where ${\cal D}^{\rm s}$ is singular,
while ${\cal D}^{\rm reg}$ is finite (regular) at $q, \ \omega \to 0$,
see e.g. Refs.~\onlinecite{woelfle84,bhatt85}.
The singular contribution is governed by the 
diffusion mode and has the form [see Eq.~(\ref{diff-prop})]
\begin{eqnarray}
&&{\cal D}^{\rm s}(\omega,\bq;\phi,\phi')
\simeq {\Psi_R(\phi,\bq)\Psi_L(\phi',\bq) 
\over
  Dq^2-i\omega}, 
  \label{dmode}\\
&&\Psi_{\nu}(\phi,\bq) =  1 - i c^{(1)}_{\nu} \cos(\phi-\phi_q) - 
i c^{(2)}_{\nu}\sin(\phi-\phi_q), \nonumber
\label{dmode-Psi-nu}
\end{eqnarray}
where 
$D=v_F^2\tau/2(1+\omega_c^2\tau^2)$ is the diffusion constant in the
presence of a magnetic field and   
\begin{eqnarray}
c^{(1)}_{R}(q)=c^{(1)}_{L}(q)&=& {qv_F \tau \over 1+\omega_c^2\tau^2}, \\
c^{(2)}_{R}(q)=-c^{(2)}_{L}(q)&=& 
{qv_F \omega_c \tau^2 \over 1+\omega_c^2\tau^2}. 
\end{eqnarray}

The leading-order contribution of the diagrams $a,b$ and $c$ (that
containing two singular diffusons $\cal D^{\rm s}$) is
exactly canceled by the part of the diagrams $d$ and $e$ with the
structure  
$\langle {\cal D}^{\rm s}n_\alpha {\cal D}^{\rm reg} 
n_\beta {\cal D}^{\rm s}\rangle$,
i.e. with one regular part of the propagator inserted between two
singular diffusons, $\langle {\cal D}^{\rm s}\rangle=(Dq^2-i\omega)^{-1}$. 
Indeed, in view of $\langle n_\alpha {\cal D}^{\rm reg} n_\beta\rangle 
= {1\over 2}T_{\alpha\beta}$,
the latter contribution reduces to 
$-{1\over 2}\langle {\cal D}^{\rm s} \rangle^2 T_{\alpha\beta}$,
while the diagrams $a,b$ and $c$ yield
\bea
&&{1\over 2}\langle {\cal D}^{\rm s}\rangle^2 T_{\alpha\gamma}
\left[ {\delta_{\gamma\delta}\over \tau}
+\omega_c \epsilon_{\gamma\delta} \right] T_{\delta\beta}\nonumber \\
&&={1\over 2}\langle {\cal D}^{\rm s} \rangle^2 T_{\alpha\beta},
\label{cancelabc}
\eea
where $\epsilon_{\alpha\beta}$ is the antisymmetric tensor,
$\epsilon_{xx}=\epsilon_{yy}=0,\ \epsilon_{xy}=-\epsilon_{yx}=1$.

It remains thus to calculate only the contribution
of the diagrams $d+e$ with three singular diffusons,  
\begin{eqnarray}
\delta\sigma_{\alpha\beta}
&=&{e^2v_F^2\over 2\pi}
\int_{-\infty}^\infty d\omega {\partial \over \partial \omega} 
\left[\omega\  {\rm coth}{\omega\over 2T}\right] \nonumber \\
&\times&
\int{d^2q\over (2\pi)^2}{\rm Im}
\frac{\langle {\cal D}^{\rm s}n_\alpha {\cal D}^{\rm s} 
n_\beta {\cal D}^{\rm s}\rangle}
{1+i\omega\langle{\cal D}^{\rm s}\rangle}\nonumber \\
&\simeq& {2e^2v_F^2\over \pi(1+\omega_c^2\tau^2)^2}\int_T^{1/\tau}
 d\omega  \nonumber \\
&\times&
\int{d^2q\over (2\pi)^2}{\rm Im}\frac{(-iq_\alpha l)(-iq_\beta l)}{Dq^2(Dq^2-i\omega)^2} 
\nonumber \\ 
&=& {e^2\over 2\pi^2}\ln(T\tau)\delta_{\alpha\beta}, 
\label{GMdiff}
\end{eqnarray}
in agreement with \cite{AA,SenGir,girvin82}.
The result for a point-like interaction differs only
by a factor $\nu V_0$.

\subsubsection{$B\to 0$, white-noise disorder}
\label{IIc2}

We allow now for arbitrary $T\tau$ but consider the limit of zero
magnetic field assuming a white-noise disorder 
($\tau=\tau_s$ and $W({\bf n},{\bf n'})=1/2\pi\nu\tau$),
which is the limit studied in Refs.~\onlinecite{ZNA-sigmaxx,ZNA-rhoxy}. 
The contribution
(\ref{hikami}) of the diagrams {\it a,b,c} takes for the white-noise
disorder the form 
\bea
&& B^{a+b+c}_{\alpha\beta} =  {1\over 2}
T_{\alpha\gamma}\left[\langle {\cal D}\rangle \langle 
{\cal D}\rangle{\delta_{\gamma\delta}\over \tau} + 
\omega_c\langle {\cal D \cal D}\rangle\epsilon_{\gamma\delta} \right.
\nonumber \\
&&\left.-{2\over \tau}\langle {\cal D}n_{\gamma} 
\rangle \langle n_{\delta}{\cal D}\rangle 
+\delta_{\gamma\delta}
\langle {\cal D}\rangle \right]T_{\delta\beta}.
\label{WNHik}
\eea
Using now the explicit form of the ballistic propagator 
for the case of white-noise disorder and $B\to 0$
[Eqs.~(\ref{prop-WN}), (\ref{g0-def}), (\ref{D0B=0}), 
(\ref{g0B=0}), and (\ref{g_0(B)})]
we recover the results for $\delta\sigma_{xx}$  and $\delta\rho_{xy}$ 
obtained in a different way in Refs.~\onlinecite{ZNA-sigmaxx} 
and ~\onlinecite{ZNA-rhoxy}, 
see Appendix \ref{A2}.

\subsubsection{$B=0$, ballistic limit $T\tau\gg 1$}
\label{IIc3}

In the ballistic limit $T\tau\gg1$ and for white-noise disorder the
result of Ref.~\onlinecite{ZNA-sigmaxx} 
(recovered in Sec.~\ref{IIc2} and Appendix~\ref{A2}) yields a
linear-in-$T$ conductivity correction,
$\delta\sigma=(2\nu V_0 e^2/\pi)T\tau$ for the point-like interaction
and $\delta\sigma=(e^2/\pi)T\tau$ for the Coulomb interaction. 
The question we address in this subsection is how this behavior
depends on the nature of disorder [i.e. on the scattering
cross-section $W(\phi)$]. 

In order to get the $T\tau$ ballistic asymptotics, it is sufficient to
keep contributions to (\ref{Bwq_general}) 
with a minimal number of scattering processes. Specifically, the
propagator ${\cal D}$ in the first and the third terms of (\ref{Bwq_general})
can be replaced by the free propagator, 
\be
{\cal D}_{\rm f}(\omega,\bq;\phi,\phi')={2\pi\delta(\phi-\phi')\over 
-i(\omega+i0)+iqv_F\cos(\phi-\phi_q)},
\label{D-free}
\ee
while in the second
term it should be expanded up to the linear term in the scattering
cross-section $W$ [the second
term produces then the same contribution as the first term in (\ref{Bwq_general})]. 
The last (fourth) term in (\ref{Bwq_general}) does
not contribute to the $T\tau$ asymptotics. We get therefore
\bea
B_{xx} &\simeq &2\pi\nu\tau^2(\langle{\cal D}_{\rm f} W {\cal D}_{\rm f}\rangle -
2\langle{\cal D}_{\rm f} n_x W n_x {\cal D}_{\rm f}\rangle)\nonumber \\
&-& 2\tau \langle n_x{\cal D}_{\rm f} n_x{\cal D}_{\rm f}\rangle. \label{Bwq_ball}
\eea
Let us consider first the case of a short-range interaction, $U_0(r)=V_0$.
The structure of Eqs.~(\ref{sigma}), 
(\ref{Bwq_ball}) implies that the interaction correction is governed by 
returns of a particle to the original point in a time $t\lesssim
T^{-1}\ll \tau$ after a single scattering
event. It follows that the coefficient in front of the linear-in-$T$
term is proportional to the backscattering probability
$W(\pi)=\tilde{w}(2k_F)$,
\be
\label{sigma_ball} 
\delta\sigma_{xx}={2\nu V_0 e^2\over \pi} 2\pi\nu W(\pi) T\tau^2.
\ee
As shown in Appendix~\ref{A3}, this result remains valid in the case of
Coulomb interaction, with the factor $2\nu V_0$ replaced by
unity. This shows that in the ballistic limit the Coulomb interaction
is effectively reduced to the statically screened form,
$U(r)=1/2\nu$ when the leading contribution
to $\delta\sigma_{xx}$ is calculated. 
According to (\ref{sigma_ball}),  in a smooth disorder with 
a correlation length $d\gg k_F^{-1}$ the $T\tau$ contribution is
suppressed by an exponentially small factor
$2\pi\nu\tau \tilde{w}(2k_F)\sim e^{-k_F d}$. In fact, for a smooth disorder
the linear term represents the leading contribution for $T\tau_s\gg 1$
only. In the intermediate range $\tau^{-1} \ll T \ll \tau_s^{-1}$ 
the dominant return processes are due to many small-angle scattering
events. However, the corresponding return probability is also 
exponentially suppressed $\sim \exp(-{\rm const}\: \tau/t)$ 
for relevant (ballistic) times $t\ll \tau$, 
yielding a contribution  
$\delta\sigma_{xx}\sim \exp[-{\rm const}(T\tau)^{1/2}]$. Thus, the 
interaction correction in the ballistic regime
is exponentially small at $B=0$ for the case of 
smooth disorder. Moreover, the same argument applies to the case of
a non-zero $B$, as long as\cite{2pi} $\omega_c \ll T$. 

In any realistic system there will be a finite concentration of
residual impurities located close to the electron gas plane and
inducing large-angle scattering processes. In other words, a
realistic random potential can be thought as a superposition of a
smooth disorder with a transport time $\tau_{\rm sm}$ and a
white-noise disorder characterized by a time $\tau_{\rm
wn}$. Neglecting the exponentially small contribution of the smooth
disorder to the linear term, we then find that the ballistic asymptotics
(\ref{sigma_ball}) of the interaction correction takes the form
\be
\label{sigma_ball_mixed}
\delta\sigma={e^2\over \pi} {\tau\over \tau_{\rm wn}} T\tau
\times\left\{ \begin{array}{ll} 2\nu V_0, & \ \ \ {\rm point-like,}\\
                                1,       & \ \ \ {\rm Coulomb,}
\end{array}\right.
\ee
where $\tau^{-1}=\tau_{\rm sm}^{-1}+\tau_{\rm wn}^{-1}$ is the
total transport scattering rate. If the transport is dominated by the
smooth disorder, $\tau_{\rm wn}\gg \tau_{\rm sm}$, the coefficient of
the $T\tau$ term is thus strongly reduced as compared to the white-noise
result of Ref.~\onlinecite{ZNA-sigmaxx}. 

Finally, it is worth mentioning that in addition to the $T\tau$
term corresponding to the lower limit $\omega\sim T$ of the
frequency integral in (\ref{sigma}), there is a much larger but
$T$-independent contribution $\delta\sigma \propto E_F\tau$
governed by the upper limit $\omega\sim E_F$. This contribution
is just an interaction-induced Fermi-liquid-type renormalization
of the bare (noninteracting) Drude conductivity.

\section{Strong B, smooth disorder}
\label{III}
\setcounter{equation}{0}

\subsection{Quasiclassical dynamics}
\label{IIIa} 

We have shown in Sec.~\ref{IIc} that due to small-angle nature of
scattering in a smooth disorder the interaction correction is
suppressed in the ballistic regime $T\tau\gg 1$ 
in zero (or weak) magnetic field. 
The situation changes qualitatively in a strong magnetic field, 
$\omega_c\tau\gg 1$ and $\omega_c \gg T$. 
The particle experiences then within the time $t\sim T^{-1}$ multiple 
cyclotron returns to the region close to the starting point. The 
corresponding ballistic propagator satisfies the equation (\ref{LB})
with the collision term (\ref{Csmooth}). 

The solution of this equation in the limit of a strong magnetic field, 
$\omega_c\tau\gg 1$, is presented in 
Appendix~\ref{A4}. For calculation of the leading order contribution to
$\delta\sigma_{xx}$ and $\delta\rho_{xx}$, the following approximate form 
is sufficient:
\bea 
&&{\cal D}(\omega,\bq;\phi,\phi') = \exp[-iqR_c(\sin\phi-\sin\phi')]
\nonumber \\
&&\times \left[ \frac{\chi(\phi)\chi(\phi')}{Dq^2-i\omega} +
 \sum_{n \neq 0}  
\frac{e^{i n(\phi-\phi')}}{Dq^2-i(\omega-n\omega_c)+n^2/\tau} \right],
\nonumber \\
&&\equiv {\cal D}^{\rm s}(\omega,\bq;\phi,\phi') +
{\cal D}^{\rm reg}(\omega,\bq;\phi,\phi'),
\label{diffuson}
\eea 
where $\chi(\phi)=1-iqR_c\cos\phi/\omega_c\tau$ and
$D\simeq{R_c^2/2\tau}$, and the polar angles of velocities
are counted from the angle of $\bq$. 
Equation (\ref{diffuson}) is valid under the assumption 
$(q R_c)^2 \ll \omega_c\tau$. We will see below that the
characteristic momenta $q$ are determined by the condition
$Dq^2 \sim \omega\sim T$, so that the above assumption is justified
in view of $\omega_c \gg T$.
Furthermore, this condition allows us to keep only the first
(singular) term  ${\cal D}^{\rm s}$
in square brackets in (\ref{diffuson}) when 
calculating $\langle{\cal D}\rangle$, 
\be 
\langle {\cal D} \rangle = {J_0^2(qR_c)\over Dq^2-i\omega}, 
\label{<D>} 
\ee 
where $J_0(z)$ is the Bessel function. 
Moreover, the formula (\ref{Bwq}) for $B_{xx}$ can be cast in a form linear in
${\cal D}$ by using
\bea
\langle {\cal D} {\cal D}\rangle &=& 
-i \frac{\partial }{\partial \omega}\langle {\cal D} \rangle,
\label{DD}\\
\langle n_\alpha {\cal D} n_\beta {\cal D}\rangle &=&
{i \over v_F} 
\frac{\partial }{\partial q_\beta}\langle n_\alpha{\cal D}\rangle,
\label{nDnD}\\
\langle {\cal D} n_x {\cal D} n_x {\cal D}\rangle &=&
-{1 \over 2 v_F^2} \frac{\partial^2 }{\partial q_x^2}\langle {\cal D} \rangle.
\label{DnDnD}
\eea
Therefore, it is again sufficient to take into account only the first term in 
(\ref{diffuson}) for calculation of $B_{xx}$ if the identities
(\ref{DD}),\ (\ref{nDnD}), and (\ref{DnDnD}) are used.
(Of course,  $B_{xx}$ can also be evaluated directly from
Eq.~(\ref{Bwq}), but then the second (regular) term ${\cal D}^{\rm reg}$
in (\ref{diffuson}) has to be included.)
Combining all four terms in (\ref{Bwq}), we get 
\bea 
B_{xx}(\omega,q)&=&{J_0^2(qR_c)\over (\omega_c\tau)^2} 
\frac{D\tau q^2 }{(Dq^2-i\omega)^3}\nonumber \\
&=&\frac{4\tau^3}{\beta^2}\frac{J_0^2(Q) Q^2 }{(Q^2-i\Omega)^3}. 
\label{Bxx} 
\eea 
In the second line we introduced dimensionless variables
$Q=qR_c, \ \Omega=2\omega\tau, \ \beta=\omega_c\tau$.

Note that Eqs.~(\ref{<D>}), (\ref{Bxx}) differ from those obtained in 
the diffusive regime by the factor ${J_0^2(qR_c)}$ only. 
This is related to the fact that the motion of the guiding center is 
diffusive even on the ballistic time scale $t\ll\tau$ (provided 
$t\gg\omega_c^{-1}$), while the additional factor 
corresponds to the averaging over the cyclotron orbit
(see Sec.~\ref{IV} below). 

We turn now to the calculation of $B_{xy}$. Substituting 
(\ref{diffuson}) in (\ref{Bwq}), we classify the obtained contributions
according to powers of the small parameter $1/\beta$. The
leading contributions are generated by the first and the last terms in
(\ref{Bwq}) and are of order $1/\beta$, i.e. larger by factor $\beta$
as compared to $B_{xx}$, Eq.~(\ref{Bxx}). (This extra
factor of $\beta$ is simply related to 
$|\sigma_{xy}|/\sigma_{xx}=\beta$.)
However, these leading contributions cancel,
\bea
&&\left.\left[{T_{xy}\over 2}\langle {\cal D} {\cal D}\rangle 
- \langle {\cal D} n_x {\cal D} n_y {\cal D}\rangle  
\right]\right|_{{\rm order}\  1/\beta} \nonumber \\ 
&=&
{\tau \over 2\beta}\langle {\cal D}^{\rm s}{\cal D}^{\rm s}\rangle-
\langle {\cal D}^{\rm s} n_x {\cal D}^{\rm reg} n_y {\cal D}^{\rm s}\rangle  
\label{1-over-beta} \\
&=& -{2 \tau^3 \over \beta}\frac{J_0^2(Q)}{(Q^2-i\Omega)^2}+
{2 \tau^3 \over \beta}\frac{J_0^2(Q)}{(Q^2-i\Omega)^2}=0,
\nonumber 
\eea
as in the diffusive limit, see the text above Eq.~(\ref{cancelabc}).

To evaluate terms of higher order in $1/\beta$, we need a more 
accurate form of the propagator (\ref{diffuson}). Since the contributions
of order $1/\beta^2$ to $B_{xy}$ turn out to cancel as well, we have to
know the propagator with the accuracy allowing to evaluate the terms of order 
$1/\beta^3$. To simplify the calculation, we use again the identities
(\ref{DD}) and (\ref{nDnD}). As to Eq.~(\ref{DnDnD}), it cannot be 
generalized onto $xy$-component of the tensor, and we use instead
\be
\langle {\cal D} n_x {\cal D} n_y {\cal D}\rangle =
{i \over v_F}\left\langle \frac{\partial{\cal D} }{\partial q_x}
  n_y{\cal D}\right\rangle.
\label{DxDyD}
\ee
It is then sufficient to calculate the propagator ${\cal D}$
up to the $1/\beta^2$ order. This is done in Appendix~\ref{A4},
see Eqs.~(\ref{Dsing-2order})-(\ref{PsiRL-n-2}). 
Substituting this result for ${\cal D}$ in Eq.~(\ref{Bxx})
and combining all terms, we get after some algebra
\bea 
B_{xy}(\omega,q)
&=& -\frac{\tau^3}{\beta^3}\left[\ \frac{7 Q^2 J_0^2(Q) + 
4 Q J_0(Q)J_1(Q)}{(Q^2-i\Omega)^2}\right.\nonumber \\
&+& \left. \frac{4 Q J_0(Q)J_1(Q))}{Q^2-i\Omega}\  \right].
\label{Bxy} 
\eea 
We see that similarly to (\ref{Bxx}) the kernel  $B_{xy}(\omega,q)$ 
has a diffusive-type structure with $Q^2-i\Omega$ in denominator
reflecting
the diffusion of the guiding center, while the Bessel functions
describe the averaging over the cyclotron orbit.
Clearly, both kernels (\ref{Bxx}) and (\ref{Bxy}) vanish at $q=0$, 
as required by
(\ref{B(q=0)}).

\subsection{Point-like interaction}
\label{IIIb}

To find the interaction correction to the conductivity, we 
have to substitute
Eqs.~(\ref{Bxx}) and (\ref{Bxy}) in the formula
(\ref{sigma}). We consider first the simplest situation, when the interaction
$U(\omega,q)$ in (\ref{sigma}) is of point-like character, $U(\omega,q)=V_0$.
Using $v_F^2 q dq = \omega_c^2 Q dQ$, we 
see that all
the $B$-dependence drops out from $\delta\sigma_{xx}$, 
and the exchange contribution reads 
\bea 
\delta\sigma_{xx}&=&-8e^2 \nu V_0 
\int_{0}^\infty 
\frac{d\Omega}{2\pi} 
\frac{\partial}{\partial \Omega} 
\left[ \Omega\  {\rm coth}{\Omega\over{4T\tau}}\right] \nonumber  \\
&\times& \int_0^\infty\frac{Q dQ}{2\pi}\  
{\rm Im} \frac{J_0^2(Q) Q^2 }{(Q^2-i\Omega)^3}. 
\label{sxxQW}
\eea 
To simplify the result (\ref{sxxQW}), it is convenient to perform
a Fourier transformation with respect to $\Omega$ (which corresponds
to switching to the time representation)
\bea
&&{\rm Im}\int_{0}^\infty\frac{d\omega}{2\pi}\: F(\omega)\: \frac{\partial}{\partial \omega} 
\left[\omega\ {\rm coth}{\omega\over{2T}}\right] \nonumber \\
&&=
\int_0^\infty dt \frac{\pi T^2 t}{{\rm sinh}^2(\pi T t)} {\tilde F}(t).
\label{Fourier}
\eea
The integral over $Q$ is then 
easily evaluated, yielding
\bea 
\delta\sigma_{xx}&=&-{e^2\over 2\pi^2}\nu V_0 G_0(T\tau),  
\label{deltaint} 
\\ 
G_0(x)&=&\pi^2 x^2 \!\! \int_0^\infty \!
\frac{du  \exp(-1/u)}{{\rm sinh}^2(\pi x u)} \label{G0(x)}\\
&\times& \left[(u-1)I_0(1/u)+I_1(1/u)\right],   
\nonumber 
\eea 
where $I_0(z)$ and $I_1(z)$ are modified Bessel functions.
The Hartree term in this case is of the opposite sign and twice larger 
due to the spin summation (we neglect here the Zeeman splitting
and will return to it later).

\begin{figure} 
\includegraphics[width=8cm]{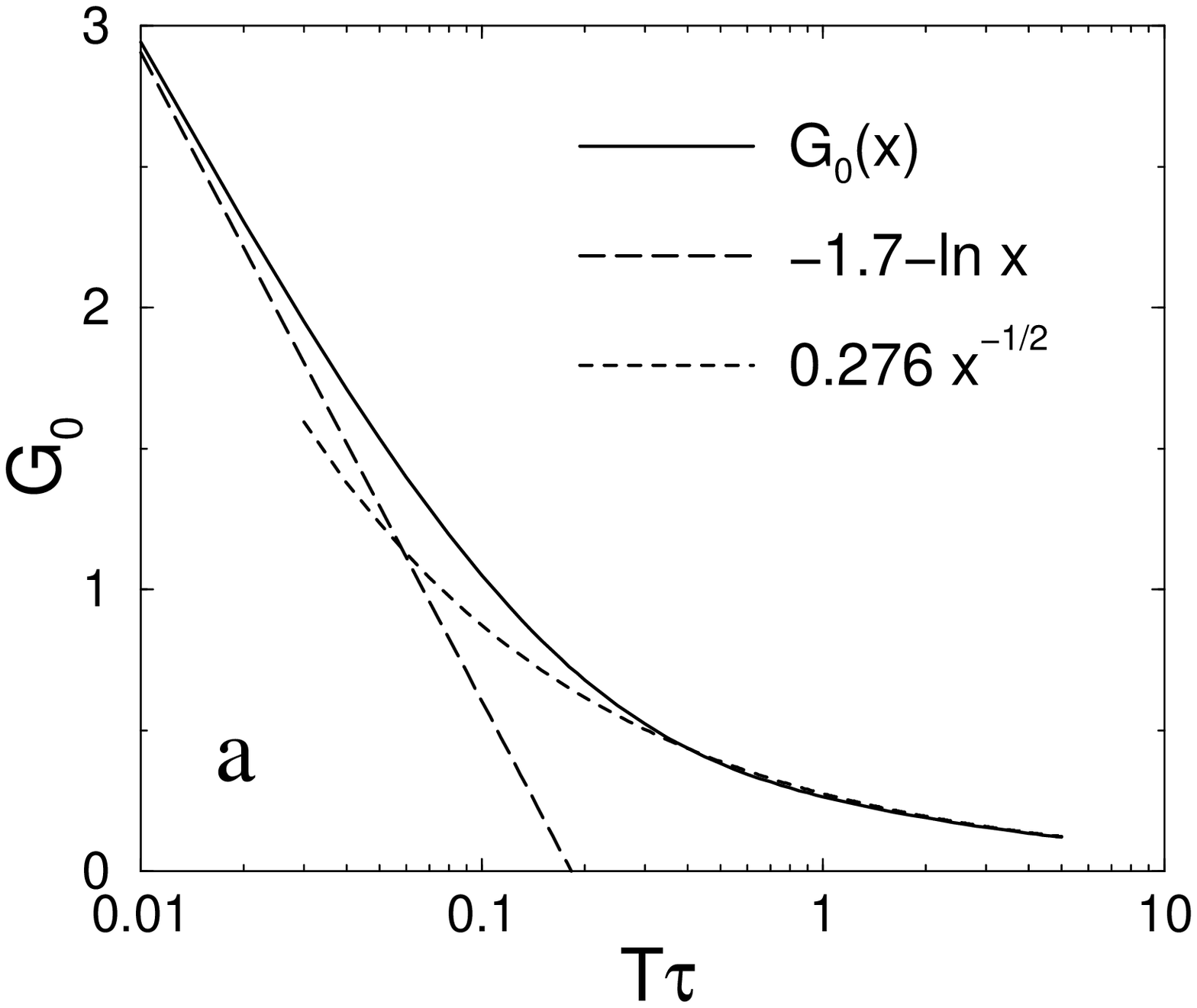} 
\vspace{2mm} 
\includegraphics[width=8cm]{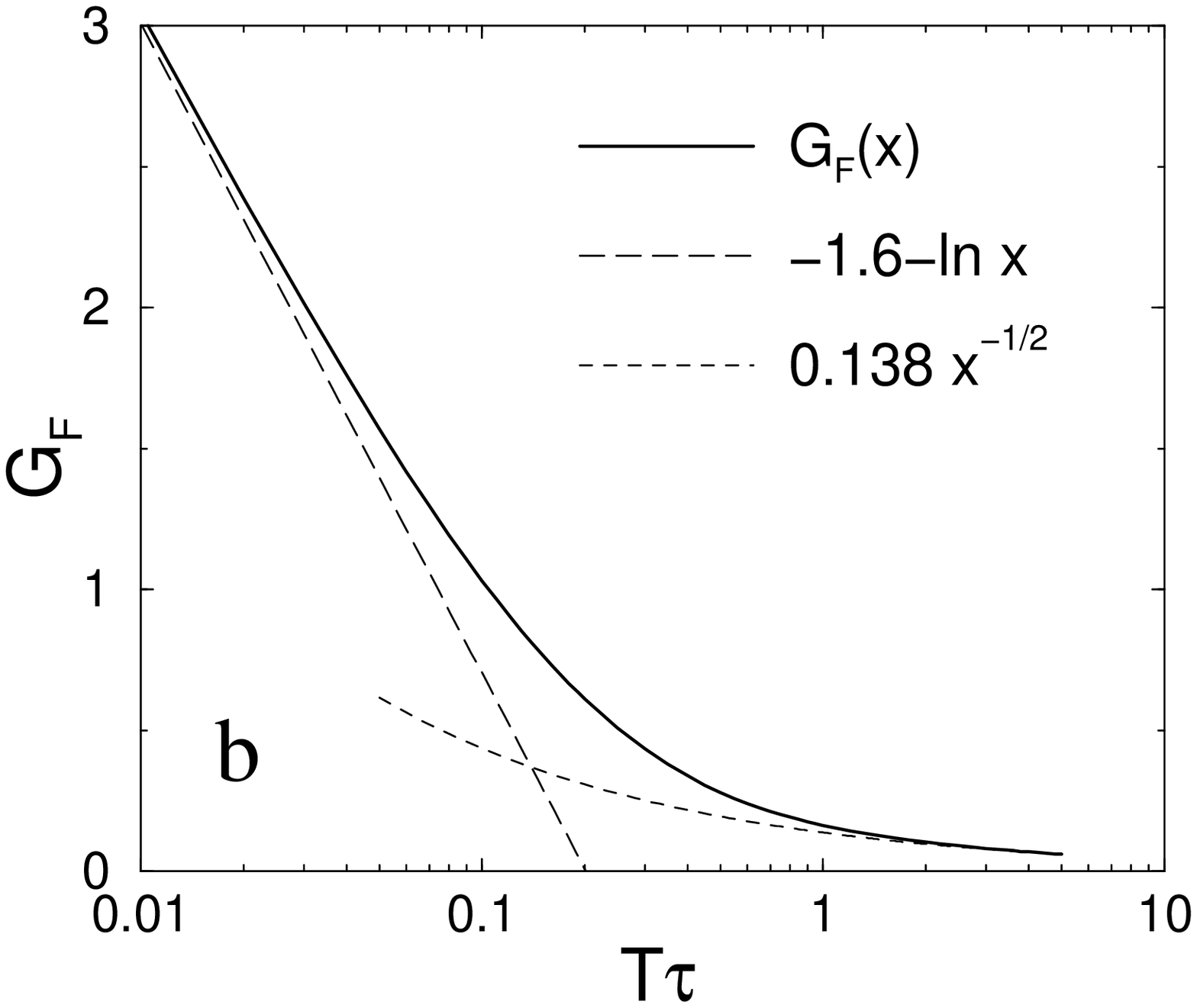} 
\vspace{2mm} 
\caption{Functions $G_0(T\tau)$ (a) and $G_{\rm F}(T\tau)$ (b)
determining the $T$-dependence of
the exchange term
for point-like, Eq.~(\ref{deltaint}), and Coulomb, Eq.~(\ref{exchange}),
interaction, respectively. 
Diffusive and ballistic asymptotics,
Eq.~(\ref{G0}) and Eq.~(\ref{GF}), are also shown.
} 
\label{fig2} 
\end{figure} 

It follows from Eqs.~(\ref{Bxx}) and (\ref{Bxy}) that the correction
to the Hall conductivity is smaller by the factor $(\omega_c\tau)^{-1}$
as compared to (\ref{deltaint}). This implies, according to
(\ref{rhoxx-sigma}) that in a strong magnetic field
the correction to the longitudinal resistivity is governed by 
$\delta\sigma_{xx}$,
\be
{\delta\rho_{xx}\over \rho_0}=
(\omega_c\tau)^2{\delta\sigma_{xx}\over \sigma_0},
\label{parabola}
\ee
similarly to the diffusive limit (\ref{MRAA}).
In fact, it turns out that the relation (\ref{parabola})
holds in a strong magnetic field, $\omega_c \gg T,$
for arbitrary disorder and interaction, see below.
On the other hand, as is seen from (\ref{rhoxy-sigma}),
contributions of both 
$\delta\sigma_{xx}$ and $\delta\sigma_{xy}$ to $\delta\rho_{xy}$
are of the same order in $(\omega_c\tau)^{-1}$.
We will return to the calculation of $\delta\rho_{xy}$
in Sec.~\ref{IIIg}.

The MR $\rho_{xx}(B)$ is thus quadratic in $B$, with the temperature 
dependence determined by the function $G_0(T\tau)$, 
which is shown in 
Fig.~\ref{fig2}a. 
In the diffusive  ($x\ll 1$)  and ballistic ($x\gg 1$) limits
the function $G_0(x)$ has the following asymptotics
\be
\label{G0}
G_0(x)\simeq \left\{\begin{array}{ll}  
-\ln x + {\rm const}, & \ \ \ x\ll 1,\\  
c_0x^{-1/2}, & \ \ \ x \gg 1,
\end{array}\right. 
\ee
with 
\be
c_0={3\zeta(3/2)\over 16\sqrt{\pi}}\simeq 0.276
\label{c0}
\ee
(here $\zeta(z)$ is the Riemann zeta-function).
Let us note that the crossover between the two limits takes 
place at numerically small values $T\tau\sim 0.1$ (a similar observation 
was made in Refs.~\onlinecite{ZNA-sigmaxx,ZNA-rhoxy}).  
This can be traced back to the fact 
that the natural dimensionless variable in (\ref{deltaint})
is $2\pi T\tau$.

\subsection{Coulomb interaction, exchange}
\label{IIIc}

For the case of the Coulomb interaction the result turns out to be 
qualitatively similar. Substituting (\ref{<D>}) in (\ref{screen})
and neglecting the first term $q \sim (T/D)^{1/2}\ll \kappa$ in the 
denominator of (\ref{screen}), 
we obtain the effective interaction
\be
U(\omega,\bq)={1\over 2\nu}\frac{Q^2-i\Omega}{Q^2-i\Omega[1-J_0^2(Q)]}.
\label{U-RPA-B}
\ee
Inserting (\ref{U-RPA-B}) and 
(\ref{Bxx}) into (\ref{sigma}), we get the exchange (Fock) 
contribution 
\bea 
\delta\sigma^{\rm F}_{xx}&=&-{e^2 \over \pi^2} \int_{0}^\infty 
d\Omega \frac{\partial}{\partial \Omega} 
\left[ \Omega\  {\rm coth}{\Omega\over{4T\tau}}\right] \nonumber  \\
&\times&{\rm Im} \int_0^\infty \! \! QdQ \
 \frac{Q^2 J_0^2(Q)  }{(Q^2-i\Omega[1-J_0^2(Q)])(Q^2-i\Omega)^2}. 
\nonumber \\
\label{CoulsxxQW}
\eea 
Using (\ref{parabola}) we find the MR
\bea 
&&{\delta\rho^{\rm F}_{xx}(B)\over \rho_0} = 
-{(\omega_c\tau)^2\over \pi k_Fl} 
G_{\rm F}(T\tau), 
\label{exchange} \\
&&G_{\rm F}(x) =32 \pi^2 x^2
\int_0^\infty dQ Q^3 J_0^2(Q) 
\label{GFx} \\ 
&&\times\sum_{n=1}^\infty  
\frac{n(12\pi xn[1-J_0^2(Q)]+[3- J_0^2(Q)]Q^2)} 
{(4\pi x n+Q^2)^3 (4 \pi x n[1-J_0^2(Q)]+Q^2)^2}.
\nonumber 
\eea 
Note that in contrast to the case of a point-like interaction,
a transformation to the time representation does not allow us to
simplify (\ref{CoulsxxQW}), since the resulting $Q$-integral can not be 
evaluated analytically. We have chosen therefore to perform 
the $\Omega$-integration, which results in an infinite sum 
(\ref{GFx}). This amounts to returning to the Matsubara 
(imaginary frequency) representation and is convenient for the purpose of
numerical evaluation of $G_{\rm F}(x)$. 
In the diffusive  ($x\ll 1$) and ballistic ($x\gg 1$) limits
this function has the asymptotics
\be
\label{GF}
G_{\rm F}(x)\simeq \left\{\begin{array}{ll}  
\displaystyle
-\ln x + {\rm const}, & \ \ \ x\ll 1,\\  
\displaystyle
{c_0\over 2}x^{-1/2}, & \ \ \ x \gg 1,
\end{array}\right.
\ee
and is shown in Fig.~\ref{fig2}b.

\subsection{Coulomb interaction, Hartree contribution}
\label{IIId}

\begin{figure}
  \medskip
\centerline{\includegraphics[width=8cm]{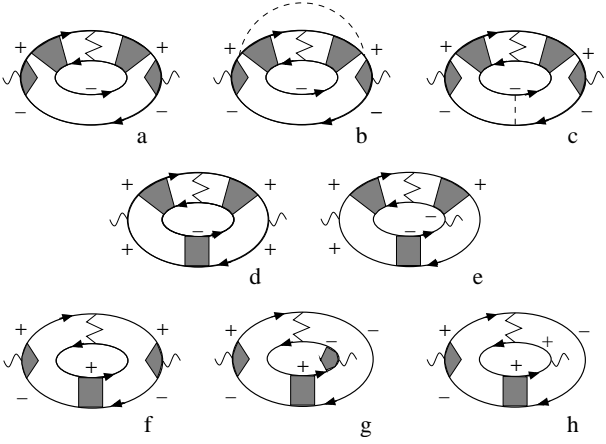}}
\caption{ Hartree diagrams for the interaction  
correction to $\sigma_{\alpha\beta}$. The diagrams are labeled
in the way as their exchange counterparts in Fig.~\ref{fig1}.
The diagrams obtained by a flip  
and/or by an exchange $+ \leftrightarrow -$ should also 
be included.} 
\label{hartree-dia} 
\end{figure}

We turn now to the Hartree contribution.
The corresponding diagrams can be generated in a way similar to exchange
diagrams (Sec.~\ref{IIa}) but in this case one should start from two
electron bubbles connected by an interaction line.
There are again two distinct ways to generate a skeleton diagram:
two current vertices can be inserted either both in the same bubble or in
two different bubbles. Then one puts signs of Matsubara frequencies in all
possible ways and insert ballistic diffusons correspondingly.
The obtained set of diagrams is shown in Fig.~\ref{hartree-dia}
There is one-to-one correspondence between these Hartree diagrams
and the exchange diagrams of Fig.~\ref{fig1}, which is
reflected in the labeling of diagrams.

As 
seen from comparison of Figs.~\ref{fig1} and \ref{hartree-dia},
the electronic part ${\cal B}_{\alpha\beta}^\mu(\phi,\phi')$ 
of each Hartree diagram is identical
to that of its exchange counterpart. The only difference is in the 
arguments of the interaction
propagator, $U(\omega,\bq)\to U(0,2k_F\sin{\phi-\phi'\over 2})$,
where $\phi$ and $\phi'$ are polar angles of the initial and final velocities
[cf. Eqs.~(\ref{Bdef}),\ (\ref{15})].
Therefore, in the first order in the interaction, the Hartree correction 
to conductivity has a form very similar to the exchange 
correction (\ref{sigma}),
\begin{eqnarray} 
\delta\sigma^{\rm H}_{\alpha\beta}&=&4e^2 v_F^2\nu \int_{-\infty}^\infty 
\frac{d\omega}{2\pi} 
\frac{\partial}{\partial \omega} 
\left[\omega\  {\rm coth}{\omega\over{2T}}\right]
\nonumber \\ 
&\times & 
\int \frac{d^2{\bf q}}{(2\pi)^2}
\int{d \phi \over 2\pi} {d \phi' \over 2\pi}\ 
\label{sigma-hartree}  \\ 
&\times &\ 
\ {\rm Im}\ \left[\ U_{\rm H}(\phi,\phi')
{\cal B}_{\alpha\beta}(\omega,{\bf q};\phi,\phi')\ \right],
 \nonumber
\end{eqnarray} 
where  
\be
U_{\rm H}(\phi,\phi')=U\left(0,2k_F\sin{\phi-\phi'\over 2}\right)
\label{UH-def}
\ee
is the Hartree interaction
and 
${\cal B}_{\alpha\beta}(\omega,{\bf q};\phi,\phi')$
is given by Eqs.~(\ref{Bwq}),\ (\ref{Bwq_general}) 
without angular brackets (denoting integration over $\phi$ and $\phi'$),
see Eq.~(\ref{15}). Clearly, for a point-like
interaction $U(\omega,\bq)=V_0$ this yields 
\be
\delta\sigma^{\rm H}_{\alpha\beta}=-2\delta\sigma^F_{\alpha\beta},
\label{sigma-Hartree-point}
\ee
as has already been mentioned in Sec.~\ref{IIIb}.

In the case of the Coulomb interaction the situation is, however,
more delicate~\cite{fink}. To analyze this case, it is convenient to split the
interaction into the singlet and triplet parts~\cite{fink,AA,ZNA-sigmaxx}.
For the weak interaction, $\kappa\ll k_F$, the conductivity correction 
in the triplet channel is then given by Eq.~(\ref{sigma-hartree}) with
an extra factor ${3\over 4}$.

As to the singlet part, 
it is renormalized by mixing with the exchange term.
The effective interaction $U_s$ in the singlet channel is therefore
determined by the equation
\bea
U^{\rm s}(\phi,\phi') &=& U_0-{1\over 2}U_{\rm H}(\phi,\phi')-
\int {d \phi_1 \over 2\pi} {d \phi_2 \over 2\pi}
\nonumber \\
&\times&[U_0-{1\over 2}U_{\rm H}(\phi,\phi_1)]\ 
{\cal P}(\phi_1,\phi_2)U^{\rm s}(\phi_2,\phi'),
\nonumber\\ \label{singlet-inter}
\eea
where 
$U_0=2\pi e^2/q$ is the bare Coulomb interaction, and
\bea
{\cal P}(\omega,\bq;\phi_1,\phi_2)=
2\nu[2\pi\delta(\phi_1-\phi_2)+i\omega {\cal D}(\omega,\bq;\phi_1,\phi_2)]
\nonumber \\
\label{calP}
\eea
describes the electronic bubble.
Solving (\ref{singlet-inter}) to the first order in $U_{\rm H}$,
we get
\be
U^{\rm s}(\omega,\bq;\phi,\phi')=U(\omega,\bq)-
U^{\rm s}_{\rm H}(\omega,\bq;\phi,\phi'),
\label{1st-order-H}
\ee
where $U(\omega,\bq)$ is the RPA-screened Coulomb interaction (\ref{screen})
which has already been considered in Sec.~\ref{IIIc}, while
the second term describes the renormalized Hartree interaction in the 
singlet channel,
\bea
U_{\rm H}^{\rm s}(\phi,\phi') &=& {1\over 2}U_{\rm H}(\phi,\phi')-
{1\over 2\Pi}\int {d \phi_1 \over 2\pi} {d \phi_2 \over 2\pi}\nonumber \\
&\times&[U_{\rm H}(\phi,\phi_1){\cal P}(\phi_1,\phi_2)
+{\cal P}(\phi_1,\phi_2)U_{\rm H}(\phi_2,\phi')]\nonumber \\
&+&{1\over 2\Pi^2} \int {d \phi_1 \over 2\pi} {d \phi_2 \over 2\pi}
{d \phi_3 \over 2\pi} {d \phi_4 \over 2\pi}\nonumber \\
&\times&{\cal P}(\phi_1,\phi_2)U_{\rm H}(\phi_2,\phi_3){\cal P}(\phi_3,\phi_4).
\label{singlet-hartree}
\eea
Here $\Pi=\langle{\cal P}\rangle $ is the polarization operator 
(\ref{polarization}), and we have used the singular nature of the bare Coulomb
interaction implying 
$|\Pi|U_0\gg 1$ for all relevant momenta.

Taking into account that the angular dependence
of leading contributions to 
${\cal B}_{xx}(\omega,\bq;\phi,\phi')$ and
${\cal D}(\omega,\bq;\phi,\phi')$ is of the form
$\exp[-iQ(\sin\phi-\sin\phi')],$
we find that the singlet Hartree correction to $\sigma_{xx}$
is given by Eq.~(\ref{sigma-hartree}) with a replacement
\be
U_{\rm H}(\phi,\phi')\to
\frac{
\langle U_{\rm H}(\phi,\phi')\rangle 
-U_{\rm H}(\phi,\phi')}
{4\: [1+i\omega\langle {\cal D}(\omega,q) \rangle]^2}. \label{replacement}
\ee
Note that in the diffusive limit ${\cal B}_{\alpha\beta}$ is independent
of $\phi,\phi'$, so that only the zero angular harmonic of the interaction
contributes.
On the other hand, the zero angular harmonic is suppressed in 
the effective singlet-channel interaction (\ref{replacement}).
Therefore, the singlet channel does not contribute
to the Hartree correction in the diffusive limit,
in agreement with Refs.~\onlinecite{AA,fink}.
The situation changes, however, in the ballistic regime, when 
${\cal B}_{\alpha\beta}$ becomes angle-dependent.

\begin{figure} 
\includegraphics[width=8cm]{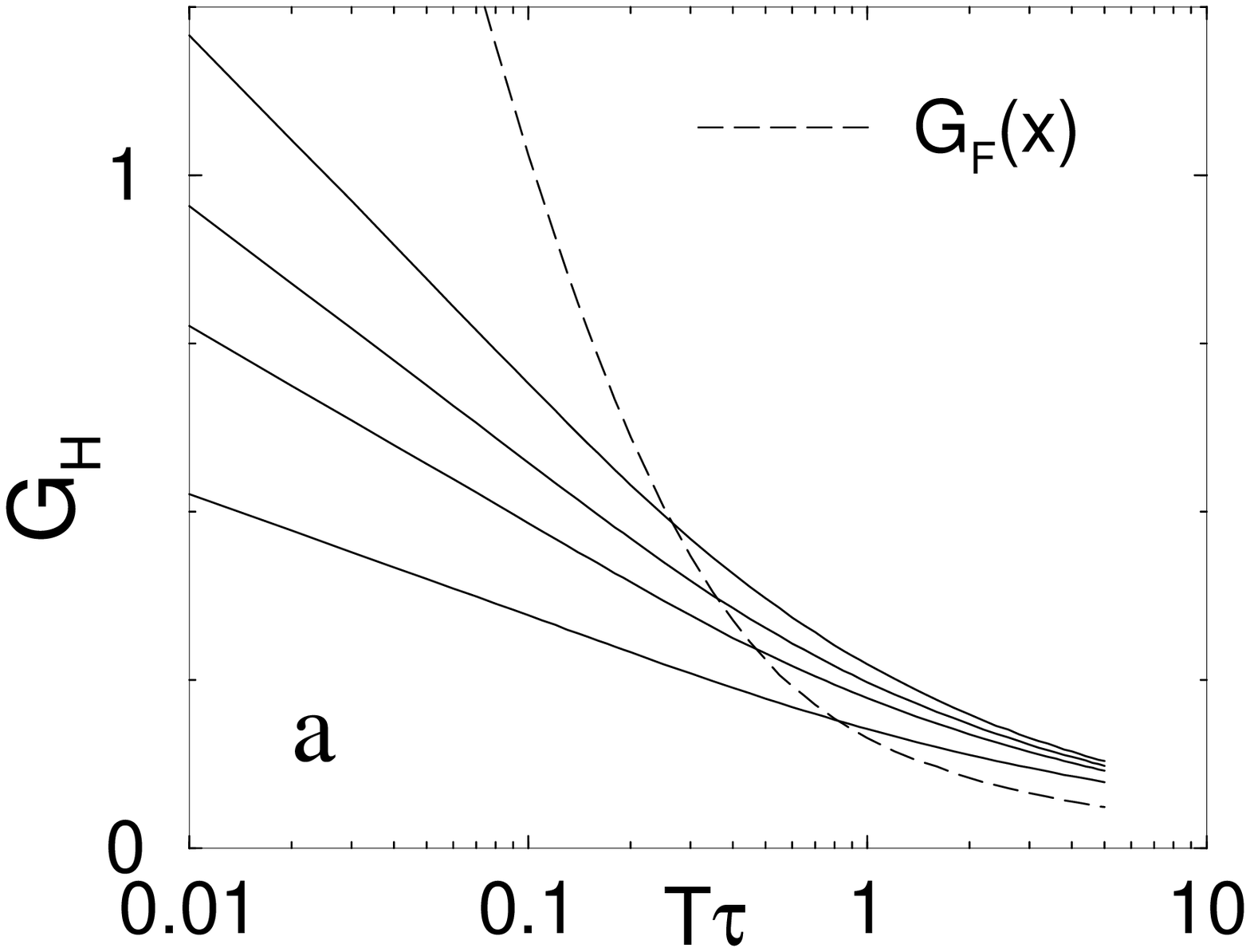} 
\vspace{1.5mm}  
\includegraphics[width=8cm]{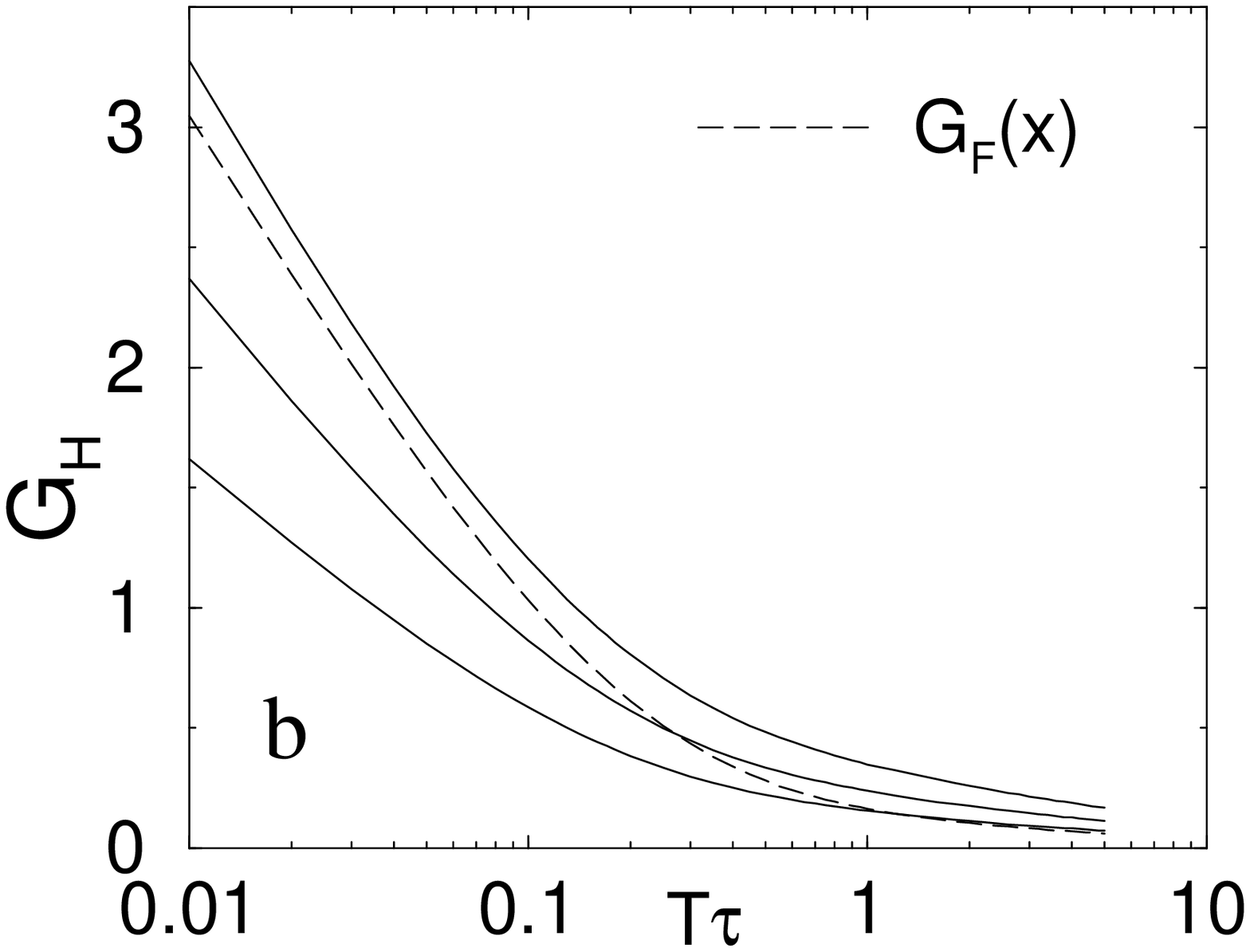}
\vspace{1.5mm} 
\caption{Hartree contribution, $ G_{\rm H}(T\tau)$, for
(a) weak interaction, $\kappa/k_F=0.1,\ 0.2,\ 0.3,\ 0.5$, and 
(b) strong interaction, $F_0=-0.3,\ -0.4,\ -0.5$
(from bottom to top). Dashed curves represent 
the exchange contribution.}
\label{fig3} 
\end{figure}

After the 
angular integration, the triplet Hartree conductivity
correction  takes the form (\ref{sxxQW}) with the replacement
$V_0 \to 1/2\nu,$ and
\be 
J_0^2(Q) 
\to -3y\int_0^\pi {d\phi \over 2\pi} {J_0(2Q \sin\phi)\over y+2\sin\phi},
\label{repl-tripH}
\ee
where $y=\kappa/k_F$.
For the singlet part we have a result similar to
(\ref{CoulsxxQW}) with a slightly different 
$Q$-integral, 
\bea 
\int_0^\infty Q dQ \  
 \frac{{\cal J}(y,Q) Q^2 }{(Q^2-i\Omega[1-J_0^2(Q)])^2(Q^2-i\Omega)},
\nonumber 
\eea 
where
\be\label{repl-singlH}
{\cal J}(y,Q) 
= - y\int_0^\pi {d\phi \over 2\pi} 
{J_0(2Q\sin\phi)-J_0^2(Q)\over y+2\sin\phi}.
\ee
This yields for the total Hartree contribution
\be
{\delta\rho^{\rm H}_{xx}(B)\over \rho_0}=
{(\omega_c\tau)^2\over \pi k_Fl}  
[G_{\rm H}^{\rm s}(T\tau,y)+
3G_{\rm H}^{\rm t}(T\tau,y)],
\label{rhohar} 
\ee
where $G_{\rm H}^{\rm s}$ and $G_{\rm H}^{\rm t}$ governing the temperature
dependence of the singlet and triplet contributions have the form
\bea
G_{\rm H}^{\rm s}(x,y)&=&32 \pi^2 x^2
\int_0^\infty dQ Q^3 {\cal J}(y,Q) \nonumber \\
&\!\!\!\!\!\times&\!\!\!\!\!\sum_{n=1}^\infty  
\frac{n(12\pi xn[1-J_0^2(Q)]+[3-2J_0^2(Q)]Q^2)} 
{(4\pi x n+Q^2)^2 (4 \pi x n[1-J_0^2(Q)]+Q^2)^3},
\nonumber \\ \label{GHs}\\
G_{\rm H}^{\rm t}(x,y)&=&{\pi x^2 y\over 4 }\int_0^\infty \!\!\!
 \frac{ du }{{\rm sinh}^2(\pi x u)} \nonumber \\
 &\!\!\!\!\!\times& \!\!\!\!\!
\int_0^\pi \! \!  d\phi \ {\exp[-2\sin^2\phi/u]  \over y+2\sin\phi} 
 \left(u-2\sin^2\phi\right). \nonumber \\
\label{GHt}
\eea 
Figure~\ref{fig3}a shows $G_{\rm H}(x,y)=G_{\rm H}^{\rm s}(x,y)+
3G_{\rm H}^{\rm t}(x,y)$ as a function of $x\equiv T\tau$ for several values 
of $y\equiv \kappa/k_F$.
The asymptotic behavior of $\delta\rho_{xx}^{\rm H}$ is as follows:
\bea
{\delta\rho^{\rm H}_{xx}(B)\over \rho_0}&=&
{(\omega_c\tau)^2\over \pi k_Fl}  
 \nonumber \\
&\times&
\left\{\begin{array}{ll}  
y\ln y\ [\ {3\over 4}\ln(T\tau)+\ln y], & \ \ T\tau\ll 1,\\[2mm]  
y\ln^2[\:y\:(T\tau)^{1/2}], & \!\!\!\!\!\!\!\! 1 \ll T\tau \ll 1/y^{2},\\[2mm] 
{\pi c_0 (T\tau)^{-1/2}}, & \ \ T\tau \gg 1/y^{2},
\end{array}\right. \nonumber\\
 \label{rhoHasympt}
\eea 
We see that at $\kappa/k_F \ll 1$ a new energy scale 
$T_{\rm H} \sim \tau^{-1}(k_F/\kappa)^2$ arises where the MR changes sign.
Specifically, at $T \ll T_{\rm H}$ the MR,
$\delta\rho_{xx}=\delta\rho_{xx}^{\rm F}+\delta\rho_{xx}^{\rm H}$, 
is dominated
by the exchange term and is therefore negative,
while at
$T \gg T_{\rm H}$ the interaction becomes effectively point-like
and the Hartree term wins,
$\delta\rho_{xx}^{\rm H}=-2\delta\rho_{xx}^{\rm F}$,
leading to a positive MR
with the same $(T\tau)^{-1/2}$ temperature-dependence, 
see Fig.~\ref{fig3}a.

\subsection{Hartree contribution for a strong interaction}
\label{IIIe}

In Sec.~\ref{IIId} we have assumed that $\kappa/k_F\ll 1,$
or, in other words, the interaction parameter 
$r_s=\sqrt{2} e^2/\varepsilon v_F$ 
(where $\varepsilon$ is the static dielectric constant
of the material) is small.
This condition is, however, typically not met
in experiments on semiconductor structures.
If $\kappa/k_F$ is not small, 
the exchange contribution (\ref{exchange})
remains unchanged, while 
the Hartree term
is subject to strong Fermi-liquid renormalization 
\cite{AA,ZNA-sigmaxx} and is determined by angular 
harmonics $F_m^{\sigma,\rho}$
of the Fermi-liquid interaction $F^{\sigma,\rho}(\theta)$ 
in the triplet ($\sigma$) and singlet ($\rho$) channels.

The effective interaction $U_{\rm eff}^{\sigma,\rho}$ replacing
$U_{\rm H}(\phi,\phi')$ in (\ref{sigma-hartree})
is then given by an equation of the type (\ref{singlet-inter})
but with $F^{\sigma,\rho}(\phi-\phi')/\nu$ substituted for
$U_{\rm H}(\phi,\phi')$ 
(and without $U_0$ in the triplet channel),
\bea
U^{\rho}_{\rm eff}(\phi,\phi') &=& U_0+{F^{\rho}(\phi-\phi')\over 2\nu}-
\int {d \phi_1 \over 2\pi} {d \phi_2 \over 2\pi}\nonumber \\
&\times&\left[U_0+{F^{\rho}(\phi-\phi_1)\over 2\nu}\right]\ 
{\cal P}(\phi_1,\phi_2)U^{\rho}_{\rm eff}(\phi_2,\phi'),
\nonumber \\
\label{strong-singlet}\\
U^{\sigma}_{\rm eff}(\phi,\phi') &=& {F^{\sigma}(\phi-\phi')\over 2\nu}-
\int {d \phi_1 \over 2\pi} {d \phi_2 \over 2\pi}\nonumber \\
&\times&{F^{\sigma}(\phi-\phi_1)\over 2\nu}
{\cal P}(\phi_1,\phi_2)U^{\sigma}_{\rm eff}(\phi_2,\phi'),
\label{strong-triplet}
\eea
A general solution of these equations requires inversion
of integral operators with the kernels 
$I-F^{\sigma}{\cal P}$ and 
$I-(U_0+F^{\rho}){\cal P}$ 
and is of little use for practical purposes.
The situation simplifies, however, in both diffusive and ballistic
limits.

In the diffusive regime, 
$T\ll 1/\tau$, the second term in the polarization bubble 
(\ref{calP}) and ${\cal B}_{\alpha\beta}$ are
independent of angles $\phi,\phi'$.
As discussed in Sec.~\ref{IIId}, this leads to
the suppression of the Hartree contribution in 
the singlet channel, while in the triplet channel only the zero angular 
harmonic survives,
\bea
U^{\sigma}_{\rm eff}(\omega,\bq)=
{1 \over 2\nu}{F_0^\sigma (Dq^2-i\omega) \over (1+F_0^\sigma )Dq^2-i\omega}
\nonumber \\
\label{Ueff-trip-diff}
\eea
We then reproduce
the known result~\cite{AA,ZNA-sigmaxx} $G_{\rm H}(T\tau)=3 G^{\rm t}_{\rm H}(T\tau)$ with
\be 
G^{\rm t}_{\rm H}(T\tau)=\left[1-{\ln(1+F^\sigma_0)\over F^\sigma_0}\right]\ln T\tau.
\label{diff-Hartree}
\ee 

In the ballistic limit,  $T\gg 1/\tau$,
the first term is dominant in (\ref{calP}), since $\langle {\cal D}\rangle$
is suppressed by a factor $J_0^2(Q) \ll 1$, according to (\ref{<D>}).
The angular harmonics then simply decouple in 
Eqs.~(\ref{strong-singlet}),\ (\ref{strong-triplet}), 
yielding effective Hartree interaction constants 
$U^{\rho}_{0,{\rm eff}}=0,$
$U^{\rho}_{m,{\rm eff}}=(2\nu)^{-1} F_m^\rho/(1+F_m^\rho), \ m\neq 0,$ 
and 
$U^{\sigma}_{m,{\rm eff}}=(2\nu)^{-1} F_m^\sigma/(1+F_m^\sigma)$. 
Therefore, the Hartree contribution reads
\be
\label{fermi-hartree-ball}
G_{\rm H}(T\tau)=-{c_0 \over 2}
\left[ 
\sum_{m\neq 0} {F^\rho_m \over 1+F^\rho_m}
+ 3 \sum_{m} {F^\sigma_m\over 1+F^\sigma_m} \right]
{1 \over \sqrt{T\tau}}.
\ee

From a practical point of view, it is rather inconvenient to describe
the interaction by an infinite set of unknown parameters 
$F_m^{\sigma,\rho}.$ For this reason, one often assumes that the 
interaction is isotropic and thus characterized by two coupling constants
$F_0^{\sigma}$ and $F_0^{\rho}$ only.
Within this frequently used (though parametrically uncontrolled)
approximation, the singlet part of the Hartree term is completely
suppressed. The Hartree contribution is then determined solely by
the triplet channel with the effective interaction
\be
U^{\sigma}_{\rm eff}(\omega,\bq)=
{1\over 2\nu}{F_0^\sigma\over 1+F_0^\sigma}
\displaystyle
\frac{Q^2-i\Omega}{Q^2-i\Omega\left[1- 
\displaystyle{F_0^\sigma\over 1+F_0^\sigma}J_0^2(Q)\right]}.
\label{Ueff-trip}
\ee
The Hartree correction to the
resistivity
takes the form of Eq.~(\ref{exchange}) with
an additional overall factor of $3$ and with 
$J_0^2(Q)$ multiplied by $F^\sigma_0/(1+F^\sigma_0)$,
\be
J_0^2(Q)\to J_0^2(Q)\ {F_0^\sigma \over 1+F_0^\sigma} 
\equiv 1-{\cal J}^\sigma(Q)\ ,
\label{calJsigmaQ}
\ee
everywhere in (\ref{GFx}); 
the result is shown in Fig.~\ref{fig3}b for several values of 
$F^\sigma_0$.

\subsection{Effect of Zeeman splitting}
\label{IIIf}

Until now we assumed that the temperature is
much larger than the Zeeman splitting $E_{\rm Z}$,
$T\gg E_{\rm Z}$.
In typical semiconductor structures this condition
is usually met in non-quantizing magnetic fields 
in the ballistic range of temperatures, allowing
one to neglect the Zeeman term.
If, however, this condition is violated,
$T\lesssim E_{\rm Z},$ 
the Zeeman splitting suppresses the triplet contributions 
with the  $z$-projection
of the total spin $S_z=\pm 1$, while 
the triplet with $S_z=0$ and singlet parts remain unchanged.

In the case of a weak interaction, $\kappa/k_F\ll1$,
the triplet contribution 
$3 G_{\rm H}^{\rm t}(T\tau,\kappa/k_F)$ in
Eq.(\ref{rhohar}) is modified in the following manner,
\be
3 G_{\rm H}^{\rm t}(x,y) \to 
G^{\rm t}_{\rm H}(x,y)+2{\rm Re}\: {\tilde G}^{\rm t}_{\rm H}(x,y;\epsilon_z),
\label{rhohar-weak-Zee} 
\ee 
where $\epsilon_z=2\tau E_{\rm Z}$, and the function 
${\tilde G}^{\rm t}_{\rm H}(x,y;\epsilon_z)$
describing the temperature dependence of the contribution with 
$\pm 1$ projection of the total spin is given by 
\bea
{\tilde G}^{\rm t}_{\rm H}(x,y;\epsilon_z)&=&{\pi x^2 y\over 4 }\int_0^\infty \!\!\!
 \frac{ \ du \exp[i \epsilon_z u]}{{\rm sinh}^2(\pi x u)} \nonumber \\
 &\times& \int_0^\pi \! \!  
 d\phi \ {\exp[-2\sin^2\phi/u]  \over y+2\sin\phi} 
 \left(u-2\sin^2\phi\right). \nonumber \\
 \label{tildeGtH}
\eea
We see that at $T\tau \ll\epsilon_z,$ the contributions
of $\pm 1$-components of the triplet saturate at 
the value given by (\ref{GHt}) with a replacement
$T\tau \to \epsilon_z$, i.e. 
at $\sim G_{\rm H}^{\rm t}(\epsilon_z,y)$.
In the opposite limit, $T\tau\gg\epsilon_z$, we have 
${\tilde G}^{\rm t}_{\rm H}(x,y;\epsilon_z)\simeq G_{\rm H}^{\rm t}(x,y)$,
and the result (\ref{rhohar}) is restored.

\begin{figure}
  \medskip
\centerline{\includegraphics[width=8cm]{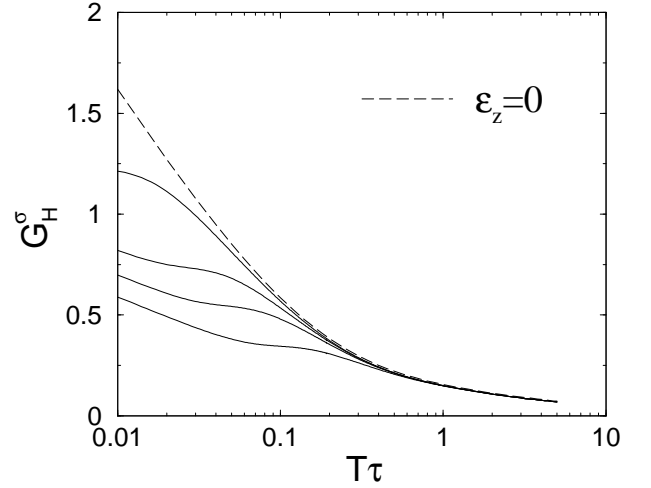}}
\caption{ The function $G_{\rm H}^\sigma(T\tau,\epsilon_z)$, Eq.~(\ref{zeeman-strong}),
describing the temperature dependence of the
triplet contribution is shown for $F_0^\sigma=-0.3$ and different values of Zeeman 
splitting, $\epsilon_z=0.1,\ 0.3,\ 0.5,\ 1.0$ (from top to bottom). 
Dashed curve represents the case $\epsilon_z=0$.} 
\label{zeeman} 
\end{figure}

The triplet contribution for strong isotropic interaction
(i.e. determined by $F_0^\sigma$ only) in the presence of Zeeman 
splitting
reads
\bea
{\delta\rho^{\rm H}_{xx}(B)\over \rho_0} = 
-{(\omega_c\tau)^2\over \pi k_Fl} 
[G_{\rm H}^{\sigma}(T\tau,0)+2{\rm Re}\: G_{\rm H}^{\sigma}(T\tau,\epsilon_z)]. 
\nonumber \\
\label{tripZee} 
\eea
The function $G_{\rm H}^{\sigma}(T\tau,\epsilon_z)$ 
is given by a formula 
similar to (\ref{GFx}),
\bea
&&G_{\rm H}^{\sigma}(x,\epsilon_z) =32 \pi^2 x^2
\int_0^\infty dQ Q^3 [1-{\cal J}^\sigma(Q) ]
\nonumber \\ 
&&\!\!\!\times\!\sum_{n=1}^\infty \! 
\frac{n(12\pi xn {\cal J}^\sigma(Q)
+[2+{\cal J}^\sigma(Q)][Q^2+i \epsilon_z])} 
{(4\pi x n+[Q^2+i \epsilon_z])^3
(4 \pi x n {\cal J}^\sigma(Q)+[Q^2+i \epsilon_z])^2},
\nonumber \\
\label{zeeman-strong}
\eea 
with ${\cal J}^\sigma(Q)$ as defined in (\ref{calJsigmaQ}).
Again, for high temperatures 
$T\tau \gg\epsilon_z,$ all the triplet components contribute, so that
the overall factor of 3 
(as in the absence of the Zeeman splitting)
restores. On the other hand,
for $T\tau \ll\epsilon_z,$ the contributions with $\pm 1$ projection 
of the spin saturate at low temperatures, and therefore 
the triplet contribution is partly suppressed, 
see Fig.~\ref{zeeman}.

\subsection{Hall resistivity}
\label{IIIg}

As discussed in Sec.~\ref{IIIb}, calculation of the correction 
$\delta\rho_{xy}$ to the Hall resistivity requires evaluation of
both $\delta\sigma_{xx}$ and $\delta\sigma_{xy}$. In fact, 
as we show below, the temperature dependence of $\delta\rho_{xy}$
in a strong magnetic field is governed by 
$\delta\sigma_{xx}$ in the diffusive limit and
by 
$\delta\sigma_{xy}$ in the ballistic limit.

Since $\delta\sigma_{xx}$ has been studied above, it remains to 
calculate $\delta\sigma_{xy}$. Using the result (\ref{Bxy}) for the corresponding
kernel $B_{xy}$, we get the exchange contribution
for the case of a point-like interaction 
\be
\label{sigmaxy-point}
\delta\sigma_{xy}=-{e^2\over 2\pi^2} {\nu V_0\over \omega_c\tau} 
\left[G_0^{(xy)}(T\tau)+G^{(xy)}_{\rm UV} \right],
\ee
where the temperature dependence of the correction is governed 
by the function
\bea
G_0^{(xy)}(x)
&=&-\pi^2 \! \int_0^\infty \!{du \over u} \exp(-1/u) \nonumber \\
&\times&\left[\frac{x^2}{{\rm sinh}^2(\pi x u)}-\frac{1}{(\pi u)^2} \right]
 \label{Gxy} \\
&\times& \left[\: (9u-3) I_0(1/u)+(3-2u) I_1(1/u)\: \right].  
\nonumber 
\eea
When writing (\ref{Gxy}), we subtracted a temperature independent 
but ultraviolet-divergent (i.e. determined by the upper limit
in frequency integral) contribution $G^{(xy)}_{\rm UV}$;
we will return to it in the end of this subsection.

\begin{figure}
  \medskip
\centerline{\includegraphics[width=8cm]{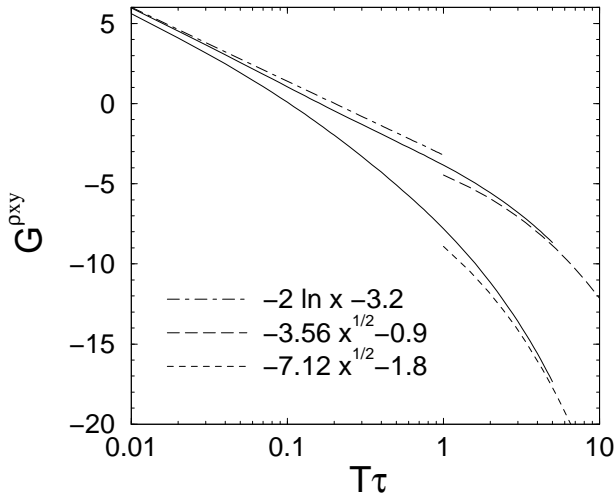}}
\caption{  Functions  $G_0^{\rho_{xy}}(T\tau)$ (lower curve)and
  $G_F^{\rho_{xy}}(T\tau)$ (upper curve) describing the temperature 
 dependence of the Hall resistivity for point-like and Coulomb interaction,
 respectively. Diffusive ($x \ll 1$) and ballistic ($x \gg 1$) 
 asymptotics, Eqs.~(\ref{G0rhoxy}) and (\ref{rhoxy-exchange}),
 are also shown. } 
\label{Grhoxy} 
\end{figure}

The function $G_0^{(xy)}(x)$ has the following asymptotics:
\be
\label{G0xy}
G_0^{(xy)}(x)\simeq \left\{\begin{array}{ll}  
9 \pi x, & \ \ \ x\ll 1,\\  
11 c_1 x^{1/2}, & \ \ \ x \gg 1,
\end{array}\right. 
\ee
with 
\be
c_1=-{\sqrt{\pi}\over 4}\zeta(1/2)\simeq 0.647.
\label{c1}
\ee
Combining (\ref{deltaint}) and (\ref{sigmaxy-point})
and using (\ref{rhoxy-sigma}), we find the correction to 
the Hall resistivity,
\be
{\delta\rho_{xy}\over \rho_{xy}}=
\frac{\nu V_0}{\pi k_F l} G_0^{\rho_{xy}}(T\tau),
\label{deltarhoxy-point}
\ee
where 
\bea
G_0^{\rho_{xy}}(x)&=&2G_0(x)-G_0^{(xy)}(x)
\nonumber \\
&\simeq& \left\{\begin{array}{ll}  
-2 \ln x +{\rm const}, & \ \ \ x\ll 1,\\  
- 11 c_1 x^{1/2}, & \ \ \ x \gg 1.
\end{array}\right. \label{G0rhoxy}
\eea
The function $G_0^{\rho_{xy}}(x)$ is shown in Fig.~\ref{Grhoxy}.
As usual, 
the Hartree term in the case of point-like interaction
has an opposite sign and is
twice larger in magnitude,
if the Zeeman splitting can be neglected.

An analogous consideration for the Coulomb interaction 
yields a similar result for the exchange correction
\bea
{\delta\rho^{\rm F}_{xy}\over \rho_{xy}}&=&
\frac{G_{\rm F}^{\rho_{xy}}(T\tau)}{\pi k_F l} ,
\label{GFrhoxy-coul} \\
G_{\rm F}^{\rho_{xy}}(x)&=&2G_{\rm F}(x)-G_{\rm F}^{(xy)}(x) \nonumber \\
&\simeq& \left\{\begin{array}{ll}  
-2 \ln x +{\rm const}, & \ \ \ x\ll 1,\\  
\displaystyle 
- {11 \over 2} c_1 x^{1/2}, & \ \ \ x \gg 1.
\end{array}\right. \label{rhoxy-exchange}
\eea
The function $G_{\rm F}^{(xy)}(x)$ is obtained by substituting 
(\ref{U-RPA-B}) and (\ref{Bxy}) in (\ref{sigma}) 
[cf. similar calculation for $\delta\sigma_{xx}^{\rm F}$
leading to Eqs.~(\ref{CoulsxxQW}) and (\ref{GFx}).]
The function $G_{\rm F}^{\rho_{xy}}(x)$ 
describing the temperature dependence of
the exchange correction to the Hall resistivity
is shown in Fig.~\ref{Grhoxy}.
In the ballistic regime, where $G_{\rm F}^{(xy)}(x)$ dominates,
the interaction becomes effectively point-like with 
$\nu V_0 = {1 \over 2}$, so that
one can simplify the calculation
using $G_{\rm F}^{(xy)}(x) \simeq {1 \over 2}G_0^{(xy)}(x).$

To analyze the Hartree contribution, we restrict ourselves 
to the
isotropic-interaction approximation.
Then, similarly to the consideration in the end
of Sec.~\ref{IIIe}, only the triplet part contributes,
and, in order to calculate $G_{\rm H}^{(xy)}(x)$, one should use Eqs.~(\ref{Bxy}) and 
(\ref{Ueff-trip}).
In the diffusive limit the Hartree correction to the Hall resistivity is
determined by (\ref{diff-Hartree}), while in the
ballistic limit we have again effectively point-like interaction
with $\nu V_0 = {3 \over 2} F_0^\sigma/(1+F_0^\sigma),$
implying that $G_{\rm H}^{(xy)}(x) \simeq 
-3G_0^{(xy)}(x)F_0^\sigma/2(1+F_0^\sigma).$
This yields
\bea
{\delta\rho^{\rm H}_{xy}\over \rho_{xy}}&=&
-\frac{G_{\rm H}^{\rho_{xy}}(T\tau)}{\pi k_F l} ,
\label{rhoxy-Hartree}\\
G_{\rm H}^{\rho_{xy}}(x) 
&\simeq& 3 \times \left\{\begin{array}{ll}  \displaystyle
2 \left[1-{\ln(1+F_0^\sigma)\over F_0^\sigma} \right]\ln x, 
& \ \ \ x\ll 1,\\[0.5cm]  \displaystyle
{11 \over 2}c_1\frac{F_0^\sigma}{1+F_0^\sigma} x^{1/2}, & \ \ \ x \gg 1.
\end{array}\right. \nonumber \\
\label{GrhoxyH-asympt}
\eea

We return now to the $T$-independent contribution $G_{\rm UV}^{(xy)}$ 
that was subtracted
in Eq.~(\ref{Gxy}). In view of the divergency of this term at $u\to 0$, 
it is determined by the short-time cut-off $u_{\rm min}=t_{\rm min}/2\tau$
\be
G_{\rm UV}^{(xy)}\propto \int_{u_{\rm min}}\frac{du}{u^{3/2}}
\sim u_{\rm min}^{-1/2}.
\label{GUV}
\ee
Since the correction we are discussing is governed by
cyclotron returns, the cut-off $t_{\rm min}$ corresponds to
a single cyclotron revolution,
$u_{\rm min}\sim \pi/\omega_c\tau$.
[On a more formal level, this is related to the assumption 
$\omega\ll \omega_c$ used for derivation of (\ref{Gxy});
see the text below Eq.~(\ref{diffuson}).]
We have, therefore, $G_{\rm UV}^{(xy)}=c^{(xy)}(\omega_c\tau)^{1/2}$,
with a constant $c^{(xy)}$ of order unity\cite{UV-term}. 
For the point-like interaction,
the considered term produces a temperature-independent
correction to the Hall resistivity of the form
\be
{\delta\rho_{xy}\over \rho_{xy}}=
\frac{\nu V_0 c^{(xy)}}{\pi k_F l}(\omega_c\tau) ^{1/2}.
\label{rhoUV}
\ee
 In the case of
Coulomb interaction, this correction
(with both, exchange and Hartree, terms included)
has the same form with
$\nu V_0 \to {1\over 2}[1-3F_0^\sigma/(1+F_0^\sigma)]$.

Finally, let us discuss the expected experimental manifestation 
of the results of this subsection. Equations (\ref{rhoxy-exchange}), 
(\ref{rhoxy-Hartree}) predict that in the presence of interaction
the temperature-dependent part of the
Hall resistivity $\rho_{xy}(B)$ in a
strong magnetic field $\omega_c\gg \tau^{-1},T$ is
linear in $B$ at arbitrary $T$,
with the $T$-dependence crossing over
from $\ln T$ in the diffusive regime to $T^{1/2}$ in the 
ballistic regime.
More specifically, if the interaction is not too
strong, the exchange contribution (\ref{rhoxy-exchange}) wins
and the slope decreases with increasing temperature, while in
the limit of strong interaction the slope increases due to the
Hartree term (\ref{rhoxy-Hartree}). In an intermediate range of $F_0^\sigma$
the slope is a non-monotonous function of temperature.
Surprisingly, this behavior of the slope of the Hall resistivity
is similar to the behavior of $\sigma_{xx}$ obtained in 
Ref.~\onlinecite{ZNA-sigmaxx} for $B=0$ and white-noise disorder.
This is a very non-trivial similarity, since the correction to 
$\rho_{xy}$ at weak fields\cite{ZNA-rhoxy} shows a completely different
behavior, vanishing as $T^{-1}$ in the ballistic regime.
In addition to the temperature-dependent linear-in-$B$ contribution, 
the interaction gives rise to a $T$-independent correction (\ref{rhoUV}),
which scales as $\delta\rho_{xy}\propto B^{3/2}$ (assuming again
that $\omega_c\gg \tau^{-1},T$)

Let us recall that 
these results are governed by multiple cyclotron returns
and thus are valid under the assumption
$\omega_c\gg T$.
In the opposite case, $\omega_c\ll T,$ the correction is suppressed in the 
ballistic regime (similarly to $\delta\rho_{xx}$, 
see Secs.~\ref{IIb}3 and \ref{IIIa}),
and the Hall resistance takes its Drude value.

\section{Qualitative interpretation:
Relation to return probability}
\label{IV}
\setcounter{equation}{0}

It was argued in Ref.~\onlinecite{AAG} by using the Gutzwiller
trace formula and Hartree-Fock approximation
that the interaction correction to conductivity
is related to a classical return probability.
The aim of this section is to demonstrate how this relation 
follows from the explicit formulas for $\sigma_{xx}$.

We begin by considering the case of smooth disorder,
when the kernel $B_{xx}(\omega,\bq)$ is given by Eq.~(\ref{Bwq}).
For simplicity, we will further assume a point-like
interaction, when only the first two terms in (\ref{Bwq})
give non-zero contributions. In fact, we know that the result for
the Coulomb interaction is qualitatively the same
[cf. Eqs.~(\ref{G0}) and (\ref{GF})].

We will concentrate on the first term in (\ref{Bwq});
the second one yields a contribution of the same order 
in the ballistic regime and is
negligible in the diffusive limit. Therefore, 
for the purpose of a qualitative discussion it is
sufficient to consider the first term. 
Using 
(\ref{DD}), the corresponding contribution can be estimated
as
\bea
{\delta\sigma_{xx} \over \sigma_{xx}} &\sim &
V_0\int_{-\infty}^\infty \!\!\! d\omega
\frac{\partial}{\partial \omega}\!
\left[\omega\ {\rm coth}{\omega\over{2T}}\right]\!\!
\int (dq) {\rm Re }\frac{\partial \langle{\cal D}(\omega,\bq)\rangle}{\partial\omega}
\nonumber \\
&\sim&V_0\int_0^\infty\!\!dt \frac{ (\pi T)^2t}{{\rm sinh}^2(\pi T t)} t 
\langle{\cal D}(r=0,t)\rangle
\nonumber \\
&\sim & V_0 \int_0^{T^{-1}}\!\!\!\!dt \: \langle{\cal D}(r=0,t)\rangle,
\label{qual-sigma}
\eea
where $\sigma_{xx}$ is the Drude conductivity in magnetic field and
we performed in the second line the Fourier transformation of 
${\cal D}$ to the coordinate-time representation (\ref{Fourier}).

\begin{figure}
  \medskip
\centerline{\includegraphics[width=8cm]{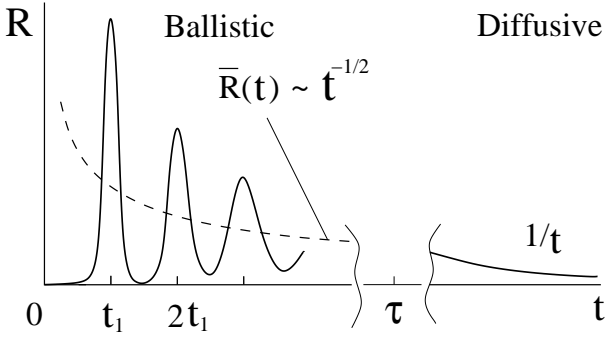}}
\caption{ Schematic plot of the return probability $R(t)$ in
a strong magnetic field and smooth disorder.
In the ballistic regime, the peaks are separated by the
cyclotron period, $t_1=2\pi/\omega_c$.
Dashed curve represents the smoothened return 
probability ${\bar R}(t)$} 
\label{peaks} 
\end{figure} 

The return probability in a strong magnetic field,
$\omega_c\tau\gg 1$,
\be
R(t)\equiv \langle{\cal D}(r=0,t)\rangle,
\label{return-prob}
\ee
is shown schematically in Fig.~\ref{peaks}.
In the diffusive time range, $t\gg \tau$, it
is given by $R(t)=1/4\pi D t$ (where $D$ is the diffusion constant
in the magnetic field, $D\simeq R_c^2/2\tau$). 
Equation (\ref{qual-sigma}) thus yields in the diffusive regime, $T\tau\ll 1$,
\be
{\delta\sigma_{xx} \over \sigma_{xx}} \sim {V_0\over D}\ln(T\tau),
\label{qual-diff}
\ee
in agreement with (\ref{AAdiff}),\ (\ref{GMdiff}).

At short (ballistic) time, $t\ll \tau$,
the return probability is governed by multiple 
cyclotron returns after $n=1,2,\dots$ revolutions,
\be
\label{R(t)}
R(t) = 
\sum_n\frac{\omega_c\tau}{4\sqrt{3}\pi^2 n R_c^2 }
\exp\left(-\frac{[t-2\pi n/\omega_c]^2\omega_c^3\tau}{12\pi n}\right).
\ee
Since $T\ll \omega_c$, the conductivity correction (\ref{qual-sigma})
is in fact determined by the smoothened return probability,
\be
{\bar R}(t)={1 \over (2\pi)^{3/2}} {1 \over R_c^2} 
\left({\tau \over t}\right)^{1/2}.
\label{Rsmooth}
\ee
Substituting (\ref{Rsmooth}) in (\ref{qual-sigma}) we find
that in the ballistic limit, $T\tau\gg 1$, the conductivity
correction scales as 
\be
{\delta\sigma_{xx} \over \sigma_{xx}} \sim {V_0\over D}(T\tau)^{-1/2},
\label{qual-ball}
\ee
in agreement with the exact results (\ref{deltaint}),\ (\ref{G0}).
As to the diffusive regime, $T\tau\ll 1$, the contribution
of short times $t \lesssim \tau$ to the integrand in (\ref{qual-sigma})
yields a subleading $T$-independent correction 
$\sim V_0/D$ to (\ref{qual-diff}).

\begin{figure}
  \medskip
\centerline{\includegraphics[width=8cm]{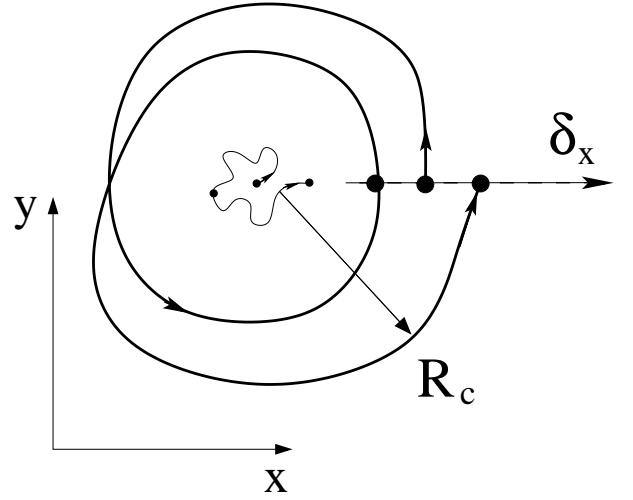}}
\caption{Schematic illustration of the ballistic dynamics
in a strong magnetic field. The thick line shows the particle
trajectory (two cyclotron revolutions disturbed
by the smooth random potential). The thin line is the diffusive
trajectory of the guiding center.} 
\label{qual}  
\medskip
\end{figure}

It is worth emphasizing that the ballistic behavior
(\ref{Rsmooth}) of the return probability ${\bar R}(t)$
corresponds to a {\it one-dimensional} diffusion. 
Consequently the {\it ballistic} result (\ref{qual-ball})
has the same form as the {\it diffusive} Altshuler-Aronov
correction in the quasi-one-dimensional geometry.
To clarify the reason for emergence of the one-dimensional diffusion, 
we illustrate
the dynamics of a particle subject to a strong magnetic field and smooth
disorder in Fig.~\ref{qual}

Let us assume that the velocity is in
$y$ direction at $t=0$. As is clear from Fig.~\ref{qual}, the return probability $R_1$
after the first cyclotron revolution (the integral of the first 
peak in Fig.~\ref{peaks}) is determined by the shift $\delta_x$ of the guiding center
in the cyclotron period $t_1=2\pi/\omega_c$. In view of the
diffusive dynamics of the guiding center, this shift has 
a Gaussian distribution with
\be
\delta_1^2\equiv
\langle \delta_x^2 \rangle = 2Dt_1={2\pi R_c^2 \over \omega_c\tau},
\label{x-shift}
\ee
yielding 
\be
R_1={(\omega_c\tau)^{1/2}\over 2\pi v_F R_c} \ .
\label{R1}
\ee
Furthermore, we have $\langle \delta_x^2 \rangle=n\delta_1^2$
after $n$ revolutions, yielding the return probability
$R_n=R_1/\sqrt{n}$. As to the $y$-component  $\delta_y$
of the guiding center shift, it only governs the width of
the peaks in Eq.~(\ref{R(t)}) and Fig.~\ref{peaks} without affecting ${\bar R}(t)$.
Therefore, the smoothened return probability is
\be
\label{R-sum-over-n}
{\bar R}(t)= \left. {R_n \over t_1}\right|_{n=t/t_1}\ ,
\ee
which reproduces Eq.~(\ref{Rsmooth}).

As mentioned in Sec.~\ref{IIIa}, the emergence of the one-dimensional diffusion
in the ballistic regime is reflected by the factor $J_0^2(Q) \sim 1/\pi Q$
in the formula (\ref{Bxx}) for the kernel $B_{xx}(\omega,\bq)$.
This factor effectively reduces the dimensionality of the $q$-integral, 
$\int d^2q \to  R_c^{-1}\int dq.$

In the above we considered a system with smooth disorder, for
which $\delta\sigma_{xx}$ at $B=0$ vanishes
exponentially in the ballistic limit.
Now we turn to the opposite case of a white-noise disorder.
We will show that the linear-in-$T$ correction~\cite{ZNA-sigmaxx,GD86} 
(Sec.~\ref{IIc}3) is again related to the return probability
but the relation is different from (\ref{qual-sigma}).
Indeed, according to (\ref{WNHik}), we have now the structure
$\langle {\cal D} \rangle \langle {\cal D} \rangle$ instead
of $\langle {\cal D}{\cal D} \rangle$ that was relevant for smooth 
disorder. On the other hand, the return probability
at ballistic times $t\ll \tau$ is clearly dominated by processes
with a single back-scattering event, implying
\be
\langle {\cal D}(r=0,t) \rangle \sim {1\over \tau}\int(dq)d\omega
\langle {\cal D}_{\rm f}(\omega,\bq)\rangle^2 e^{i\omega t}.
\label{WNRt}
\ee
Therefore, the contribution of the first term in (\ref{WNHik})
can be cast in the form
\bea
{\delta\sigma_{xx} \over \sigma_{xx}} &\sim &
V_0 \tau \int_{E_F^{-1}}^{T^{-1}}{dt \over t} 
\langle{\cal D}(r=0,t)\rangle,
\nonumber \\
&\sim & V_0\tau [\: \rm{const}-\langle{\cal D}(0,t\sim T^{-1})\rangle].
\label{WN-qual-sigma}
\eea
It is easy to see that the probability of a ballistic return after
a single scattering event is
\be
\label{WN-ret-prob}
\langle{\cal D}(r=0,t)\rangle \sim {1 \over \tau}\int d^2 r_1 
{\delta(t-2 r_1/v_F) \over (v_F r_1)^2}
\sim {1 \over v_F^2 \tau t}.
\ee
Substituting (\ref{WN-ret-prob}) in  (\ref{WN-qual-sigma}), we
reproduce the linear-in-$T$ correction (\ref{sigma_ball}),
\be
\delta\sigma_{xx}(T) \sim e^2 \nu V_0 T\tau
\label{qual-lin-T}
\ee
The constant term in (\ref{WN-qual-sigma}) comes from the lower
limit of the time integral, which is of the order of $E_F^{-1}$.
This constant merely renormalizes the bare value of the Drude conductivity.

On the diffusive time scale 
$\langle {\cal D} \rangle \langle {\cal D} \rangle 
\simeq \langle {\cal D}{\cal D} \rangle,$
so that there is no difference between white-noise and smooth disorder.
Therefore, in the diffusive limit the result (\ref{qual-sigma})
applies, yielding the usual logarithmic correction (\ref{qual-diff}).
In fact the contribution of the type (\ref{qual-sigma}) arises also
in the ballistic regime when all terms
in (\ref{Bwq_general}) are taken into account.
According to (\ref{WN-ret-prob}), it has the form
\be
{\delta\sigma_{xx} \over \sigma_{xx}} \sim 
{V_0\over D}[\:  \ln(T\tau)-{\rm const}], 
\ee
which is a subleading correction to the linear-in-$T$ term (\ref{sigma_ball}), \
(\ref{qual-lin-T}).

In the ballistic regime, $T\tau\gg 1$, 
the above qualitative arguments for a white-noise 
disorder can be re-formulated in terms of the interaction-induced
renormalization of the differential 
scattering cross-section on a single impurity. 
Specifically, the renormalization occurs due to the interference 
of two waves, one scattered off the impurity and another scattered off
the Friedel oscillations
created by the impurity\cite{RAG-ball-DOS,ZNA-sigmaxx}. 
The interference contribution is proportional to the
probability $W(\pi)$ of backscattering off the impurity
(see Appendix~\ref{A3}) and hence, to the
return probability after a single-scattering event, 
as discussed above. 

On the other hand, this implies that the scattering 
cross-section around $\phi\sim \pi$ is itself 
modified 
by the Friedel oscillations (in other words, the impurities
are seen by electrons as composite scatterers with an anisotropic
cross-section). The renormalization of the bare impurity 
depends on the energy of the scattered waves, 
which after the thermal averaging 
translates into 
the $T$-dependence of the effective transport scattering 
time,~\cite{ZNA-sigmaxx} $\tau(T)$ 
[this corresponds to setting $t\sim T^{-1}$ 
in the return probability, see Eq.~(\ref{WN-qual-sigma})].
This mechanism provides a systematic microscopic
justification of the concept of temperature-dependent 
screening~\cite{GD86}.

We recall that, in addition to the linear-in-$T$ term, 
the conductivity
correction contains a $T$-independent contribution 
determined by the ultraviolet frequency cut-off $\sim E_F$.
In the case of strong interaction this 
term can be of the same order as
the bare (non-interacting) Drude conductivity.
The coefficient in front of this term cannot 
be calculated within
the quasiclassical approach because it 
is governed by
short-distance physics at scales
of the order of $\lambda_F$. 
At the same time, according to the above picture,
this $T$-independent correction 
also modifies the impurity scattering cross-section 
around $\phi=\pi$. The corresponding
correction $\delta W(\phi)$ may thus
be comparable to the bare isotropic scattering
probability $W_0$.
An interesting consequence of this fact
is a possible situation when the total relaxation rate
$\tau_s^{-1}\propto \int d\phi [W_0+\delta W(\phi)]$
is {\it smaller} than the transport relaxation rate
$\tau^{-1}\propto \int d\phi[W_0+\delta W(\phi)] (1-\cos\phi)$.

In smooth disorder (small-angle scattering), 
the backscattering amplitude vanishes exponentially with $k_F d$,
and so does the amplitude of Friedel oscillations.
This leads to the suppression of the  
$T\tau$-contribution to the conductivity [see Sec.~\ref{IIc}3;
this fact was realized within the $T$-dependent screening
picture already in Ref.~\onlinecite{GD86} for the case 
of scattering on long-range interface roughness].
We note, however, that the understanding of the
interaction effects in terms of scattering off
Friedel oscillations is only possible in the ballistic regime.
Indeed, the diffusive correction in a {\it smooth} random 
potential is {\it not} exponentially small and is related to
random (having no $2k_F$-oscillating structure) fluctuations of the 
electron density, as was pointed out in Refs.~\onlinecite{AA,AAG}. 
The correlations of these
fluctuations (which reduce to the Friedel oscillations 
on the ballistic scales) are described by the return 
probability at arbitrary scales. 

Finally, we use the interpretation of the interaction
correction in terms of return probability to estimate
the MR in the white-noise random potential
and at sufficiently weak magnetic fields, $\omega_c\ll T$.
Note that the zero-$B$ ballistic correction (\ref{qual-lin-T})
does not imply any dependence of resistivity on magnetic field.
Indeed, as follows from (\ref{drude}), a temperature
dependence of the transport time $\tau(T)$ is not sufficient
to induce any non-trivial MR,
$$\Delta\rho_{xx}(B,T)\equiv\rho_{xx}(B,T)-\rho_{xx}(0,T)=0,$$
if $\tau$ is $B$-independent. 

In order to obtain the $B$-dependence of the resistivity,
we thus have to consider the influence of the magnetic field
on the return probability determining the correction 
to the transport time.
Since in the ballistic regime the characteristic length
of relevant trajectories is $L\sim v_F/T \ll l$, 
their bending by the magnetic 
field modifies only slightly the return probability 
for $\omega_c\ll T$.  The relative
correction to the return probability is thus of the order of
$(L/R_c)^2\sim \omega_c^2/T^2$ independently of the relation 
between $\omega_c$ and $\tau^{-1}$.
Therefore, to estimate the MR in the white-noise potential
for $\omega_c,\tau^{-1}\ll T$,
one can simply multiply the result (\ref{qual-lin-T}) for $B=0$ by a 
factor $(\omega_c/T)^2$, yielding
\be
{\Delta \rho_{xx} \over \rho_0} 
\sim {(\omega_c\tau)^2 \over k_F l}{1\over {T\tau}}, \quad \omega_c\ll T.
\label{rho-wn-qual}
\ee
A formal derivation of this result is presented in
Sec.~\ref{Vb}. In a stronger magnetic field,
$\omega_c\gg T$, the situation changes dramatically
due to multiple cyclotron returns, see above.
This regime is considered in Sec.~\ref{Va} below.

\section{Mixed disorder model}
\label{V}
\setcounter{equation}{0}

\subsection{Strong $B$}
\label{Va}

In Sec.~\ref{III}, we studied the interaction correction for a system
with a small-angle scattering induced by smooth disorder with
correlation length $d\gg k_F^{-1}$. This is a typical situation
for high-mobility GaAs structures with sufficiently large spacer $d$.
It is known, however, that with further increasing width of the spacer
the large-angle scattering on residual impurities and interface 
roughness becomes important and limits the mobility.
Furthermore, in Si-based structures the transport relaxation
rate is usually governed by scattering on short-range impurities.

This motivates us to analyze the situation when resistivity is 
predominantly due to large-angle scattering. We thus consider the
following two-component model of disorder (``mixed disorder''):
white-noise random potential with a mean free time $\tau_{\rm wn}$
and a smooth random potential with a transport relaxation time
$\tau_{\rm sm}$ and a single particle relaxation time
$\tau_{{\rm sm,}s}$ 
[$\tau_{\rm sm}/\tau_{{\rm sm,}s}\sim (k_F d)^2 \gg 1$].
We will further assume that while the transport relaxation rate 
$\tau^{-1}=\tau^{-1}_{\rm wn}+\tau^{-1}_{\rm sm}$ is 
governed by short-range disorder, $\tau_{\rm wn}\ll \tau_{\rm sm}$,
the damping of SdHO is dominated by smooth random potential,
$\tau_{{\rm sm,}s}\ll \tau_{\rm wn}.$ This allows us to consider the range
of classically strong magnetic fields,
$\omega_c\tau_{\rm wn}\gg 1,$ neglecting at the same time Landau
quantization (which is justified provided $\omega_c\tau_{{\rm sm,}s}/\pi\ll 1$).

To calculate the interaction corrections, we have to find the corresponding
kernel $B_{\alpha\beta}(\omega,\bq)$ determined by the classical dynamics.
Naively, one could think that under the assumed condition  
$\tau_{\rm wn}\ll \tau_{\rm sm}$ the smooth disorder can simply be neglected in the
expression for the classical propagator. While this is true in diffusive limit, 
the situation in the ballistic regime is much more nontrivial. To demonstrate 
the problem, let us consider the kernel $B_{xx}^{(\rho)}$ in the ballistic limit
$T\tau \simeq T\tau_{\rm wn} \gg 1$ and in a strong magnetic field 
$\omega_c\gg T \gg \tau^{-1}$. If the smooth random potential is completely neglected in
classical propagators, 
we have [see Appendix~\ref{A2}; 
 the second term in Eq.~(\ref{Brhoxx-WN-ball}) can be
neglected for $\omega_c\gg T$]
\be
B_{xx}^{(\rho)}\simeq 
{1 \over \tau_{\rm wn}}
\left[i{\partial g_0 \over \partial \omega} + g_0^2- 
{1 \over 4} \left({\partial g_0 \over \partial Q}\right)^2 \right],
\label{Brho-ball-wn}
\ee
where $g_0(\omega,\bq)$ is the angle-averaged propagator with only out-scattering
processes included,
\be
g_0(\omega,\bq)={i\pi \over \omega_c} \frac{J_\mu(qR_c) J_{-\mu}(qR_c)}{\sin{\pi\mu}},
\label{g0-wn}
\ee
and $\mu=(\omega+i\tau_{\rm wn}^{-1})/\omega_c$. If characteristic 
frequencies satisfy $\omega\ll \omega_c$ (which is the case for $T\ll \omega_c$),
Eq.~(\ref{g0-wn}) can be further simplified,
\be
g_0=\frac{J_0^2(Q)}{-i\omega +\tau_{\rm wn}^{-1}}.
\label{g0-wn-1}
\ee
Substituting (\ref{Brho-ball-wn}) and (\ref{g0-wn-1}) in (\ref{rho}),
we see that momentum- and frequency-integrations decouple and that the first
term in (\ref{Brho-ball-wn}) generates a strongly ultraviolet-divergent $q$-integral
$\sim \int dQ$. 

The physical meaning of this divergency is quite transparent. The contribution of the
first term in (\ref{Brho-ball-wn}) to $\delta\rho_{xx}$ is proportional
to the time-integrated return probability $\int dt g_0(r=0,t)$, similarly to
(\ref{qual-sigma}). For $t\ll \tau_{\rm wn}$ the propagator
$g_0(r,t)$ describes the ballistic motion
in the absence of scattering, which is merely the undisturbed cyclotron rotation in
the case of a strong magnetic field. Since at $t=2\pi n/\omega_c$ the particle returns
exactly to the original point, the integral $\int dt g_0(r=0,t)$ diverges.

The encountered divergency signals that the neglect of smooth disorder is not 
justified, even though $\tau_{\rm wn}\ll \tau_{\rm sm}$. Indeed, with smooth disorder
taken into account, the particle does not return exactly to the
original point after a cyclotron revolution, see Sec.~\ref{IV}.
The return probability is then described by Eqs.~(\ref{R(t)}), (\ref{Rsmooth})
with $\tau$ replaced by $\tau_{\rm sm}$. It is worth mentioning
a similarity with the problem of memory effects in a system with strong 
scatterers, where even a weak smooth disorder turns out to be crucially important
\cite{Antidots,polyakov01}.

To demonstrate the role of the smooth disorder
on a more formal level, we write down the angle-averaged propagator in the
ballistic regime, $T\tau_{\rm wn}\gg 1$, for the mixed-disorder model,
\be
\displaystyle
\langle{\cal D}(\omega,\bq)\rangle=
\frac{J_0^2(Q)}{Q^2/2\tau_{\rm sm} - i\omega +\tau_{\rm wn}^{-1}}.
\label{D-mixed}
\ee
Clearly, in both limits $\tau_{\rm sm}=\infty$ and $\tau_{\rm wn}=\infty$
this formula reduces to (\ref{g0-wn-1}) and (\ref{<D>}), respectively.
In view of $\omega\tau_{\rm wn}\gg 1$ the last term in the denominator
of (\ref{D-mixed}) can be neglected, and we return to the expression for
solely smooth disorder. The presence of the term $Q^2/2\tau_{\rm sm}$
regularizes the $Q$-integrals, thus solving the problem of 
ultraviolet-divergencies discussed above. The characteristic momenta
are thus determined by $Q^2\sim T\tau_{\rm sm}$. Therefore, despite the weakness
of the smooth disorder, $\tau_{\rm sm}\gg \tau_{\rm wn},$ it is the first
($Q$-dependent)
rather than the third term which has to be retained in the denominator
of (\ref{D-mixed}). In other words, in the ballistic regime 
and in a strong magnetic field the 
dynamics in the considered model is governed by smooth 
disorder.

The above discussion demonstrates that at 
$\omega_c\gg T \gg \tau_{\rm wn}^{-1}$ 
the kernel $B_{\alpha\beta}(\omega,\bq)$ for the mixed-disorder model 
is given by (\ref{Bwq_general}) with propagators ${\cal D}$ calculated
in smooth random potential (i.e. with white-noise disorder neglected).
The time $\tau_{\rm wn}$ enters the result only through the matrices 
$T_{\alpha\beta}$ (determined by the transport time 
$\tau\simeq \tau_{\rm wn}$) and ${\cal S}_{\alpha\beta}$. 
Using $\tau_{\rm sm}/\tau_{\rm wn}\gg 1,$
we find then that the resulting
expression,
\bea
B_{xx}&\simeq&
{1\over 2\omega_c^2\tau}\left[1+{\tau\over\tau_{\rm wn}}\right]\langle {\cal D D}\rangle 
\nonumber \\
&-&{1\over 2\omega_c^2\tau_{\rm wn}}[\langle {\cal D}\rangle \langle {\cal D}\rangle
-2 \langle {\cal D} n_y\rangle \langle n_y {\cal D}\rangle ]\nonumber \\
&-&{1\over 2\omega_c^2}[\langle {\cal D}\rangle
-2 \langle  n_y {\cal D} n_y \rangle ]\nonumber \\
&+&{2\over \omega_c}\langle n_y{\cal D}n_x{\cal D}\rangle-
\langle {\cal D}n_x{\cal D}n_x{\cal D}\rangle \label{Bxxtotal}
\eea
is dominated by the first term 
corresponding to the first term in Eq.~(\ref{Brho-ball-wn}).
This yields for $Q\equiv q R_c \gg 1$
\be
B_{xx}(\omega,\bq)\simeq {1\over \omega_c^2\tau}\langle {\cal D D}\rangle \simeq 
\frac{4\tau_{\rm sm}^2}{\omega_c^2\tau}\frac{J_0^2(Q)}{(Q^2-i\Omega)^2},
\label{Bxx-mixed}
\ee
where $\Omega=2\omega\tau_{\rm sm}$. 

As in previous sections, we first calculate the conductivity correction
for a point-like interaction. Substituting (\ref{Bxx-mixed})
in (\ref{sigma}), we get 
\be
\delta\sigma_{xx}=-{e^2 \over 2\pi^2}\: \nu V_0\: 
\left({\tau_{\rm sm}\over \tau}\right)^{1/2}{4c_0\over (T\tau)^{1/2}},
\qquad T\gg 1/\tau_{\rm wn},
\label{sigma-mixed-ball}
\ee
with the constant $c_0$ as defined in Eq.~(\ref{c0}).
For an arbitrary (not necessarily small) value of
the ratio $\tau/\tau_{\rm sm}$ the coefficient $4$
in (\ref{sigma-mixed-ball}) is replaced by 
$4 - 3\tau/\tau_{\rm sm}$. For 
$\tau=\tau_{\rm sm}$ (i.e. without white-noise disorder)
we then recover the ballistic asymptotics of Eq.~(\ref{G0}).

As in the case of purely smooth disorder,
the resistivity correction $\delta\rho_{xx}$ is related
to $\delta\sigma_{xx}$ via Eq.~(\ref{parabola}).
Comparing (\ref{sigma-mixed-ball}) with (\ref{G0}),
we see that the correction $\delta\rho_{xx}$ is enhanced 
in the mixed-disorder model by a factor 
$\sim 4(\tau_{\rm sm}/\tau)^{1/2}\gg 1$
as compared to the purely smooth-disorder case. On the other hand, 
the scaling with temperature and magnetic field,
$\delta\rho_{xx}\propto B^2T^{-1/2}$, remains the same.

Let us analyze now the crossover from the
ballistic to the diffusive regime.
Setting $T\tau\sim 1$ in (\ref{sigma-mixed-ball}), we
find that the correction is parametrically large,
$\delta\sigma_{xx}\sim(\tau_{\rm sm}/\tau_{\rm wn})^{1/2}$.
Clearly, this does not match the diffusive contribution
(\ref{GMdiff}), yielding $\delta\sigma_{xx}\sim 1$
at $T\tau\sim 1$. This indicates that returns without 
scattering on white-noise disorder continue to
govern the correction in certain temperature window
below $T\sim1/\tau$, which normally belongs to the diffusive regime.

\begin{figure}
  \medskip
  \includegraphics[width=8cm]{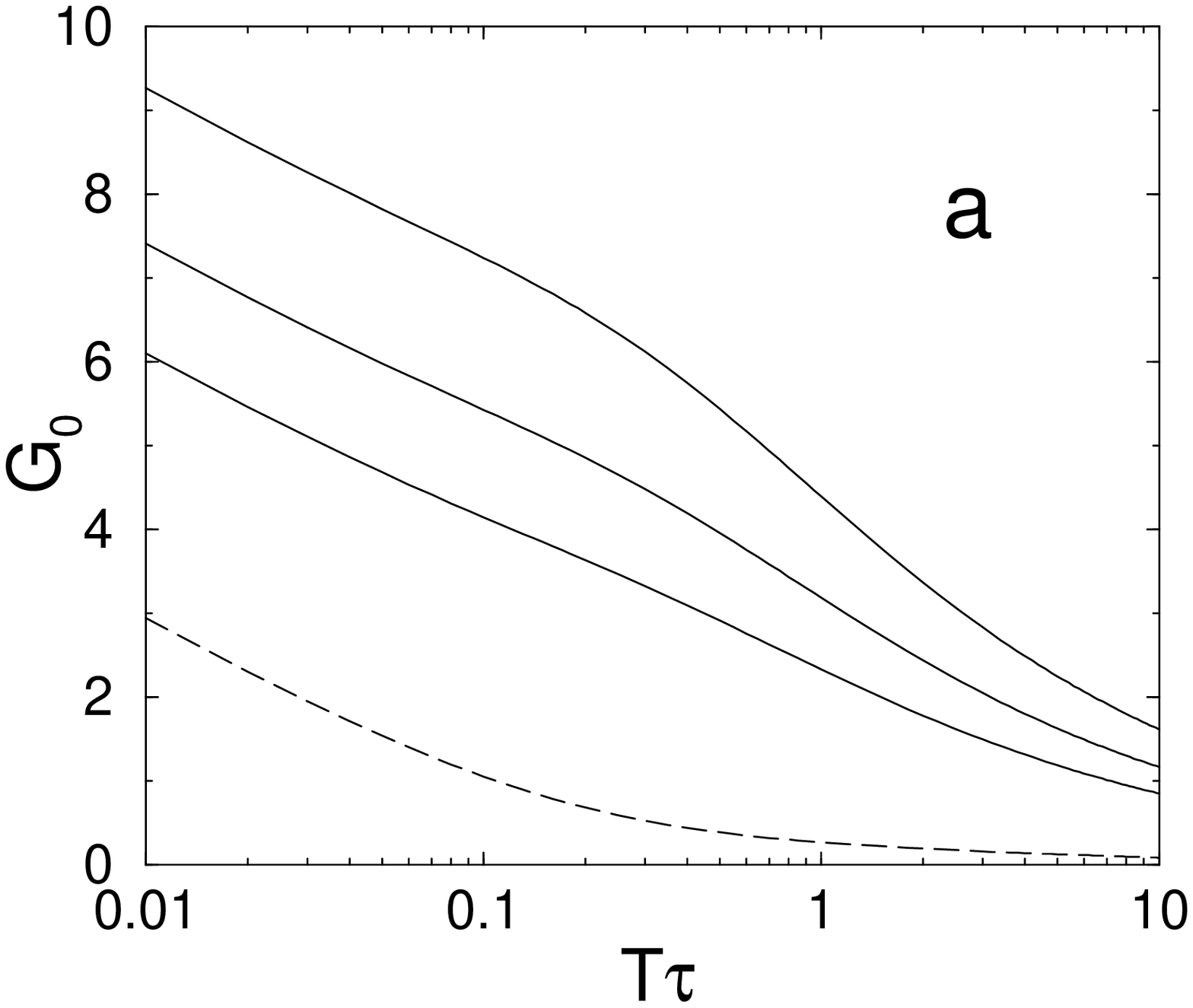}
  \vspace{2mm}
\includegraphics[width=8cm]{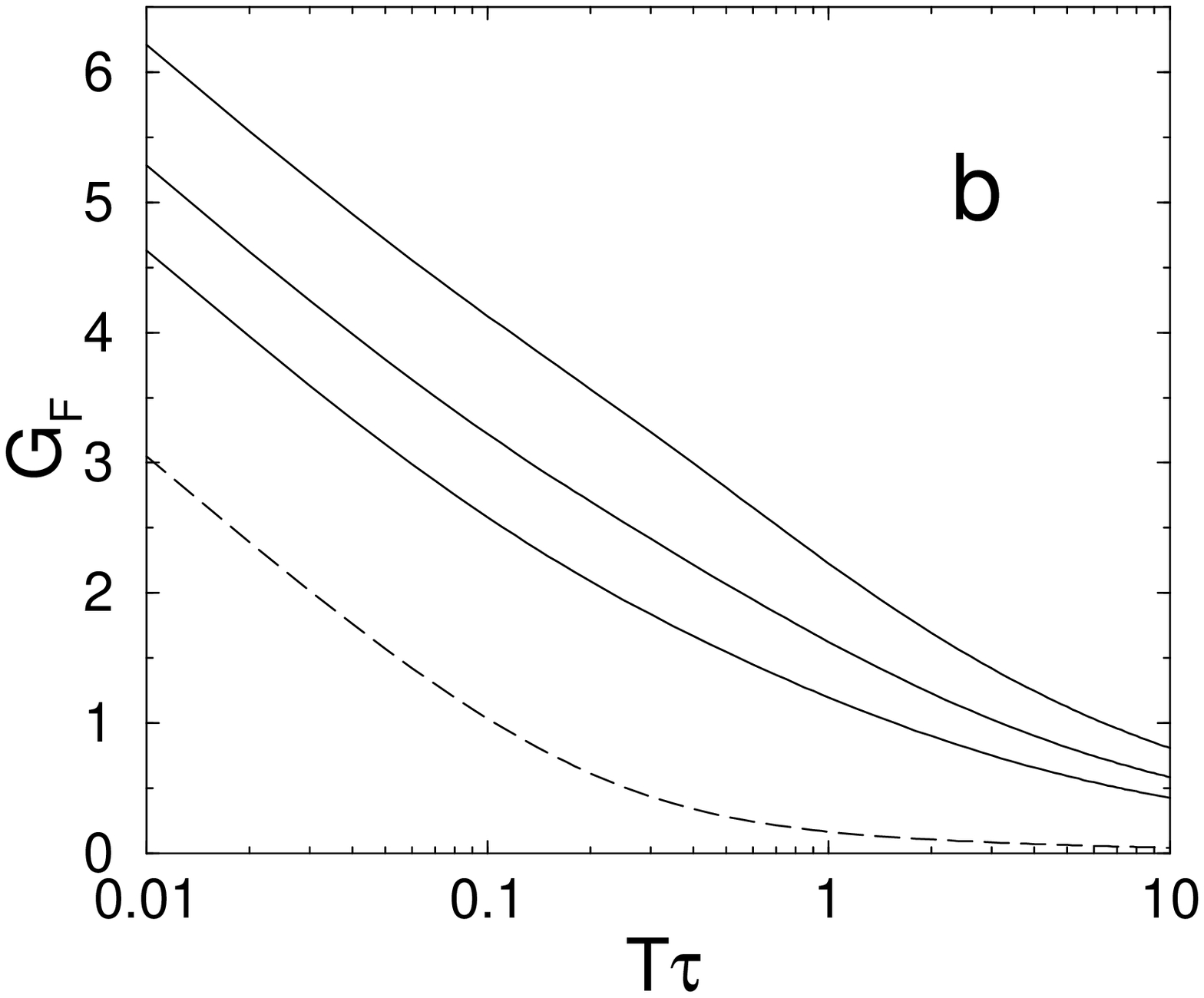}
\vspace{2mm}
\caption{ Functions $G_0^{\rm mix}(T\tau)$ (a) and  $G_F^{\rm mix}(T\tau)$ (b)
describing the temperature 
 dependence of the resistivity correction due to point-like and Coulomb (exchange)
 interaction, respectively, in the mixed-disorder model
 for different values of parameter $\gamma\equiv \tau_{\rm sm}/\tau=20,\ 10,\ 5$
 (from top to bottom).  Dashed curves represent these functions for 
 purely smooth disorder ($\gamma=1$).
} 
\label{mix} 
\end{figure}

To find the corresponding contribution, one should take into 
account the scattering-out term $\tau_{\rm wn}^{-1}$ in the denominator
of (\ref{D-mixed}), which yields 
\bea
G_1(x,\gamma)&=&{2 \over \pi}\left({\gamma\over 2x} \right)^{1/2}
\int_0^\infty \! \frac{dz z^{3/2} \exp[-z/\pi x]}{{\rm sinh}^2 z} 
\nonumber 
\\
\label{G1}
&=&\left\{\begin{array}{ll}  
(2\gamma)^{1/2}, & \ \ \ x\ll 1,\\  
4 c_0 \gamma^{1/2} x^{-1/2}, & \ \ \ x \gg 1,
\end{array}\right.
\eea
where $\gamma=\tau_{\rm sm}/\tau\gg 1$ and $x=T\tau$ .
To describe the temperature dependence of the interaction
correction for all $T$, we have to add here the diffusive contribution,
which has the form (\ref{GMdiff}) for $T\tau\ll 1$ and vanishes
for $T\tau\gg 1$. This contribution corresponds to long times
$t\gg \tau$ and describes the trajectories
multiply scattered off white-noise disorder.
Since at $T\tau\sim 1$ the sum of the ballistic and diffusive
contributions will be dominated by $G_1(1,\gamma)\sim \gamma^{1/2} \gg 1$,
the precise way of vanishing of the diffusive contribution 
at $T\tau\sim 1$ is inessential. Therefore, we can describe it by 
the function $G_0(x)$, Eq.~(\ref{deltaint}). The resistivity correction
for a system with mixed disorder and point-like interaction has thus 
the following form:
\be
{\delta\rho_{xx}(B)\over \rho_0} = 
-{(\omega_c\tau)^2\over \pi k_Fl} \nu V_0
G^{\rm mix}_0(T\tau,\tau_{\rm sm}/\tau), 
\label{rho-mix} 
\ee
where
\bea
G^{\rm mix}_0(x,\gamma) &=& G_1(x,\gamma) + G_0(x)
\label{G-mix} \\
&=&\left\{\begin{array}{ll}  
-\ln x + (2\gamma)^{1/2}, & \ \ \ x\ll 1,\\  
4 c_0 \gamma^{1/2} x^{-1/2}, & \ \ \ x \gg 1,
\end{array}\right.
\nonumber 
\eea 
This result is illustrated in Fig.~\ref{mix}a.

In the case of Coulomb interaction, 
we have as usual a similar
result for the exchange contribution
\be
{\delta\rho^{\rm F, mix}_{xx}(B)\over \rho_0} = 
-{(\omega_c\tau)^2\over \pi k_Fl} 
G_{\rm F}^{\rm mix}(T\tau,\tau_{\rm sm}/\tau), 
\label{rho-mix-F} 
\ee
with
\bea
G_{\rm F}^{\rm mix}(x,\gamma) &=& {1 \over 2} G_1(x,\gamma) + G_{\rm F}(x)
\label{G-mix-F} \\
&=&\left\{\begin{array}{ll}  \displaystyle
-\ln x + (\gamma/2)^{1/2}, & \ \ \ x\ll 1,\\[0.2cm]  
2 c_0 \gamma^{1/2} x^{-1/2}, & \ \ \ x \gg 1.
\end{array}\right.
\nonumber 
\eea 
This function is shown in in Fig.~\ref{mix}b.
In fact, here the diffusive contribution can be described either 
by the function $G_{\rm F}(x)$ or by $G_0(x)$ because in the diffusive limit
they coincide up to a small constant. 
Since in the intermediate and 
ballistic regimes [where $G_{\rm F}(x)$ and $G_0(x)$ differ]
the contribution ${1 \over 2} G_1(x,\gamma)$ 
is dominant, the behavior of the diffusive contribution 
is of no importance, as in the case of point-like interaction.
Note that the ballistic contribution corresponds to
the point-like interaction with $\nu V_0 ={1\over 2}$, yielding
a factor ${1 \over 2}$ in front of  $G_1(x,\gamma)$ as 
compared to (\ref{G-mix}).
This is because the dynamical part of screening is suppressed
for all relevant $Q \sim T\tau_{\rm sm} \gg 1$ in the whole range of 
temperatures, even for $T\tau<1$, where this contribution is important.

This also applies to the Hartree contribution to 
the resistivity. Within the ``$F_0^\sigma$-approximation''
we have again an effectively point-like interaction 
with $\nu V_0 \simeq {3 \over 2} F_0^\sigma/(1+F_0^\sigma)$
in the ballistic term. The result thus reads
\be
{\delta\rho^{\rm H, mix}_{xx}(B)\over \rho_0} = 
3{(\omega_c\tau)^2\over \pi k_Fl} 
 G_{\rm H}^{\rm mix}(T\tau,\tau_{\rm sm}/\tau), 
\label{rho-mix-H} 
\ee
where
\bea
&&G_{\rm H}^{\rm mix}(x,\gamma) = 
{1\over 2} {F_0^\sigma \over (1+F_0^\sigma)} G_1(x,\gamma) + G^{\rm t}_{\rm H}(x)
\nonumber\\
&&\!\!= \left\{\begin{array}{ll}  \displaystyle
\left[1-{\ln(1+F_0^\sigma)\over F_0^\sigma} \right]\ln x + 
{F_0^\sigma\over 1+F_0^\sigma}\left({\gamma\over 2}\right)^{1/2} 
\!\!\!, \!\!\!& \\
& x\ll 1,\\[0.3cm]
\displaystyle
- 2 c_0{F_0^\sigma \over (1+F_0^\sigma)}\gamma^{1/2}x^{-1/2}, & x \gg 1.
\end{array}\right.
\nonumber\\
\label{G-mix-H}
\eea 

Before closing this subsection, we briefly discuss the 
Hall resistivity in the mixed disorder model.
Repeating the steps
described above, we find that the ballistic contribution
to $\rho_{xy}$ also contains an extra factor 
$(\tau_{\rm sm}/\tau)^{1/2}$,
similarly to $\rho_{xx}$. 
For an arbitrary (not necessarily small) value of
the ratio $\tau/\tau_{\rm sm}$ the coefficient $11$
in Eqs.~(\ref{G0rhoxy}),\ (\ref{rhoxy-exchange}) is replaced by 
$[6 + 5\tau/\tau_{\rm sm}](\tau_{\rm sm}/\tau)^{1/2}$.

\subsection{Weak $B$}
\label{Vb}

In the case of a purely smooth disorder (Sec.~\ref{III}) 
the resistivity correction in the ballistic regime
is exponentially suppressed
for $\omega_c\ll T$ because the particle cannot return 
to the origin. When the short-range potential is present, the
situation changes and the return probability is determined
for $T\tau\gg 1$ by the single-backscattering processes.
The interaction-induced
MR arises then due to the influence 
of the magnetic field on the probability of such return, 
as discussed in the end of Sec.~\ref{IV}.
In this case, there is no need to take the smooth potential
into account and the 
MR is determined solely by the
white-noise disorder.
Let us calculate the corresponding correction using the
ballistic form (\ref{deltaBrhoxx-ball}) of the 
kernel $\Delta B_{xx}^{(\rho)}$. 

For a point-like interaction, substituting 
(\ref{deltaBrhoxx-ball}) in (\ref{rho}), we
find the following ballistic ($T\tau\gg 1$) asymptotics of
the longitudinal MR,
\be
{\Delta\rho_{xx} \over \rho_0}=
-\frac{(\omega_c\tau)^2 \nu V_0}{\pi k_F l} \frac{\pi }{72\: T\tau}.
\label{drhoxx-weak-wn-point}
\ee
In the case of Coulomb interaction,
$\Delta B_{xx}^{(\rho)}$ is 
multiplied by the ballistic asymptotics of the interaction,
Eq.~(\ref{U-ball}). Substituting this product in
Eq.~(\ref{rho}), we get the Fock contribution to the
MR in the form
\be
{\Delta\rho_{xx}^{\rm F}\over \rho_0}=
-\frac{(\omega_c\tau)^2 }{\pi k_F l} \frac{17 \pi }{192\: T\tau}, \quad T\tau\gg 1.
\label{drhoxx-weak-wn-coul}
\ee
The corresponding Hartree term also scales as $B^2/T$.
It is worth noting that there is another contribution 
to the MR in this
regime, which comes from the suppression of the triplet
channel due to Zeeman splitting $E_{\rm Z}$ rather than from the orbital
effects. This contribution is identical to that found
in Ref.~\onlinecite{ZNA-MRpar} for the ballistic 
magnetoresistance in a parallel magnetic field. 
It also scales as $B^2/T$ in a weak magnetic field;
however, it contains an extra factor $(E_{\rm Z}/\omega_c)^2$,
as compared to (\ref{drhoxx-weak-wn-coul}).
This factor is small in typical experiments
on semiconductor heterostructures where the 
effective mass of the carriers
is much smaller that the bare electron mass.

We now turn to the Hall resistivity.
Using 
(\ref{Bxy-ZNA-ball}) and (\ref{DeltaBrhoxy}),
we find for $\omega_c,\tau^{-1}\ll T$ and for
arbitrary relation between
$\omega_c$ and  $\tau^{-1}$
\be
{\delta\rho_{xy} \over \rho_{xy}}=
\frac{\nu V_0}{\pi k_F l} \frac{\pi }{12\: T\tau}
\label{drhoxy-weak-wn-point}
\ee
for the point-like interaction, 
and 
\be
{\delta\rho_{xy}^{\rm F}\over \rho_{xy}}=
\frac{1}{\pi k_F l} 
\left[1-{49 (\omega_c\tau)^2 \over 330}\right]
\frac{11 \pi }{96\: T\tau}
\label{drhoxy-weak-wn-coul}
\ee
for the Coulomb interaction.
The result (\ref{drhoxy-weak-wn-coul}) reduces in the limit
$B\to 0$ to that
obtained in Ref.~\onlinecite{ZNA-rhoxy} from the quantum
kinetic equation.
We see that in view of a relatively small
value of the numerical coefficient
$49/330$, the first ($B$-independent) term in square brackets in 
(\ref{drhoxy-weak-wn-coul}) dominates for 
$\omega_c\tau\alt 1$, so that the results of Ref.~\onlinecite{ZNA-rhoxy} 
are applicable in sufficiently broad range of magnetic fields.
For the corresponding Hartree-correction to $\delta\rho_{xy}$
calculated within the ``$F_0^\sigma$-approximation'',
we refer the reader to Ref.~\onlinecite{ZNA-rhoxy}.

\section{Anisotropic systems}
\label{VI}
\setcounter{equation}{0}

\subsection{Qualitative discussion}
\label{VIa}

In the preceding consideration, we assumed 
that the 2D system is isotropic.
While this is true for the majority of magnetotransport experiments
we have in mind, there exists a number of important 
situations when the transport
is anisotropic, $\sigma_{xx}\neq \sigma_{yy}$. First, such an anisotropy
can be induced by the orientation of the 2D electron gas 
plane with respect to
the crystal axes, see e.g. Ref.~\onlinecite{Bishop84} for a measurement of 
the quantum correction for the (110) surface of the Si-MOSFET. 
Second, the electron-electron interaction may lead to spontaneous
formation of a charge-density wave. Finally, the anisotropy
may be induced by a one-dimensional periodic modulation 
(lateral superlattice). The latter example is of special
interest in view of emergence of commensurability
oscillations (known as Weiss oscillations)\cite{weiss89},
and we will discuss it in more detail in Sec.~\ref{VIc}.

The interaction-induced correction to the conductivity 
tensor of an anisotropic system was calculated for the diffusive regime and
$B=0$ by Bhatt, W\"olfle, and Ramakrishnan \cite{bhatt85}.
They showed, in particular, that the quantum correction
preserves the anisotropy of the quasiclassical (Boltzmann) conductivity.
Below we will generalize their result onto the case of a classically
strong magnetic field, and, furthermore, will extend the
consideration to the ballistic regime.

We begin by presenting a simple argument allowing one to
estimate the conductivity correction in an anisotropic system;
we will confirm it by a formal calculation in Sec.~\ref{VIb}.
According to Eq.~(\ref{qual-sigma}), the relative correction
to a diagonal component $\sigma_{\mu\mu} \ (\mu=x,y)$
of the conductivity tensor is determined by the return probability
(and is, thus, the same for $\mu=x$ and $\mu=y$).
This implies, in the diffusive regime
\be
\label{sigma-mumu}
\frac{\delta\sigma_{\mu\mu}}{\sigma_{\mu\mu}}
\sim -{\rm Re}\; {1\over \nu}\left.
\int(dq)\frac{1}{D_{\alpha\beta}q_\alpha q_\beta -i\omega}
 \right|^{\omega=1/\tau}_{\omega=T},
\ee
yielding
\be
\delta\sigma_{xx}\sim 
e^2\left({\sigma_{xx}\over \sigma_{yy}}\right)^{1/2}\ln T\tau
\label{anis-diff-qual}
\ee
and analogously for $\delta\sigma_{yy}$.
In the ballistic regime the time-integrated return
probability $\int^{T^{-1}}dt\langle{\cal D}(t)\rangle$
scales as $(T\tau)^{-1/2}$ [see Eqs.~(\ref{Rsmooth}) and 
(\ref{qual-ball})], so that we have instead of 
(\ref{anis-diff-qual}),
\be
\delta\sigma_{xx}\sim 
e^2\: {\cal K}\left({\sigma_{xx}\over \sigma_{yy}}\right) \;(T\tau)^{-1/2}.
\label{anis-ball-qual}
\ee
The explicit form of the function ${\cal K}(x)$
will be calculated below. 
Since the conductivity corrections (\ref{anis-diff-qual}) and 
(\ref{anis-ball-qual}) are only determined by the 
anisotropic diffusion,
we expect that they do not depend on the particular 
source of anisotropy, in analogy with Ref.~\onlinecite{bhatt85}.
An important feature of the results
(\ref{anis-diff-qual}) and 
(\ref{anis-ball-qual}) is that they mix the components 
$\sigma_{xx}$ and $\sigma_{yy}$ of the conductivity
tensor. This will play a central role in our analysis of the
interaction effect on the magnetoresistivity
of modulated systems in Sec.~\ref{VIc}.

It is worth mentioning that the validity of the
formula (\ref{anis-ball-qual}) for the ballistic regime
is restricted on the high-temperature side by the condition
$T\lesssim T_{\rm ad}$, where $T_{\rm ad}^{-1}$ is the time
scale on which the anisotropic diffusion of the
guiding center sets in. The value of $T_{\rm ad}$
depends on the particular microscopic mechanism of 
anisotropy. We will estimate $T_{\rm ad}$ and the behavior
of the conductivity correction at $T\gg T_{\rm ad}$
for a modulated system in Sec.~\ref{VIc}.

\subsection{Calculation of the interaction-induced correction
to resistivity}
\label{VIb}
We proceed now with a formal calculation of
the quantum correction to the conductivity of an anisotropic
system in a strong magnetic field. As a model of anisotropy, we will
assume anisotropic impurity scattering, with a cross-section
$W(\phi,\phi')\neq W(\phi-\phi')$.
Repeating the derivation performed in Secs.~\ref{IIa} and \ref{IIb},
we find that the result (\ref{sigma}),\ (\ref{Bwq}) remains valid
in the anisotropic case, with the matrix $T_{\alpha\beta}$
proportional to the corresponding (anisotropic)
diffusion tensor $D_{\alpha\beta}$,
\be
\label{t_ab-anis}
T_{\alpha\beta}=
{2D_{\alpha\beta}\over v_F^2}
= {1\over 1+\omega_c^2\tau_x\tau_y} \left(
\begin{array}{cc} \tau_x      &  -\omega_c \tau_x\tau_y\\
             \omega_c \tau_x\tau_y& \tau_y
\end{array}
\right),
\ee
where $\tau_x$ and $\tau_y$ are the relaxation times 
for the corresponding 
components of the momentum.
We begin by considering the diffusive limit, when the leading
contribution comes from three-diffuson diagrams, 
Fig.~\ref{fig1}{\it d} and {\it e} (see Sec.~\ref{IIc}1),
which are represented by the last term in Eq.~(\ref{Bwq}).
The singular contribution to the propagator ${\cal D}$,
governed by the diffusion mode, has a form analogous
to (\ref{dmode}),
\be
{\cal D}^{\rm s}(\omega,\bq;\phi,\phi')
\simeq {\Psi_R(\phi,\bq)\Psi_L(\phi',\bq) 
\over D_{\alpha\beta} q_\alpha q_\beta-i\omega},
\label{dmode-anis}
\ee
see Appendix~\ref{A5}  for the derivation of (\ref{dmode-anis})
and explicit expressions of $\Psi_{R,L}$.
Using (\ref{dmode-anis}) and (\ref{PsiRL-anis}), we get
\be
\langle {\cal D}\rangle \simeq \langle {\cal D}^{\rm s}\rangle=
{1 \over D_{\alpha\beta} q_\alpha q_\beta-i\omega}
\label{<D>-anis-diff}
\ee
and
\bea
B_{xx}(\omega,\bq)&\simeq& - \langle {\cal D}^{\rm s}n_\alpha {\cal D}^{\rm s} 
n_\beta {\cal D}^{\rm s}\rangle \nonumber \\
&=&{4 \over v_F^2}\frac{D_{xx}^2q^2_x}{(D_{xx}q^2_x+D_{yy}q^2_y-i\omega)^3}.
\label{Bxx-anis-diff}
\eea
The result (\ref{Bxx-anis-diff}) can also be obtained with making
use of the identity (\ref{DnDnD}); then it is sufficient to keep only
the leading term (unity) in the expressions for functions
$\Psi_{R,L}$ entering (\ref{dmode-anis}).
Substituting (\ref{Bxx-anis-diff}), \ (\ref{<D>-anis-diff}), \ (\ref{screen})
in (\ref{sigma}), we obtain the final result for the conductivity
correction in the diffusive regime,
\bea
\delta\sigma_{xx}&=& 
{e^2\over 2\pi^2}\left({D_{xx}\over D_{yy}}\right)^{1/2}\ln T\tau, 
\label{anis-diff-xx}\\
\delta\sigma_{yy}&=& 
{e^2\over 2\pi^2}\left({D_{yy}\over D_{xx}}\right)^{1/2}\ln T\tau,
\label{anis-diff-yy}
\eea
in full agreement with a qualitative consideration of Sec.~\ref{VIa}
[Eq.~(\ref{anis-diff-qual})].
The correction to the Hall conductivity is zero in the leading
($\ln T\tau$) order, as in the isotropic case.
For the point-like interaction, the result remains the same, 
up to a factor $\nu V_0$.

We now extend the consideration beyond the diffusive limit
(thus allowing for $qR_c\gtrsim 1$), assuming first the smooth 
disorder and concentrating on longitudinal  
components of the conductivity and resistivity tensors.
In analogy with (\ref{diffuson}), 
the singular contribution ${\cal D}^{\rm s}$
to the propagator acquires then the form (see Appendix~\ref{A5})
\bea 
{\cal D}(\omega,\bq;\phi,\phi')\! &=&\!\exp\{-iqR_c[\sin(\phi-\phi_q)-\sin(\phi'-\phi_q)]\}
\nonumber \\
&\times& \frac{\chi(\phi)\chi(\phi')}{ D_{\alpha\beta} q_\alpha q_\beta-i\omega} 
\label{anis-ball-diffuson}
\eea 
where 
\be
\chi(\phi)=1-{iq v_F \over \omega_c^2}
\left({1 \over \tau_x}\cos\phi\cos\phi_q+ 
{1 \over \tau_y}\sin\phi\sin\phi_q\right).
\label{chi-anis}
\ee
This yields
\be
\langle {\cal D}\rangle=
{J_0^2(qR_c) \over D_{xx}q^2_x+D_{yy}q^2_y-i\omega }
\label{<D>-anis-ball}
\ee
and
\be
B_{xx}(\omega,\bq) 
={4 \over v_F^2}
\frac{J_0^2(qR_c) D_{xx}^2q^2_x}{(D_{xx}q^2_x+D_{yy}q^2_y-i\omega)^3},
\label{Bxx-anis-ball}
\ee
which differs from (\ref{<D>-anis-diff}), (\ref{Bxx-anis-diff})
by the factor $J_0^2(qR_c)$ only. In the ballistic limit 
$T\tau_x, T\tau_y\gg 1$ the relevant values 
of $q R_c$ are large, $q R_c\gg 1$,
so that the screening is effectively static and the 
interaction is effectively point-like with $V_0=1/2\nu$.
Substituting then (\ref{Bxx-anis-ball}) in
(\ref{sigma}) and rescaling the integration variables
$q_x=D_{xx}^{-1/2}{\tilde q}_x, \ q_y=D_{yy}^{-1/2}{\tilde q}_y$,
we find
\bea
\delta\sigma_{yy}&=& 
-{e^2\over 4\pi^2}c_0(T\tau_y)^{-1/2}\nonumber \\
&\times& 
{2\over \pi}{\bf K}\left(\sqrt{1-D_{xx}/D_{yy}}\right), 
\label{anis-ball-yy}\\
\delta\sigma_{xx}&=& 
{D_{yy}\over D_{xx}}\ \delta\sigma_{yy},
\label{anis-ball-x}
\eea
where ${\bf K}$ is the elliptic integral,
\be
\int_0^{\pi/2}\frac{d x}{\sqrt{\cos^2 x + q^2 \sin^2 x}}=
{\bf K}(\sqrt{1-q^2}), \ \ 0<q<1,
\label{K-elliptic}
\ee
and we assumed that $y$ is the easy-diffusion axis, 
$D_{yy}>D_{xx}$.

Let us analyze the obtained results in the limits of weak and strong 
anisotropy. It is convenient to set $D_{xx}=D_0$,
$\tau_x=\tau_0$, $D_{yy}=D_0+\Delta D$, and
to introduce a dimensionless anisotropy parameter 
$\alpha=\Delta D/D_0$. Using the asymptotics of the elliptic integral,
\be
{\bf K}(s)\simeq \left\{\begin{array}{ll}  \displaystyle
{\pi\over 2} \left(1+{s^2 \over 4}\right), & \ \ \ s\ll 1,\\[0.2cm]
\displaystyle
\ln{4\over \sqrt{1-s}}, & \ \ \ 1-s \ll 1,
\end{array}\right.
\label{asympt-elliptic-K}
\ee
we find 
\bea
\delta\sigma_{xx}&\simeq& -{e^2\over 4\pi^2}c_0(T\tau_y)^{-1/2} 
\nonumber \\
&\times&\left\{\begin{array}{ll}  \displaystyle
1-{\alpha\over 4}, & \ \ \ \alpha\ll 1,\\[0.2cm]
\displaystyle
{\ln(16\alpha)\over \pi \alpha^{1/2}}, & \ \ \ \alpha \gg 1,
\end{array}\right.\label{limits-anis-ball}
\eea
and $\delta\sigma_{yy}=(1+\alpha)\delta\sigma_{xx}$
Equations (\ref{anis-ball-yy}),(\ref{anis-ball-x}) and 
(\ref{limits-anis-ball}) confirm the qualitative arguments 
in Sec.~\ref{VIa} (based on the behavior of
the return probability) which led to 
Eq.~(\ref{anis-ball-qual}).

\subsection{Modulated systems}
\label{VIc}

In this subsection, we apply the results of
Sec.~\ref{VIb} to a particularly important class of
anisotropic 2D systems, namely, 2D electron gas subject to
a periodic potential varying in one direction.
Such systems (lateral superlattices) have been
intensively investigated experimentally during the last 
fifteen years. In a pioneering work\cite{weiss89},
Weiss, von Klitzing, Ploog and Weimann discovered
that even a weak one-dimensional periodic modulation
with a wave vector $\bk \parallel{\bf e}_x$ may
induce strong oscillations of the magnetoresistivity
$\rho_{xx}(B)$ [while showing almost no effect on $\rho_{yy}(B)$
and $\rho_{xy}(B)$], with the minima satisfying the condition
$2R_c/a=n-1/4$.  Here $n=1,2,\dots$ and $a=2\pi/k$ is
the modulation period.
The quasiclassical nature of these commensurability 
oscillations was demonstrated by Beenakker\cite{beenakker89},
who showed that the interplay of the cyclotron motion
and the superlattice potential induces a drift of the guiding
center along $y$ axis, with an amplitude squared oscillating
as $\cos^2(kR_c-\pi/4)$ [this is also reproduced
by a quantum-mechanical calculation, see 
Refs.~\onlinecite{gerhardts89,winkler89,vasilopoulos89}].
While Ref.~\onlinecite{beenakker89} assumed isotropic
impurity scattering (white-noise disorder), it was shown later
that the character of impurity scattering affects crucially
the dependence of the oscillation amplitude on the magnetic field.
The theory of commensurability oscillations in the 
situation of smooth disorder characteristic for high-mobility
2D electron gas was 
worked out in Ref.~\onlinecite{MW98} (see also numerical solution
of the Boltzmann equation in Ref.~\onlinecite{menne98})
and provided a quantitative description of the experimentally
observed oscillatory magnetoresistivity
$\Delta\rho_{xx}(B)$.
The result has the form\cite{MW98}
\be
\label{weiss-smooth}
{\Delta\rho_{xx}\over \rho_0}=
\frac{\pi \eta^2 k^2 l R_c}{4\ {\rm sinh}(\pi\lambda)}J_{i\lambda}(kR_c)J_{-i\lambda}(kR_c),
\ee
where $\eta$ is the dimensionless amplitude of the
modulation potential [$V(x)=\eta E_F \cos(kx)$], and
\be
\lambda={1 \over \omega_c\tau_s}
\left\{1-\left[1+{\tau_s \over \tau}(k R_c)^2\right]^{-1/2}\right\}.
\label{mu-weiss}
\ee
In the range of sufficiently strong magnetic fields
Eq.~(\ref{weiss-smooth}) describes the commensurability oscillations
with an amplitude proportional to $B^3$,
\be
\label{weiss-cos-smooth}
{\Delta\rho_{xx}\over \rho_0}\simeq
\eta^2 { (\omega_c \tau)^2 \over \pi k R_c}\cos^2(kR_c-\pi/4),
\ee
For precise conditions of validity of (\ref{weiss-cos-smooth}),
as well as for an analysis of the result (\ref{weiss-smooth})
in the whole range of magnetic fields, the reader is referred to 
Ref.~\onlinecite{MW98}.

As to the modulation-induced corrections 
$\Delta\rho_{xy}, \Delta\rho_{yy}$ to the other components of the 
resistivity tensor, they are exactly zero within the quasiclassical 
(Boltzmann equation) approach, independently of the form
of the impurity collision integral\cite{beenakker89,MW98,menne98}.
Such corrections appear in the quantum-mechanical treatment
of the problem~\cite{gerhardts90,PV} and are related to the 
de Haas-van Alphen oscillations of the density of states
induced by the Landau quantization of spectrum.
As a consequence, these oscillations are exponentially 
damped by disorder, with the damping factor 
$\sim \exp[-2\pi/\omega_c\tau_s]$. 
The phase of such quantum oscillations $\Delta\rho_{yy}^{({\rm DOS})}$
is opposite to that of quasiclassical commensurability oscillations
in $\Delta\rho_{xx}$, Eqs.~(\ref{weiss-smooth}), (\ref{weiss-cos-smooth}).
Indeed, oscillations in $\Delta\rho_{yy}$ that are much weaker
than those in $\Delta\rho_{xx}$, have the opposite phase, and vanish
much faster with decreasing $B$, were observed in Ref.~\onlinecite{weiss89}.
We will neglect these oscillations, 
which are exponentially weak in the range of magnetic
fields considered in the present paper, $\omega_c\tau_s/\pi \ll 1$.
We are going to show that the interaction-induced correction 
to resistivity also generates oscillations in $\rho_{yy}$,
which are, however, unrelated to the DOS oscillations of
a non-interacting system and become dominant with
lowering temperature.

To demonstrate this, we apply the result of Sec.~\ref{VIb}
for the interaction-induced correction in an anisotropic system.
The anisotropy parameter is governed by the quasiclassical
correction to $\rho_{xx}$ due to
modulation,
\be
\alpha = {\sigma_{yy}\over \sigma_{xx}}-1
\simeq {\rho_{xx}\over \rho_{yy}}-1={\Delta \rho_{xx}\over \rho_0}
\label{beta-weiss}
\ee
and is given by Eq.~(\ref{weiss-smooth}). For
simplicity we will assume that the effect of 
modulation is relatively weak, $\alpha\ll 1$.
(Generalization to the large-$\alpha$ case with making use of
the corresponding formulas of Sec.~\ref{VIb} is completely
straightforward.) Using (\ref{anis-diff-xx}) and (\ref{anis-ball-x}),
we find the oscillatory correction to $\rho_{yy}$ as a 
combined effect of the modulation and the Coulomb interaction,
\bea
{\delta\rho_{yy}\over \rho_0}&=&\frac{(\omega_c\tau)^2}{2\pi k_F l}
{\Delta\rho_{xx}\over \rho_0}\nonumber   \\
&\times& \left\{\begin{array}{ll}  \displaystyle
-\ln T\tau, & \ \ \ T\tau\ll 1,\\[0.2cm]
\displaystyle
{c_0\over 4}(T\tau)^{-1/2}, & \ \ \ T\tau \gg 1.
\end{array}\right.\label{interact-weiss}
\eea
In the presence of strong scatterers (mixed disorder model),
the result for the ballistic regime is enhanced by the factor 
$4(\tau_{\rm sm}/\tau)^{1/2}\gg 1$, as discussed in Sec.~\ref{V}.

Let us remind the reader that the result (\ref{interact-weiss}) is valid
for temperatures $T\ll T_{\rm ad}$, where 
$T_{\rm ad}^{-1}\ll \tau$ is the characteristic time
on which the motion of the guiding center
takes the form of anisotropic
diffusion (see Sec.~\ref{VIa}). For the case of a modulated system
with a smooth random potential we find 
$T_{\rm ad}^{-1}\sim \tau(a/R_c)^2$. This is because on
a scale shorter than $T_{\rm ad}^{-1}$ the modulation leads to a drift
of the guiding center along $y$ axis
with the velocity depending on the coordinate $X$ of the
guiding center~\cite{beenakker89},
\bea
v_d(X)&=&-{\eta v_F\over 2}kR_c J_0(kR_c) 
\sin(kX) \label{vdrift-mod} \\
&\simeq& -\frac{\eta v_F}{\sqrt{2\pi kR_c}}\cos(kR_c-\pi/4)\sin(kX).
\nonumber
\eea
In a time $a^2/D\equiv T_{\rm ad}^{-1}$ the position $X$ of the
guiding center is shifted by a distance of the order of the
modulation period $a$ due to the small-angle impurity scattering.
Therefore, the drift velocity $v_d$ 
typically changes sign on this time scale, so that
the drift is transformed to an additional diffusion process,
with $\Delta D_{yy} \sim \langle v_d^2 \rangle T_{\rm ad}^{-1}$,
in agreement with (\ref{weiss-cos-smooth}).
To estimate the resistivity correction $\delta\rho_{yy}$
in the ultra-ballistic regime $T\gg T_{\rm ad}$, we use
the relation between the conductivity correction
and the return probability (Sec.~\ref{IV}).
The return probability $R_n$ after $n$ revolutions
(introduced in Sec.~\ref{IV}) is modified by the 
modulation-induced drift in the following way:
\be
R_n^{\rm mod}=R_n
\left(1-n\:\omega_c\tau{\pi \langle v_d^2 \rangle \over 2v_F^2}
\right).
\label{Rn-mod}
\ee
According to (\ref{qual-sigma}), this yields an oscillatory
correction to resistivity
suppressed by a factor $\sim T_{\rm ad}/T$ as compared to the second
line (ballistic regime) of Eq.~(\ref{interact-weiss}).

\begin{figure}
  \medskip
  \includegraphics[width=8cm]{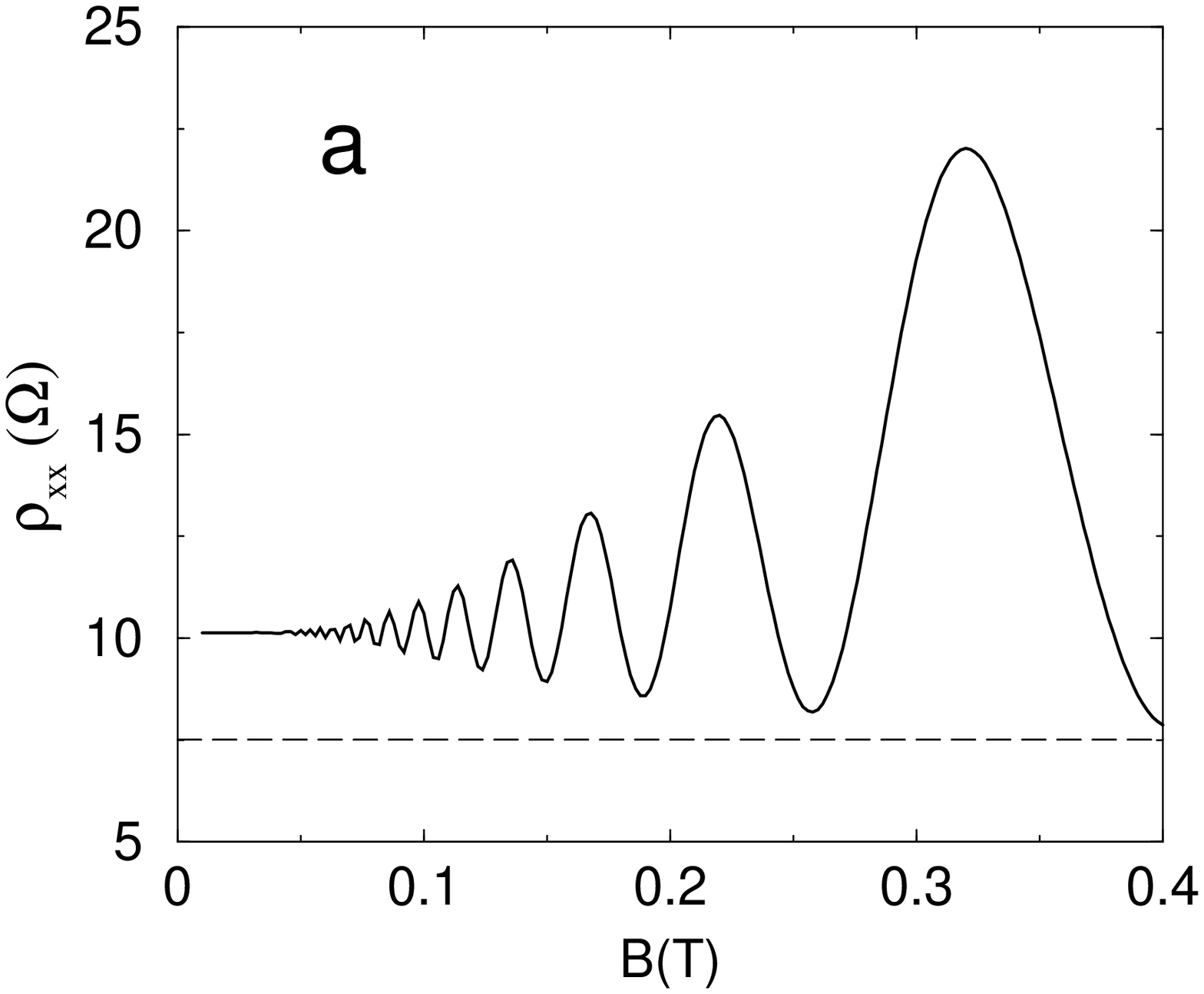}
  \vspace{2mm}
\includegraphics[width=8cm]{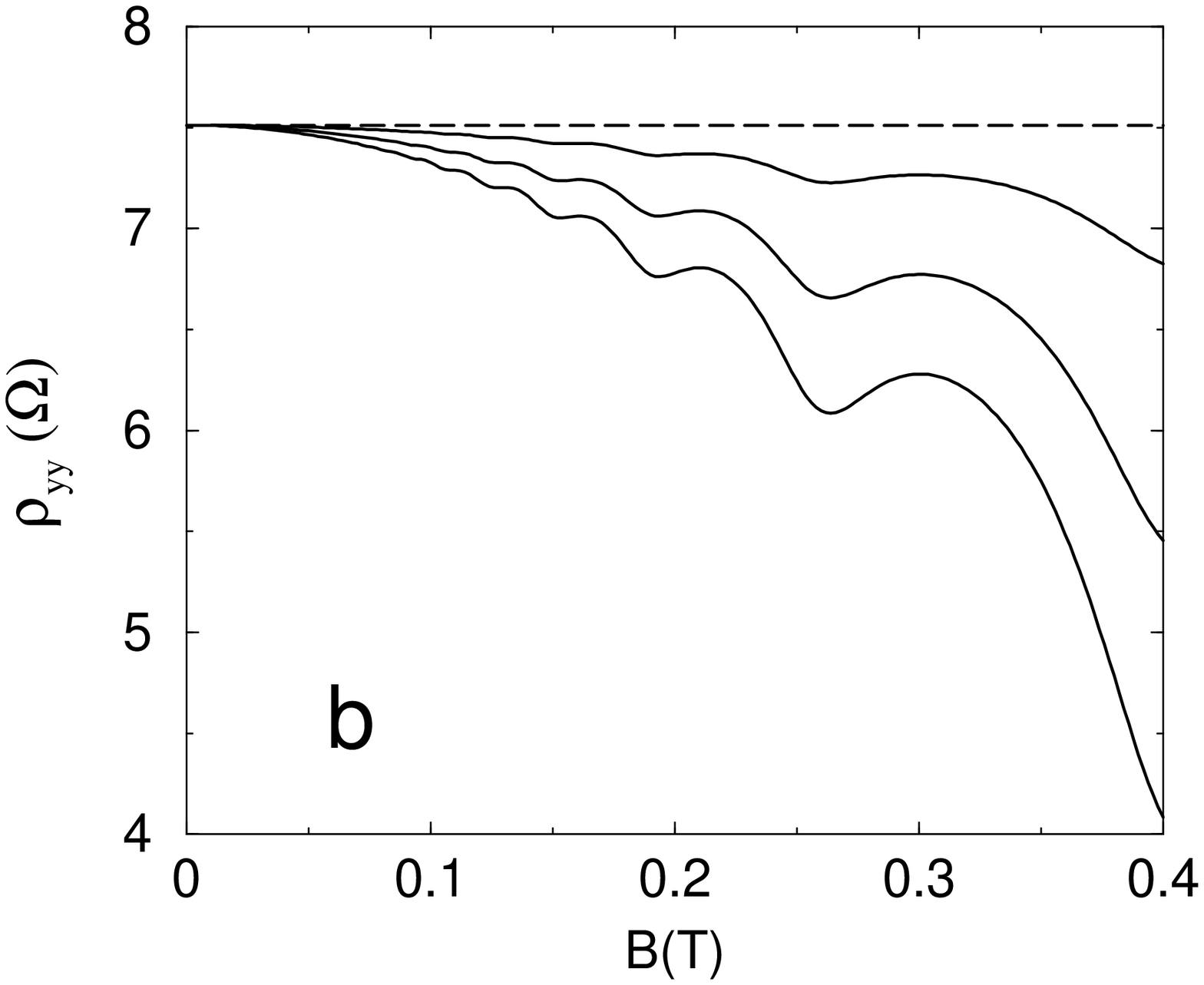}
\vspace{2mm}
\caption{ Magnetoresistivity in a lateral
superlattice with modulation wave vector ${\bf k}\parallel {\bf e}_x$.
(a) Quasiclassical Weiss-oscillations; dashed line shows the
resistivity in the absence of modulation. (b)
Interaction-induced quantum oscillations in $\rho_{yy}$ for
three temperatures. The curves correspond to the
values of the parameter $2c_0(\tau_{\rm sm}/T\tau^2)^{1/2}=0.1,\ 0.3,\ 0.5$
(from top to bottom), assuming mixed disorder. 
Dashed line represents the resistivity of
the non-interacting system. Typical experimental parameters
are used: effective mass $m=0.067\times 9.1\times10^{-28}$ g,
electron density $n_e=3.16\times10^{11}\ {\rm cm}^{-2}$,
modulation strength $\eta=0.05$, modulation period
$a=260$nm, momentum and single-particle
relaxation times $\tau=100$pc and $\tau_s=5$pc, 
respectively.
} 
\label{weiss} 
\end{figure}

Let us summarize the results obtained in this subsection.
We have shown that in a periodically modulated system
the interaction induces, in addition to the quadratic MR studied in
Secs.~\ref{III} and \ref{IV}, an oscillatory contribution
to the component $\rho_{yy}$ of the resistivity tensor,
which is not affected by modulation (and thus shows no
oscillations) within the Boltzmann theory. When the parabolic MR
is negative (meaning that the exchange contribution dominates),
which is the case under typical experimental conditions
and for not too high temperatures, these quantum 
interaction-induced oscillations in $\rho_{yy}$ are {\it in phase}
with classical oscillations in $\rho_{xx}$, as follows immediately
from Eq.~(\ref{interact-weiss}) [see Fig.~\ref{weiss}]. In other words, their phase is
opposite to that of the above-mentioned contribution
$\Delta\rho_{yy}^{({\rm DOS})}$ induced by the DOS oscillations.

We come therefore to the following conclusion concerning the phase of
the total oscillatory contribution to $\rho_{yy}$.
While at sufficiently high temperatures the 
$\rho_{yy}$-oscillations have, due to the contribution 
$\Delta\rho_{yy}^{({\rm DOS})}$ [and possible due to the Hartree counterpart
of Eq.~(\ref{interact-weiss})], the phase opposite to 
$\Delta\rho_{xx}$, with lowering temperature the exchange contribution
Eq.~(\ref{interact-weiss}) starts to dominate, implying that $\rho_{yy}$
oscillates in phase with $\rho_{xx}$. Furthermore, the both contributions
are damped differently by disorder: the high-temperature out-of-phase
oscillations $\Delta\rho_{yy}^{({\rm DOS})}$ vanish with lowering $B$ much faster 
that the low-temperature in-phase interaction-induced 
oscillations $\delta\rho_{yy}$.

Our results are in qualitative agreement with a recent
experiment\cite{Mitzkus02}. It was observed there
that at sufficiently high temperature, $T\agt 2.5~K$,
the oscillations in $\rho_{yy}$ have the opposite phase with
respect to $\rho_{xx}$, in accord with earlier experimental
findings\cite{weiss89}. However, when the temperature was lowered,
the phase has changed and $\rho_{yy}$ started to oscillate
in phase with $\rho_{xx}$, with an amplitude increasing with 
decreasing $T$. In addition to these novel oscillations, a smooth
negative MR was seen to develop in the same temperature
range. The authors of Ref.~\onlinecite{Mitzkus02} emphasized a puzzling
character of the temperature dependence of the observed oscillations,
which cannot be explained by earlier 
theories\cite{beenakker89,gerhardts89,winkler89,vasilopoulos89,MW98} discarding
the interaction effects.
Our theory leading to Eq.~(\ref{interact-weiss}) provides a plausible
explanation of these experimental findings. Quantitative comparison
of the theory and experiment requires, however, a more 
systematic experimental study of the temperature  dependence
of the amplitude of $\rho_{yy}$-oscillations in a broader 
temperature range.

\section{Conclusions}
\label{VII}
\setcounter{equation}{0}

\subsection{Summary of main results}
\label{VIIa}

Let us summarize the key results of the present paper. 
We have derived a general formula (\ref{sigma}), (\ref{Bwq_general})
for the interaction-induced quantum correction 
$\delta\sigma_{\alpha\beta}$ to the conductivity tensor of 2D  
electrons valid for arbitrary temperature, magnetic field
and disorder range.  
It expresses $\delta\sigma_{\alpha\beta}$ in terms of 
classical propagators in random potential 
(``ballistic diffusons''). In the appropriate limiting cases, it
reproduces all previously known results on the interaction correction
(see Sec.~\ref{IIc}).

Applying this formalism, we 
have calculated the interaction contribution to the MR 
in strong $B$ in systems with various types of disorder and for
arbitrary $T\tau$.
In the diffusive limit, $T\tau\ll 1$, the result does not depend on the
type of disorder, as expected. Specifically, the 
MR scales with magnetic field and temperature as
follows, $\delta\rho_{xx}\propto B^2\ln(T\tau)$  and
$\delta\rho_{xy}\propto B\ln T\tau$, in agreement with
Refs.~\onlinecite{SenGir} and \onlinecite{girvin82}. 

In the ballistic limit, $T\tau\gg 1$, the result is strongly affected by
the character of disorder. In Sec.~\ref{III} we have performed a
detailed study of the case of smooth disorder characteristic for
high-mobility GaAs heterostructures. We have found that the
temperature-dependent MR scales
at $\omega_c\gg T$ as $\delta\rho_{xx}\propto B^2(T\tau)^{-1/2}$ and 
$\delta\rho_{xy}\propto B(T\tau)^{1/2}$. In addition, there is a
temperature-independent (but larger) contribution $\propto B^{3/2}$ 
to the Hall resistivity. In the opposite limit
$\omega_c\ll T$ the MR is suppressed.

We have further considered a mixed disorder model, with strong
scatterers (modeled by white-noise disorder) 
superimposed on a smooth random potential
(Sec.~\ref{V}). A qualitatively new situation arises when
the momentum relaxation rate $\tau_{\rm sm}^{-1}$ due to smooth
disorder is much less than
the total momentum relaxation rate $\tau^{-1}$. 
Such a model is believed to be relevant to Si-based
structures, as well as to GaAs structures with very large spacer. 
We have shown that in the ballistic limit and at $\omega_c\gg T$ 
the corrections to both longitudinal and Hall resistivities are enhanced
(as compared to the case of smooth disorder)
by a factor $\sim (\tau_{\rm sm}/\tau)^{1/2}\gg 1$. 
In the range of weaker magnetic fields, $\omega_c\ll T$, the
interaction-induced MR scales in the ballistic regime as
$\Delta\rho_{xx}\propto B^2(T\tau)^{-1}$
and $\delta\rho_{xy}\propto B(T\tau)^{-1}
[1-{\rm const}\:(\omega_c\tau)^2]$. 

For a weak interaction ($\kappa\ll k_F$) the correction is dominated
by the exchange contribution, implying that $\Delta\rho_{xx}$ is
negative and that the slope of $\rho_{xy}$ decreases with increasing
temperature. This is true up to a temperature $T_{\rm H} \gg \tau^{-1}$
(defined in Sec.~\ref{IIId}) where the sign changes. 
In the case of a strong interaction the magnitude of the Hartree
contribution (and thus the sign of the total correction)
depends on angular harmonics $F_n^{\rho,\sigma}$ of the Fermi-liquid
interaction (Sec.~\ref{IIIe}). It is worth emphasizing that in
contrast to the diffusive limit where only $F_0^\sigma$ is relevant,
in the ballistic regime all the Fermi-liquid parameters are, strictly
speaking, important, see Eq.~(\ref{fermi-hartree-ball}).  
Therefore, predictions of the ``$F_0^\sigma$-approximation'' (with
only one Fermi-liquid parameter retained) should be treated with
caution. 

We have further applied our formalism to anisotropic systems
(Sec.~\ref{VI}) and
demonstrated that the correction mixes the components $\rho_{xx}$ and
$\rho_{yy}$ of the resistivity tensor. This result is of special
interest in the case of systems subject to a one-dimensional periodic
modulation (lateral superlattice; wave vector 
${\bf k}\parallel {\bf e}_x$). Specifically, we have shown that
the interaction induces novel oscillations in $\rho_{yy}$, which are
in phase with quasiclassical commensurability (Weiss) oscillations in
$\rho_{xx}$.  

\subsection{Comparison with experiment}
\label{VIIb}

Our results for $\rho_{xx}$ in the case of smooth disorder (published
in a brief form in the Letter \cite{prl}) have been confirmed by a
recent experiment on $n$-GaAs system \cite{Sav}, which was performed
in the broad temperature range, from the diffusive to the ballistic
regime. Specifically, Li {\it et al.} \cite{Sav} found that the
MR 
scales as $\Delta\rho_{xx} \propto B^2$ in strong magnetic fields.
The obtained temperature-dependence of the proportionality coefficient
$G(T\tau)$ was in good agreement with our predictions.

Very recently, Olshanetsky {\it et al.} \cite{kvon}
studied the MR in
the ballistic regime in a Si/SiGe
structure of $n$-type, where both short- and long-range potential are expected to
be present. They found the interaction-induced correction
to $\rho_{xx}$ larger by a factor $\sim 5$ as compared to our prediction
\cite{prl} for the case of smooth disorder. This conforms
with the results of the present paper for the mixed-disorder model, where
we find an enhancement of $\Delta\rho_{xx}$ by a
factor $4(\tau_{\rm sm}/\tau)^{1/2}\gg 1$.

As has been mentioned in Introduction, the interaction-induced
MR in the ballistic 
regime was measured for the first time as early as in 
1983, by Paalanen, Tsui, and Hwang\cite{PTH83}, who studied GaAs
structures. Again, a parabolic temperature-dependent  
MR $\Delta\rho_{xx}$ was obtained, in
agreement with our findings. However, its magnitude was
considerably (roughly an order of magnitude) larger compared with our
theoretical result for the case of smooth disorder, as well as with
the recent experiment \cite{Sav}. We speculate that samples used in
Ref.~\onlinecite{PTH83} probably contained an appreciable concentration of
background impurities, which has led to an enhancement of the
interaction-induced contribution to resistivity, similarly to the
recent work \cite{kvon}. (Indeed,
the results for the mixed-disorder model
shown in Fig.~\ref{mix} may create an impression
that the $\log T$ behavior extends up to $T\tau\sim 10$,
as was concluded in Ref.~\onlinecite{PTH83}.) 
Remarkably, the interaction-induced quantum
correction to conductivity may serve as an indicator of the dominant
type of disorder.

To the best of our knowledge, no experimental study of the interaction
effects on Hall resistivity $\rho_{xy}$ has been published. This part
of our predictions therefore awaits its experimental verification.

Finally, our results for systems with one-dimensional periodic
modulation are in qualitative agreement with the recent work by
Mitzkus {\it et al.} \cite{Mitzkus02}, as we discussed in detail 
in Sec.~\ref{VIc}. Quantitative comparison of the theory and experiment
requires an experimental study of the temperature-dependence of novel
oscillations (found experimentally in Ref.~\onlinecite{Mitzkus02} and
theoretically in the present paper) in a broader temperature range.

\subsection{Outlook}
\label{VIIc}

Before closing the paper, we list a few further applications of our 
formalism and its possible generalizations.  First, our results 
can be generalized to frequency-dependent (rather than
temperature-dependent) MR. 
Second, the interaction effects in systems of other 
dimensionality, as well as in macroscopically 
inhomogeneous systems, can be investigated by our general
method. Third, the formalism can be
used to calculate the phonon-induced contribution to resistivity,
which becomes larger than that due to Coulomb interaction at
sufficiently high temperatures.  
Further, thermoelectric phenomena  
in the full range of magnetic fields and temperatures can be studied
in a similar way. 
Finally, our approach can be generalized to the
regime of still stronger magnetic fields, where the Landau
quantization can not be neglected anymore; the work in this 
direction is in progress \cite{in-progress}.

\section{Acknowledgments}
We thank P.T.~Coleridge, A.V.~Germanenko, Z.D.~Kvon, G.M.~Minkov,
C.~Mitzkus, A.K.~Savchenko, and D.~Weiss for informing
us on experimental results prior to publication and
for stimulating discussions.
This work was supported by the Schwerpunktprogramm 
``Quanten-Hall-Systeme'' and the SFB195 der Deutschen 
Forschungsgemeinschaft, and 
by the RFBR.

\appendix

\section{Cancellation of the inelastic term}
\label{A1}
\renewcommand{\theequation}{A.\arabic{equation}}
\setcounter{equation}{0}

\begin{figure}
  \medskip
  \includegraphics[width=8cm]{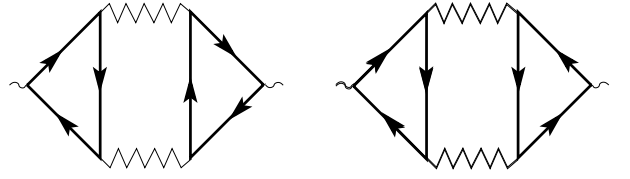}
 \caption{ Aslamazov-Larkin-type diagrams describing the Coulomb-drag 
 contribution to the resistivity, which cancels the ``inelastic''
 part of the diagrams {\it f},\ {\it g} from Fig.~\ref{fig1}. } 
\label{drag} 
\end{figure}

As discussed in Sec.~\ref{IIa}, diagrams {\it f} and {\it g}
give rise, in addition to the contribution (\ref{sigmaf+g}),
to a term of the type (\ref{sigma-inel}), characteristic for
inelastic effects. This yields at $B=0$ a disorder-independent
correction to resistivity $\delta\rho \sim (T/eE_F)^2$,
see below. Note that such a contribution to
resistivity would be obtained if one substitutes the
inelastic relaxation rate of a clean 2D electron gas,
$\tau_{\rm inel}^{-1}\sim T^2/E_F$ in the Drude formula (\ref{drude}).
However, such a procedure clearly makes no sense. Indeed, in a
translationally invariant system electron-electron collisions
conserve the total momentum and thus give
no contribution to resistivity. 
Therefore, the correction (\ref{sigma-inel}) ought to be
canceled by some other contribution.
Below we show explicitly that this is indeed the case,
and that this second contribution is of the Coulomb-drag type,
described by the diagrams in Fig.~\ref{drag}.

For simplicity, we restrict our consideration
here to the case of zero $B$ and white-noise disorder,
which allows us to use the results of Ref.~\onlinecite{kamenev-oreg}
for the Coulomb drag. 
Note that while Ref.~\onlinecite{kamenev-oreg} considered the drag
between two layers, we refer to the ``self-drag'' within a single layer.
As we will see below, the characteristic momenta $q$ determining the
contribution (\ref{sigma-inel}) are large, $q\sim k_F$.
For this reason, there is no need to take into account
impurity-line ladders while evaluating this term, similarly to the
calculation of drag in Ref.~\onlinecite{kamenev-oreg} for 
a small inter-layer distance.
We thus have
\bea
\delta B_{xx}^{\rm f}(\omega,\bq)&=&
{1\over 2\pi \nu v_F^2 }\int {d^2p\over (2\pi)^2} \nonumber \\
&\times& {\rm Re}\left[ 
2 p_x^2 G^2_{\rm R}(\epsilon,p) 
G_{\rm R}(\epsilon-\omega,p-q) G_{\rm A}(\epsilon,p)\right.
\nonumber \\
&+&p_x (p_x-q_x) G_{\rm R}(\epsilon,p)G_{\rm A} (\epsilon,p)\nonumber \\
&\times&\left.
G_{\rm R}(\epsilon-\omega,p-q)G_{\rm A} (\epsilon-\omega,p-q)\right],
\label{deltaBf}
\eea
where $G_{\rm R,A}(\epsilon,p)=(E_F+\epsilon - p^2/2m \pm i/2\tau)^{-1}$
are the disorder-averaged retarded and advanced Green's functions.
Using the identity 
$$G_{\rm R}(\epsilon,p)G_{\rm A}(\epsilon,p)=
i\: \tau\: [\: G_{\rm R}(\epsilon,p)-G_{\rm A}(\epsilon,p)\: ],$$
we reduce (\ref{deltaBf}) to the form
\bea
\delta B_{xx}^{\rm f}(\omega,\bq)&=&
-{\tau^2 \over \pi v_F^2\nu}\int {d^2p\over (2\pi)^2} p_x q_x \nonumber \\
&\times& {\rm Re}[G_{\rm R}(\epsilon,p)G_{\rm A}(\epsilon-\omega,p-q)]\nonumber \\
&=&-{\tau^2 q_x^2 \over 2 v_F^2\nu \omega} {\rm Im}\Pi(\omega,\bq),
\label{Bf-ImPi}
\eea
where $\Pi(\omega,\bq)$ is the polarization operator (\ref{polarization}),
\bea
{\rm Im}\Pi(\omega,\bq)&=&
{\omega\over \pi} 
\int {d^2p\over (2\pi)^2} G_{\rm R}(\epsilon,p)G_{\rm A}(\epsilon-\omega,p-q)\nonumber \\
&\simeq&2\nu {\omega \over q v_F}\theta(qv_F-\omega),
\label{ImPi-ball}
\eea
where $\theta(x)$ is the step function.
Furthermore, the imaginary part of the interaction propagator
within the RPA is proportional to ${\rm Im}\Pi(\omega,\bq)$
\be
{\rm Im}U(\omega,\bq)=-|U(\omega,\bq)|^2 {\rm Im}\Pi(\omega,\bq).
\label{ImU}
\ee
Substituting  (\ref{Bf-ImPi}) and (\ref{ImU})
in (\ref{sigma-inel}), we finally obtain
\bea
\delta\sigma^{\rm inel}_{xx}&=&
-{e^2 \tau^2\over m^2}
\int_{-\infty}^\infty {d\omega \over 2\pi}{1\over 2T\sinh^{2}(\omega/2T)}
\int {d^2q\over (2\pi)^2}\nonumber \\
&\times&q_x^2|U(\omega,\bq)|^2[{\rm Im}\Pi(\omega,\bq)]^2,
\label{sigma-in-ImPi}
\eea
This expression is identical, up to a sign, to the result 
of Ref.~\onlinecite{kamenev-oreg} for Coulomb drag.
This demonstrates that two contributions indeed
cancel each other, 
\be
\delta\sigma^{\rm inel}+\delta\sigma^{\rm drag}=0.
\label{nol}
\ee
Using the explicit form of ${\rm Im}\Pi(\omega,\bq)$, Eq.~(\ref{ImPi-ball}),
and of $U(\omega,q)$, Eq.~(\ref{screen}), in the ballistic
regime, it is easy to estimate
$\delta\sigma^{\rm inel}$ (we assume here $\kappa\sim k_F$ for simplicity),
\be
\delta\sigma^{\rm inel} 
\sim -e^2 \tau^2 T \int_0^T d\omega \int^{k_F} {q dq\over k_F^2}
\sim -e^2(T\tau)^2.
\label{sigma-inel-estim}
\ee
As has been stated above, the $q$-integral
is determined by the ultraviolet cutoff.

Finally, we note that in double-layer system
the inter-layer interaction does give rise to
a correction $\delta\sigma^{\rm inel}$ to the driving-layer
conductivity, which is equal in magnitude and opposite in sign
to the transconductivity.
This effect is, however, reduced by a factor $\sim (k_F \xi)^{-4}$
(where $\xi$ is the interlayer distance), as compared to 
(\ref{sigma-inel-estim}), see Ref.~\onlinecite{kamenev-oreg}.

\section{Propagator and kernels $B_{\alpha\beta}$ for white-noise disorder}
\label{A2}
\renewcommand{\theequation}{B.\arabic{equation}}
\setcounter{equation}{0}

In this appendix we will derive the general expressions
(valid for arbitrary magnetic field) 
for the kernels $B_{xx}^{(\rho)}$ and 
$B_{xy}^{(\rho)}$ in terms of the quasiclassical propagator
for a white-noise random potential.
This will allow us to reproduce the results of 
Refs.~\onlinecite{ZNA-sigmaxx,ZNA-rhoxy}, where
the interaction-induced corrections to $\sigma_{xx}$ and
$\rho_{xy}$ were studied for a white-noise disorder
in the limit $B\to 0$. We will further apply the formalism to calculate 
the longitudinal MR and the Hall resistivity
in a {\it finite} magnetic field with $\omega_c\ll T$.
The resistivity tensor in
yet stronger magnetic field, $\omega_c\gg T$, 
is studied, in the more general framework of
a mixed disorder model in Sec.~\ref{V}.

Using Eqs.~(\ref{Bwq_general}) and (\ref{Brho}), we get
\bea
B_{xx}^{(\rho)}&=&{1 \over 2\tau}\langle {\cal D}\rangle^2-
{1 \over \tau}\langle {\cal D} n_x\rangle\langle n_x {\cal D}\rangle
\nonumber\\
&+&{1\over 2}\langle {\cal D}\rangle
-\langle n_x {\cal D} n_x\rangle\nonumber\\
&-&{2\over \tau}\langle n_x {\cal D} n_x {\cal D}\rangle
+2\omega_c\langle n_x {\cal D} n_y {\cal D}\rangle\nonumber\\
&-&{1-\omega_c^2\tau^2\over \tau^2}
\langle {\cal D} n_x {\cal D} n_x {\cal D}\rangle\nonumber\\
&+&{2\omega_c\over \tau}\langle {\cal D} n_x {\cal D} n_y {\cal D}\rangle
\label{Brhoxx-WN}
\eea
for the kernel describing the longitudinal resistivity,
and 
\bea
B_{xy}^{(\rho)}&=&{\omega_c\over 2}\langle {\cal D}{\cal D}\rangle-
{1 \over \tau}\langle {\cal D} n_x\rangle\langle n_y {\cal D}\rangle
-\langle n_x {\cal D} n_y\rangle\nonumber\\
&-&{2\over \tau}\langle n_x {\cal D} n_y {\cal D}\rangle 
- 2\omega_c\langle n_x {\cal D} n_x {\cal D}\rangle
\nonumber\\
&-&{1-\omega_c^2\tau^2 \over \tau^2}
\langle {\cal D} n_x {\cal D} n_y {\cal D}\rangle\nonumber\\
&-&{2\omega_c\over \tau}\langle {\cal D} n_x {\cal D} n_x {\cal D}\rangle
\label{Brhoxy-WN}
\eea
for the Hall resistivity.

The propagator ${\cal D}(\phi,\phi')$ in the case of white-noise disorder
can be expressed through the  propagator
${\cal D}_0(\phi,\phi')$, obeying the Liouville-Boltzmann equation
with only scattering-out term present in the collision integral,
\bea
{\Big [}-i\omega + i q v_F \cos(\phi-\phi_q)
&+&\omega_c{\partial\over\partial\phi} 
+{1\over \tau}\ {\Big ]}
{\cal D}_0(\phi,\phi')\nonumber \\
 &=& 2\pi\delta(\phi-\phi'). 
\label{LB-out-WN} 
\eea
As in a zero magnetic field, the total propagator 
is given by the sum of the ladder-diagrams
(thus including the scattering-in processes), yielding 
\be
{\cal D}(\phi,\phi')={\cal D}_0(\phi,\phi')+
\int {d\phi_1\over 2\pi} {d\phi_2\over 2\pi}
\frac{{\cal D}_0(\phi,\phi_1){\cal D}_0(\phi_2,\phi')}{\tau - \langle {\cal D}_0 \rangle},
\label{prop-WN}
\ee
which we write symbolically as follows
\be
{\cal D}={\cal D}_0+
\frac{{\cal D}_0 \rangle\langle {\cal D}_0}{\tau - g_0}.
\label{prop-WN-symb}
\ee
Here we introduced a short-hand notation
\be
g_0(\omega,\bq)\equiv \langle {\cal D}_0 \rangle = 
\int {d\phi\over 2\pi} {d\phi'\over 2\pi}{\cal D}_0(\omega,\bq;\phi,\phi')
\label{g0-def}
\ee
for the angle-averaged scattering-out propagator.
It turns out that for a white-noise disorder
the kernels $B_{xx}^{(\rho)}$ and 
$B_{xy}^{(\rho)}$ can be expressed in terms 
of $g_0$ (and its derivatives with respect to $q$ and $\omega$).
The solution of (\ref{LB-out-WN}) is given by
\bea
{\cal D}_0(\omega,\bq;\phi,\phi')
&=&\exp\{i q R_c[\sin(\phi'-\phi_q)-\sin(\phi-\phi_q)]\}
\nonumber \\
&\times&\sum_{n=-\infty}^{\infty}\frac{\exp[i n(\phi-\phi')]}
{-i(\omega - n\omega_c)+1/\tau}.
\label{D0-WN-solution}
\eea
It is worth mentioning that in the mixed-disorder model
introduced
in Sec.~\ref{V} with both, white-noise and smooth disorder present, 
the solution of the Liouville-Boltzmann 
equation also has the form (\ref{prop-WN}).
In that case, the propagator ${\cal D}_0$
satisfies the Liouville-Boltzmann equation for a purely
smooth disorder (considered in Appendix~\ref{A4}) 
with the replacement $\omega \to \omega+i/\tau_{\rm wn}$,
where $\tau_{\rm wn}$ is relaxation time due to white-noise potential.

Using (\ref{D0-WN-solution}) and a series representation 
for the Bessel functions,
we find [see, e.g. Ref.~\onlinecite{MW98}]
\bea
g_0(\omega,\bq)&=&{i\over \omega_c}\sum_n\frac{J^2_n(qR_c)}{\mu-n}\nonumber \\
&=&{i\pi \over \omega_c} \frac{J_\mu (qR_c) J_{-\mu}(qR_c)}{\sin{\pi\mu}},
\label{g0wq-WN}
\eea
where $J_\mu(z)$ is the Bessel function and
\be
\mu = {\omega\over \omega_c} + {i\over \omega_c\tau}.
\label{mu-def}
\ee
In the absence of magnetic field ($\omega_c=0, \ R_c=v_F/\omega_c = \infty$)
the propagators ${\cal D}_0(\omega,\bq;\phi,\phi')$ 
and $g_0(\omega,\bq)$ acquire a simple form
\bea
&&{\cal D}_0(\omega,\bq;\phi,\phi')=
\frac{2\pi\delta(\phi-\phi')}{-i\omega+qv_F\cos(\phi-\phi_q)+1/\tau},
\nonumber \\
\label{D0B=0}\\
&&g_0(\omega,\bq)={1 \over \sqrt{q^2v_F^2 + 
(-i\omega+1/\tau)^2}}\equiv {1 \over S(\omega,q)}.\nonumber \\
\label{g0B=0}
\eea
To proceed further, we first reduce [using (\ref{prop-WN-symb})]
the ``matrix elements'' appearing in (\ref{Brhoxx-WN}) and
(\ref{Brhoxy-WN}) to the form containing only the propagators
${\cal D}_0$,
\bea
\langle {\cal D}\rangle
&=&{\langle{\cal D}_0\rangle\tau  \over \tau - g_0}, 
\label{<D>-WN} \\
\langle {\cal D\cal D}\rangle
&=&\frac{\tau^2 \langle {\cal D}_0{\cal D}_0\rangle}{(\tau-g_0)^2}, 
\label{<DD>-WN} \\
\langle {\cal D} n_x\rangle\langle n_{\beta} {\cal D}\rangle 
&=&
\frac{\tau^2\langle {\cal D}_0 n_x\rangle\langle n_{\beta} {\cal D}_0\rangle }{(\tau-g_0)^2}, 
\label{DnnD-WN}\\ 
\langle n_x {\cal D} n_{\beta}\rangle 
&=& \langle n_x {\cal D}_0 n_{\beta}\rangle \nonumber \\ 
&+& 
{\langle n_x {\cal D}_0 \rangle\langle {\cal D}_0 n_{\beta} \rangle \over \tau-g_0},
\label{nDn-WN}\\
\langle n_x {\cal D} n_\beta {\cal D}\rangle &=& \langle n_x {\cal D}_0 n_\beta {\cal D}_0\rangle
\nonumber \\
&+&\frac{\tau  \langle n_x {\cal D}_0 \rangle 
\langle {\cal D}_0 n_{\beta} {\cal D}_0\rangle }{(\tau-g_0)^2}, 
\label{nDnD-WN}\\
\langle {\cal D} n_x {\cal D } n_{\beta} {\cal D}\rangle &=&
\frac{\tau^2 \langle {\cal D}_0 n_x {\cal D }_0 n_{\beta} {\cal D}_0\rangle
}{(\tau-g_0)^2} \nonumber \\
&+&\frac{\tau^2 \langle {\cal D}_0 n_x {\cal D}_0\rangle
\langle {\cal D}_0 n_{\beta} {\cal D}_0\rangle }{(\tau-g_0)^3},
\label{DnDnD-WN}
\eea
where $\beta=x,y$.
Next, using (\ref{D0-WN-solution}) and (\ref{LB-out-WN}) 
and performing the 
averaging over $\phi_q$,
we can express the matrix elements involving ${\cal D}_0$ via the propagator
$g_0$. Introducing the notation 
${\cal W}=-i\omega+1/\tau$, we get the 
following $\phi_q$-averaged matrix elements,
\bea
\langle {\cal D}_0{\cal D}_0\rangle&=&-i{\partial g_0 \over \partial \omega},
\label{<D0D0>}\\
\langle {\cal D}_0 n_x\rangle\langle n_{x} {\cal D}_0\rangle &=&
-{1\over 2 q^2v_F^2}\left[1-{\cal W} g_0\right]^2 \nonumber \\
&+ & {\omega_c^2\over 8 v_F^2}\left({\partial g_0\over \partial q}\right)^2,
\label{<D0nx><nxD0>}\\
\langle n_x {\cal D}_0 n_{x}\rangle &=&
{\omega_c^2\over 4 v_F^2}\left({\partial^2 g_0\over \partial q^2} 
+ {1\over q}{\partial g_0\over \partial q}\right)
+ {g_0\over 2}, \nonumber \\ \label{<nxD0nx>}\\ 
\langle n_x {\cal D}_0 n_x {\cal D}_0\rangle 
&=&-{{\cal W}\over 2 q v_F^2}{\partial g_0\over \partial q},
\label{<nxD0nxD0>}\\ 
\langle n_x {\cal D}_0\rangle
\langle {\cal D}_0 n_{x} {\cal D}_0\rangle &=&
{[1-{\cal W} g_0]\over 2 q v_F^2}{\partial g_0\over \partial q},
\label{<nxD0><D0nxD0>}\\ 
\langle {\cal D}_0 n_x {\cal D }_0 n_{x} {\cal D}_0\rangle
&=&-{1 \over 4v_F^2}\left({\partial^2 g_0\over \partial q^2} 
+ {1\over q}{\partial g_0\over \partial q}\right)\!,\nonumber \\ 
\label{<D0nxD0nxD0>}\\
\langle {\cal D}_0 n_x {\cal D}_0\rangle^2
&=&-{1\over 2 v_F^2}\left({\partial g_0\over \partial q} \right)^2\!,
\label{<D0nxD0>2}
\eea
for the ``longitudinal correlators'', 
and 
\bea
\langle {\cal D}_0 n_x\rangle\langle n_{y} {\cal D}_0\rangle &=&
{\omega_c \over 2 q v_F^2}[1-{\cal W} g_0]{\partial g_0\over \partial q},
\label{<D0nx><nyD0>}\\
\langle n_x {\cal D}_0 n_{y}\rangle &=& 
\omega_c {{\cal W}\over 2 q v_F^2}{\partial g_0\over \partial q},
\label{<nxD0ny>}\\
\langle n_x {\cal D}_0 n_y {\cal D}_0\rangle 
&=& {\omega_c \over 4 v_F^2}\left({\partial^2 g_0\over \partial q^2} 
+ {1\over q}{\partial g_0\over \partial q} \right)\!\!,
\label{<nxD0nyD0>}\\ 
\langle n_x {\cal D}_0\rangle
\langle {\cal D}_0 n_{y} {\cal D}_0\rangle &=&
{\omega_c \over 4 v_F^2}\left({\partial g_0\over \partial q} \right)^2\!\!,
\label{<nxD0><D0nyD0>}\\ 
\langle {\cal D}_0 n_x {\cal D }_0 n_{y} {\cal D}_0\rangle
&=& {i\over 2\omega_c}\left({\partial g_0 \over \partial \omega} +
{i{\cal W}\over q v_F^2}{\partial g_0\over \partial q}  \right),\nonumber \\
\label{<D0nxD0nyD0>}
\eea
for the ``Hall correlators''.

Substituting Eqs.~(\ref{<D>-WN})-(\ref{<D0nxD0nyD0>}) in 
(\ref{Brhoxx-WN}) and (\ref{Brhoxy-WN}),
we obtain the kernels $B_{xx}^{(\rho)}$ and $B_{xy}^{(\rho)}$
averaged over $\phi_q$,
\bea
B_{xx}^{(\rho)}(\omega,q)&=&\left(\tau\over \tau-g_0\right)^2 
\nonumber\\
&\times&\left\{ \,
{2\tau-g_0\over 2\tau^2} 
\left[g_0^2+\frac{(1-{\cal W}g_0)^2}{q^2v_F^2}\right]\right.\nonumber \\
&+&{i\over \tau}{\partial g_0 \over \partial \omega}
-{1\over qv_F^2\tau^2}{\partial g_0\over \partial q}
+{1 \over 4v_F^2\tau^3}\left({\partial g_0\over \partial q} \right)^2\nonumber \\
&\times& \left[(1-\omega_c^2\tau^2){2\tau \over \tau-g_0}+
\omega_c^2\tau^2\left(1+{g_0\over 2\tau}\right)\right]\nonumber \\
&+&\left.{1\over 4v_F^2\tau^2}\left({\partial^2 g_0\over \partial q^2} 
+ {1\over q}{\partial g_0\over \partial q} \right)
\left[1-\omega_c^2 g_0^2\right]\, \right\}\nonumber \\
\label{Brhoxx-WN-g0}\\
B_{xy}^{(\rho)}(\omega,q)&=&\left(\tau\over \tau-g_0\right)^2 
\nonumber \\
&\times&\left\{ \,
-{i\over 4\omega_c\tau^2}\left({\partial g_0 \over \partial \omega} +
{i{\cal W}\over q v_F^2}{\partial g_0\over \partial q}  \right) \right.\nonumber \\
&+&{\omega_c\over  2q v_F^2 \tau} \left[1-{\cal W}g_0-{g_0\over 2\tau} \right]
{\partial g_0\over \partial q} \nonumber \\
&+&{\omega_c g_0\over 4 v_F^2\tau^2}\left({\partial^2 g_0\over \partial q^2} 
+ {1\over q}{\partial g_0\over \partial q} \right)\nonumber \\
&+&\left.{\omega_c \over 4 v_F^2\tau^2}
{\tau+g_0 \over \tau-g_0}
\left({\partial g_0\over \partial q} \right)^2 \,
\right\}.
\label{Brhoxy-WN-g0}
\eea
In zero magnetic field, we put $\omega_c=0$
and substitute $g_0=1/S$ in (\ref{Brhoxx-WN-g0}).
After some algebra, we reduce the obtained expression
for the kernel $B_{xx}$ to the form
\bea
B_{xx}^{(\rho)}(\omega,q)&=&{(qv_F)^2 \over 2\tau^3 S^3(S-1/\tau)^3}+
{3(qv_F)^2 \over 4\tau^2 S^3(S-1/\tau)^2} \nonumber \\
&+&\frac{S-{\cal W}}{\tau S(S-1/\tau)^2} +
\frac{(2S-1/\tau)[S-{\cal W}]^2}{2\tau (q v_F)^2 S(S-1/\tau)^2},\nonumber \\
\label{Bxx-ZNA}
\eea
which agrees with Eq.~(16b) of Ref.~\onlinecite{ZNA-rhoxy} up to an overall
factor $1/2\tau$ related to different normalization.
In the ballistic limit, $T\tau\gg 1$, expanding (\ref{Bxx-ZNA})
in $\tau^{-1}$, one finds the leading contribution [${\cal O}(1/\tau)$] 
given by the last two terms in
(\ref{Bxx-ZNA}),
\bea
B_{xx}^{(\rho)}(\omega,q)&\simeq&
\frac{S_0+i\omega}{\tau S_0^3} + \frac{[S_0+i\omega]^2}{\tau (q v_F)^2 S_0^2},
\nonumber \\
&=&\frac{S_0+i\omega}{\tau S_0^2}\left[{1\over S_0}+
{1\over S_0-i(\omega+i0)}\right],\nonumber \\
 \label{Bxx-ZNA-ball}
\eea
where $S_0=[q^2v_F^2-(\omega+i0)^2]^{1/2}$.
Substituting (\ref{Bxx-ZNA-ball}) in (\ref{rho}) and using
(\ref{U-ball}) for exchange interaction, we reproduce the 
linear-in-$T$ correction to the resistivity
in the ballistic regime,
\be
{\delta\rho_{xx}^{\rm F}\over\rho_0}=-{T\over E_F}.
\label{rhoF-ball-B0}
\ee
Within the approximation of isotropic interaction
used in Ref.~\onlinecite{ZNA-sigmaxx}, 
the Hartree term 
is determined by the triplet channel and is given by 
\be
{\delta\rho_{xx}^{\rm H}\over\rho_0}=
-{3\: F_0^\sigma\over 1+F_0^\sigma}{T\over E_F}.
\label{rhoH-ball-B0}
\ee
It is worth noting that one should exercise a certain caution
when comparing the experimental data with the results 
(\ref{rhoF-ball-B0}) and (\ref{rhoH-ball-B0}), even in systems
with short-range impurities.
First, the higher angular harmonics $F_{n\neq 0}^{\rho,\sigma}$ 
of the interaction~\cite{ceperley} (neglected in the above 
approximation) may change the numerical coefficient in front
of the Hartree term (see discussion in Sec.~\ref{IIIe} and 
in Ref.~\onlinecite{ZNA-sigmaxx}).
Second, anisotropy of the impurity scattering 
introduces an extra factor $2\pi\nu W(\pi)\tau \neq 1$
(where $W(\pi)$ is the effective impurity-backscattering probability) 
in both exchange and Hartree terms 
(see Sec.~\ref{IIc}3 and Appendix~\ref{A3}).
The anisotropy may arise due to some amount of smooth disorder present
in any realistic system, due to a finite range of scatterers, or due to
the screening of originally point-like impurities (see Sec.~\ref{IV}).
Therefore, the interaction parameter $F_0^\sigma$ extracted 
from the measured linear-in-$T$ resistivity with the use of
(\ref{rhoF-ball-B0}),\ (\ref{rhoH-ball-B0}) may differ
considerably from that found from a measurement of other
quantities (e.g. the resistivity correction in the diffusive limit or 
the spin susceptibility).

To find the leading contribution to $B_{xy}^{(\rho)}$ in the limit
of vanishing magnetic field,
we have to expand the propagator $g_0$ up to the second order in $\omega_c$ 
in the first term in curly brackets in (\ref{Brhoxy-WN-g0}).
This can be easily done by treating 
the term $\omega_c{\partial/\partial\phi}$ in (\ref{LB-out-WN})
as a perturbation, which yields
\bea
g_0(B\to 0)&=&g_0(\omega,q;B=0)+\omega_c^2 h(\omega,q) 
\nonumber \\
&=&{1\over S}-\omega_c^2\frac{q^2v_F^2(S^2-5{\cal W}^2)}{8 S^7}.
\label{g_0(B)}
\eea
After a simple algebra, we find  $B_{xy}^{(\rho)}$ in the following form,
\bea
{B_{xy}^{(\rho)}(\omega,q)\over \omega_c} &=&{(qv_F)^2 \over \tau^2 S^3(S-1/\tau)^3}+
{(qv_F)^2 [2S -5 {\cal W}]\over 4 \tau^2 S^5(S-1/\tau)^2} \nonumber \\
&+& \frac{{\cal W}[S-{\cal W}]^2}{ 2 \tau^2 S^4(S-1/\tau)^2},
\eea
which agrees with Eq.~(16a) of Ref.~\onlinecite{ZNA-rhoxy}.
In the ballistic limit, $T\tau\gg 1$, the leading contribution 
[${\cal O}(1/\tau^2)$] 
to $B_{xy}^{(\rho)}$ has the form
\be
B_{xy}^{(\rho)}(\omega,q)\simeq 
{\omega_c (S_0+i\omega) \over 4\tau^2 S_0^7}[6S_0^2-3iS_0\omega +5 \omega^2].
\label{Bxy-ZNA-ball}
\ee

In arbitrary magnetic field, 
Eq.~(\ref{Brhoxx-WN-g0}) can be also significantly simplified 
when the condition of the ballistic regime, $T\tau\gg 1$, is assumed.
Then the leading contribution to the longitudinal MR,
$\Delta\rho_{xx}=\rho_{xx}(B)-\rho_{xx}(0),$
is determined by the kernel
\bea
\tau B_{xx}^{(\rho)}(\omega,q)&\simeq& g_0^2+\frac{(1-{\cal W}g_0)^2}{q^2v_F^2} 
\nonumber \\
&+& i{\partial g_0 \over \partial \omega} 
-{\omega_c^2 \over 4 v_F^2}\left({\partial g_0\over \partial q} \right)^2.
 \label{Brhoxx-WN-ball}
\eea
The remaining terms in (\ref{Brhoxx-WN-g0}) yield the contributions
to the MR which are smaller at least by an additional 
factor $(T\tau)^{-1}$. 
Using (\ref{g_0(B)}) [which tells us that 
for $\omega_c\ll T$ the magnetic-field-induced corrections to the 
propagator $g_0$ are small by a factor  $(\omega_c/T)^2$], we find
that the MR for not very strong magnetic fields,
$\omega_c\ll T$, is determined by a quadratic in $\omega_c$
correction to the kernel $B_{xx}^{(\rho)}$,
\bea
\Delta B_{xx}^{(\rho)}(\omega,q) 
&=& {\omega_c^2\over \tau}\left[ \, {2 h(\omega,q)\over S} 
-{2h(\omega,q) \over q^2v_F^2}\frac{{\cal W}(S-{\cal W})}{S}\right. \nonumber \\
&+& \left. i{\partial h(\omega,q) \over \partial \omega}
-{1 \over 4v_F^2 S^4}\left({\partial S\over \partial q} \right)^2 
\, \right]\nonumber \\
&=&-\omega_c^2 \frac{S-{\cal W}}{4 \tau S^6}  
\left[\frac{S^2-5{\cal W}^2}{S} \right.  \nonumber   \\
&-&\left.  
\frac{5{\cal W}(S+{\cal W})(3S^2-7{\cal W}^2)}{2 S^3} 
+S+{\cal W} \, \right] \nonumber \\
&\simeq&-\omega_c^2\frac{S_0+i\omega}{8 \tau S_0^9}
\left(4S_0^4+13 i\omega S_0^3 \right.\nonumber   \\ 
&+&\left. 25 \omega^2 S_0^2-35 i \omega^3 S_0+35\omega^4\right),
\label{deltaBrhoxx-ball}
\eea
independently of the relation between $\omega_c$ and $\tau^{-1}$.
Similarly, using (\ref{g_0(B)}), one can find the correction to 
Eq.~(\ref{Bxy-ZNA-ball})
in a finite magnetic field $\omega_c\ll T$,
\be
\Delta B_{xy}^{(\rho)}(\omega,q)
=-\omega_c^3\frac{ i\omega (S_0^2+\omega^2)}{4 S_0^9}[3S_0^2+7\omega^2].
\label{DeltaBrhoxy}
\ee
Again, this correction is independent of the relation between 
$\omega_c$ and $\tau^{-1}$.
The results (\ref{deltaBrhoxx-ball}) and (\ref{DeltaBrhoxy})
are used for calculation of the 
interaction-induced corrections to $\rho_{xx}$ and $\rho_{xy}$
for the white-noise disorder and
$\omega_c\ll T$ in Sec.~\ref{Vb}.

\section{Linear-in-$T$ term in the ballistic limit at $B=0$}
\label{A3}
\renewcommand{\theequation}{C.\arabic{equation}}
\setcounter{equation}{0}

\begin{figure}
  \medskip
\centerline{\includegraphics[width=7cm]{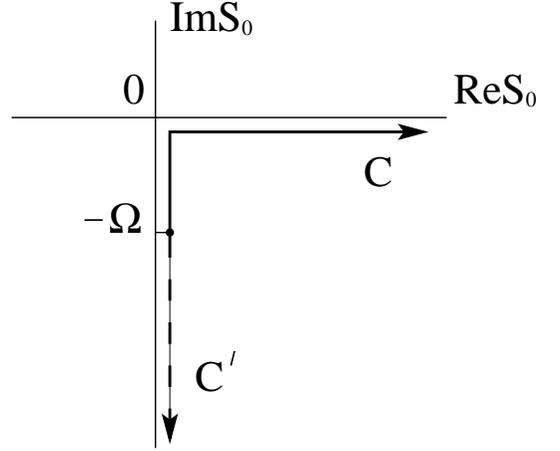}}
\caption{The contours ${\bf C}$ and ${\bf C'}$ of integration 
over $S_0$ in Eq.~(\ref{sigma1}) } 
\label{contour} 
\end{figure}

In this appendix we calculate the leading ballistic correction to the
conductivity at $B=0$ for a generic scattering cross-section $W(\phi-\phi')$
in the case of the Coulomb interaction. As explained in Sec.~\ref{IIc}3,
this term (proportional to $T\tau$) is obtained by substituting the
ballistic asymptotics (\ref{Bwq_ball}) of $B_{xx}$ in the general formula
(\ref{sigma}). Likewise, the interaction propagator $U(\omega,\bq)$
entering (\ref{sigma}) has to be replaced by
\be
U(\omega,\bq)={1\over 2\nu} \frac{1}{1+i\omega\langle{\cal D}_{\rm f}\rangle}
\label{U-ball}
\ee
with the free propagator ${\cal D}_{\rm f}$ given by Eq.~(\ref{D-free}).
Performing the angular integration $\langle...\rangle$, we get
\bea
&&\delta\sigma_{xx}\simeq -{e^2\over 2\pi^2} T\tau \ {\rm Im} \int_0^\infty
d\Omega
{\partial \over \partial \Omega}\left(\Omega {\rm coth}{\Omega\over 2}\right)
\nonumber \\
&&\times \int_0^\infty Q dQ 
 {[Q^2-(\Omega+i0)^2]^{1/2}\over 
[Q^2-(\Omega+i0)^2]^{1/2} +i(\Omega+i0)} \nonumber \\
&&\times \left[ \int {d\phi\over 2\pi}
\frac{(-i\Omega)\ {\tilde W}(\phi)(1-\cos\phi)}{[Q^2\cos^2{\phi\over 2}
-(\Omega+i0)^2][Q^2-(\Omega+i0)^2]^{1/2}} \right. \nonumber \\
&&\left. - 
\frac{-i\Omega}{[Q^2-(\Omega+i0)^2]^{3/2}}
\right],
\label{sigma-Ttau}
\eea
where we introduced the dimensionless variables 
$\Omega=\omega/T, \ \ Q=qv_F/T,$ and ${\tilde W}(\phi)=2\pi\nu\tau W(\phi).$
It is convenient to split the interaction propagator as follows
\be
 2\nu U(\Omega,Q)=\frac{S_0}{S_0+i(\Omega+i0)}=
\left(1-\frac{i\Omega}{S_0+i(\Omega+i0)}\right),
\label{U-split}
\ee
where $S_0=[Q^2-(\Omega+i0)^2]^{1/2}.$ The first term corresponds to a 
statically screened
interaction and is equivalent to a point-like interaction with $V_0=1/2\nu,$
the second term results from the dynamical weakening of screening.
As discussed in Sec.~\ref{IIc}3, the contribution $\delta\sigma_{xx}^{(1)}$
of the first (constant) term  is proportional to
the backscattering probability $W(\pi),$ see Eq.~(\ref{sigma_ball}).
Let us show that this follows also from Eq.~(\ref{sigma-Ttau}).
Performing the variable change $Q\to S_0,$ we find
\bea
\delta\sigma_{xx}^{(1)}&\simeq& -{e^2\over 2\pi^2} T\tau \ {\rm Im} 
\int_0^\infty d\Omega
{\partial \over \partial \Omega}\left(\Omega {\rm coth}{\Omega\over 2}\right) 
{\tilde \Phi}(\Omega)\nonumber \\
{\tilde \Phi}(\Omega)&=&
\int_{\bf C} dS_0 {(-i\Omega)(S_0^2+\Omega^2)\over S_0^2}\nonumber \\
&\times&  \int {d\phi\over 2\pi}
\frac{2\sin^4(\phi/2){\tilde W}(\phi)}
{S_0^2\cos^2{\phi\over 2}-(\Omega+i0)^2\sin^2{\phi\over 2}} .
\label{sigma1}
\eea
The contour ${\bf C}$ of integration over $S_0$ in Eq.~(\ref{sigma1})
is shown in Fig.~\ref{contour}.
Interchanging the order of integration over $\phi$ and $S_0$, 
we see that for any 
$\phi\neq \pi$ (i.e. $\cos{\phi\over 2} \neq 0$) the $S_0$-integral 
converges. Furthermore, transforming the integration contour
${\bf C} \to {\bf C'}$ as shown in Fig.~\ref{contour},
it is straightforward to reduce ${\tilde\Phi}(\Omega)$ to an explicitly real form.
Therefore, only the singular point $\phi=\pi$ 
(where the result of $S_0$-integration diverges as $1/|\cos{\phi\over 2}|$,
implying that the imaginary part of  ${\tilde\Phi}(\Omega)$ is determined
by a delta-function in $\phi$-integral)
contributes to (\ref{sigma1}), so that 
$\delta\sigma_{xx}^{(1)}\propto W(\pi)$. To find the corresponding 
coefficient, one can consider the isotropic scattering $W(\phi)={\rm const}$
and to integrate over $\phi$ first, yielding
\be
\delta\sigma_{xx}^{(1)}={e^2\over \pi}{\tilde W}(\pi) T\tau,
\label{sigma1pi}
\ee
in agreement with (\ref{sigma_ball}). Note that the integral over $\Omega$
is formally divergent at the upper limit. It should be cut off at
$\Omega \sim E_F/T$ yielding a temperature independent contribution 
$\sim e^2{\tilde W}(\pi) E_F\tau$ which renormalizes the 
value of the Drude conductivity.

We now turn to the contribution $\delta\sigma_{xx}^{(2)}$ 
of the second (dynamical) term in the interaction propagator
(\ref{U-split}), which differs from Eq.~(\ref{sigma1}) by an extra factor
$-i\Omega/[S_0+i(\Omega+i0)].$ Rotating at $\phi\neq \pi$ 
the integration contour as before, 
we reduce the $S_0$-integral to the form  ($S_0\to -iY$)
\be
\Omega^2\int_\Omega^\infty {d Y \over Y^2} 
\frac{Y+\Omega}{(Y^2\cos^2{\phi\over 2}+\Omega^2\sin^2{\phi\over 2})},
\ee
which is again real and thus
yields no contribution to $\delta\sigma_{xx}^{(2)}$.
Though the point $\phi=\pi$ is singular in this case as well,
the singularity is only logarithmic ($\sim \ln|\cos{\phi\over 2}|$), 
so that no contribution
proportional to $W(\pi)$ arises. This can be easily checked by assuming 
$W(\phi)={\rm const}$ and performing the $\phi$-integration first. Therefore
$$\delta\sigma_{xx}^{(2)}=0,$$
and the linear-in-$T$ term is given by Eq.~(\ref{sigma1pi}).

In the above consideration we have expanded the ballistic propagator
${\cal D}$ up to terms with one scattering event. In the case of small-angle
scattering this is justified provided $T\tau_s\gg 1,$  while in the 
intermediate temperature range $\tau^{-1}\ll T \ll \tau_s^{-1}$ processes with 
many scattering events dominate (though the particle motion is typically
close to the straight line).
The term $\delta\sigma_{xx}^{(1)}$ which is governed by anomalous processes
of returns in a time $t\lesssim T^{-1} \ll \tau$ is exponentially
small in this case, see Sec.~\ref{IIc}3.
As to the $\delta\sigma_{xx}^{(2)}$ contribution to the linear-in-$T$
term, it remains zero in this case as well. To demonstrate this, we use 
Eq.~(\ref{Bwq}). In the first and the third terms we can replace ${\cal D}$ 
by the free propagator (\ref{D-free}), the fourth term gives no $T\tau$
contribution, while in the second term we should take into account
the angular diffusion (\ref{Csmooth}) around the straight trajectory,
\be
{1\over 2}\langle{\cal D}\rangle-\langle n_x{\cal D}n_x\rangle
\to \frac{-i\omega}{2\tau[q^2v_F^2-(\omega+i0)^2]^{3/2}}.
\ee
Combining the contributions to $B_{xx}$ of all the three
terms, we get
\be
B_{xx}(\omega,q)=\frac{-i\omega}{[q^2v_F^2-(\omega+i0)^2]^{3/2}}
\left({1\over 2}+{1\over 2}-1 \right)=0,
\label{Bxx-ball-smooth}
\ee
so that the coefficient of the $T\tau$-term indeed vanishes.

\section{Solution of Liouville-Boltzmann equation for a smooth disorder}
\label{A4}
\renewcommand{\theequation}{D.\arabic{equation}}
\setcounter{equation}{0}

In this appendix, we will solve the classical equation for
a propagator of a particle moving in a smooth random potential in a 
magnetic field,
\bea
\left[-i\omega + i q v_F \cos\phi+\omega_c{\partial\over\partial\phi} \right.
&-&\left.{1\over \tau}{\partial^2\over \partial \phi^2}\right]
{\cal D}(\phi,\phi') \nonumber \\
&=& 2\pi\delta(\phi-\phi'). 
\label{LBsmooth} 
\eea
Here the polar angle of the velocity is counted from the 
angle of $\bq$, $\phi-\phi_q \to \phi$.

We first consider the diffusive limit, $Dq^2,\ \omega\ll 1/\tau$,
and solve this equation perturbatively in $q$ for arbitrary magnetic field.
Putting $q=0$, we obtain the solution in the form 
\be
{\cal D}(\phi,\phi')=
\sum_n {e^{i n(\phi-\phi')}\over -i\omega + in\omega_c +n^2/\tau},
\label{q=0}
\ee
which is just a standard expansion in eigenfunctions of the  
Liouville-Boltzmann operator. 
We now treat the term 
$\delta {\cal L} = iv_F q \cos\phi$
as a perturbation.
The first-order  correction to the
eigenvalues $\lambda_n^{(0)}=-i\omega + in\omega_c +n^2/\tau$ 
vanishes, while the second order correction 
is 
\be
\lambda_n^{(2)}={q^2v_F^2\over 2}{1\over 1-(i\omega_c\tau+2n)^2}
\label{second-corr-eigen}
\ee
and can be neglected along with $-i\omega$ in all terms except
for $n = 0$ in the diffusive limit. 
The first order correction to the right eigenfunction for $n=0$
reads 
\be
\Psi_{0,R}^{(1)}(\phi)=-{i q v_F \tau \over 1+\omega_c^2\tau^2}
[\cos\phi+\omega_c\tau\sin\phi],
\label{eigen}
\ee
while the left eigenfunction differs from (\ref{eigen}) by 
a replacement $\omega_c \to 
-\omega_c$.
Thus, in the diffusive limit the propagator has the form
\begin{eqnarray}
{\cal D}(\omega,\bq;\phi,\phi')
&\simeq& {1 \over Dq^2-i\omega} \nonumber \\
&\times& \left[1-{i\: q v_F\: \tau\ (\cos\phi\: +\omega_c\tau\sin\phi\: )
\over 1+\omega_c^2\tau^2}
  \right] \nonumber \\
&\times& \left[1-{i\: q v_F\: \tau\ (\cos\phi'-\omega_c\tau\sin\phi')
\over 1+\omega_c^2\tau^2}
 \right]\nonumber  \\
&+& \sum_{n\neq 0} {e^{i n(\phi-\phi')}\over in\omega_c +n^2/\tau}.
\label{diff-prop}
\end{eqnarray}

In a strong magnetic field ($\omega_c\tau \gg 1$) one can go beyond the 
diffusion approximation. In this case one can represent the propagator
in the form 
\be
{\cal D}(\omega,\bq;\phi,\phi')=d(\omega,\bq;\phi,\phi')
\exp[-iqR_c(\sin\phi-\sin\phi')],
\label{sin-trans}
\ee
and solve the equation for $d(\phi,\phi'),$ 
\bea
\left[-i\omega - i { q v_F\over \omega_c\tau} \sin\phi \right.
&+&\left\{\omega_c + 2i{ q v_F\over \omega_c\tau}\cos\phi\right\}
{\partial\over\partial\phi} 
\nonumber \\
 + {1\over \tau}\left({ q v_F\over \omega_c}\right)^2 \cos^2\phi
&-&\left.{1\over \tau}{\partial^2\over \partial \phi^2}\right]
d(\phi,\phi')\nonumber \\ 
&=& 2\pi\delta(\phi-\phi')
\label{rot-prop} 
\eea
perturbatively in $q$.
At $q=0$ we have the same solution (\ref{q=0}) as in the diffusive limit.
The first order correction to the eigenvalues is now produced 
by the $q^2$-term in (\ref{rot-prop}), $\lambda_n^{(1)}=Dq^2,$
with $D=R_c^2/2\tau$ the diffusion constant in a strong magnetic field.
The second order corrections $\lambda_n^{(2)}$ turn out to be small
compared to $\lambda_n^{(1)}$ for $(qR_c)^2\ll \omega_c\tau$.
As in the diffusive limit, for calculation 
of $B_{xx}$
the corrections to the eigenfunctions $\Psi_n$
with $n\neq 0$ 
can be neglected.  
The first-order correction to $\Psi_0$
is found to be (we drop the term $\propto \sin 2\phi$, since it does not
contribute to $B_{xx}$ in the leading order)
\be
\Psi_0^{(1)}(\phi)\simeq -{iqv_F\tau \cos\phi\over (\omega_c\tau)^2},
\label{Psi-0-pert}
\ee
leading to Eq.~(\ref{diffuson}).

To calculate the kernel $B_{xy}$, we need a more accurate form 
of the propagator. Therefore, we should analyze the 
corrections to the
eigenvalues and eigenfunctions of the Liouville-Boltzmann
operator to the next order in $(qR_c)^2/\omega_c\tau$.
To do this, it is convenient to perform the transformation
\bea
{\cal D}(\omega,\bq;\phi,\phi')&=&{\tilde d}(\omega,\bq;\phi,\phi')\nonumber \\
&\times&\exp\left\{-i{qR_c\over 1+\beta^2}
[\beta^2(\sin\phi-\sin\phi')\right.\nonumber \\
&+&\beta(\cos\phi-\cos\phi')] {\Big \}},
\label{sin-cos-trans}
\eea
and introduce the dimensionless variables
$\beta=\omega_c\tau,\ {\tilde Q}=qR_c \beta/(1+\beta^2)^{1/2}, \ 
\Omega=2\omega\tau$. 
The equation for ${\tilde d}(\omega,\bq;\phi,\phi')$
takes then the form
\bea
\left[-i{\Omega\over 2} +  {\tilde Q}^2\cos^2{\tilde\phi} \right.
&+&\beta{\partial\over\partial{\tilde\phi}} 
\nonumber \\+ 
2i{\tilde Q}\cos{\tilde\phi}{\partial\over\partial{\tilde\phi}}
 &-&\left.{\partial^2\over \partial {\tilde\phi}^2}\right]
{\tilde d}(\omega,\bq;{\tilde\phi},{\tilde\phi}')\nonumber \\ 
&=& 2\pi\tau\delta(\phi-\phi'),
\label{tilde-g}
\eea
where we performed a rotation ${\tilde \phi}= \phi+\phi_\beta,
\ \phi_\beta={\rm arccot}\beta$.
Treating for ${\tilde Q}^2\ll {\rm max}[1,\beta]$ 
(i.e. $(qR_c)^2\ll \omega_c\tau$ in a strong
magnetic field, $\beta\gg 1$) the term
\be
\delta {\hat L}={{\tilde Q}^2\over 2}\cos2{\tilde\phi}+
2i{\tilde Q}\cos{\tilde\phi}{\partial\over\partial{\tilde\phi}}
\label{deltaL}
\ee
as a perturbation to the operator
\be
{\hat L}_0=-i{\Omega\over 2} +  {{\tilde Q}^2\over 2} +
\beta{\partial\over\partial{\tilde\phi}} 
-{\partial^2\over \partial {\tilde\phi}^2},
\label{L-unperturb}
\ee
we find the unperturbed
solution 
\be
\displaystyle
{\tilde d}_0({\tilde\phi},{\tilde\phi}')=
2\tau\sum_n\frac{e^{i n({\tilde\phi}-{\tilde\phi}')}}
{-i\Omega+{\tilde Q}^2+2in\beta+2 n^2},
\label{tilde-g0}
\ee
and the first-order correction to the eigenvalues
${\tilde \lambda}_n^{(1)}=0$.
Calculating the $n=0$ eigenfunctions and eigenvalue
up to the second order in the perturbation (\ref{deltaL})
we finally obtain the singular part of the propagator
for $\beta\gg 1$ with required accuracy,
\bea
&&{\cal D}^{\rm s}(\omega,\bq;\phi,\phi')=
2\tau \exp[-i Q(\sin\phi-\sin\phi')]\nonumber \\
&&\times\frac{\chi_R(\phi,Q)\chi_L(\phi',Q)}
{Q^2[1-(1-Q^2/4)/\beta^{2}]-i\Omega} 
\label{Dsing-2order},
\eea
where $Q=qR_c$ and the functions $\chi_{R,L}(\phi,Q)$ 
are given by
\bea
\chi_{R,L}(\phi,Q)&=&1-{1\over \beta}
\left[\ iQ\cos\phi \pm {Q^2\over 4}\sin2\phi\ \right] \nonumber \\
&+&{1\over \beta^2}\left[\ {Q^2\over 4}\pm iQ(1+{5 Q^2\over 8})\sin\phi 
\right. 
\nonumber \\
&-& {5 Q^2\over 4}\cos2\phi \pm {7iQ^3\over 24}\sin3\phi  \nonumber \\
&+& \left.{Q^4\over 64}(1-\cos4\phi)\ \right].
\label{chiRL-2}
\eea
As to the regular part of the propagator,
for $n\neq 0$
it is sufficient to calculate the eigenfunctions
to the first order in the perturbation, which yields
\bea
&&{\cal D}^{\rm reg}(\omega,\bq;\phi,\phi')=
2\tau \exp[-i Q(\sin\phi-\sin\phi')]\nonumber \\
\displaystyle
&&\times\sum_{n\neq 0}\frac{\Psi_R(\phi',Q;n) \Psi_L(\phi',Q;n)}
{-i\Omega+ Q^2+2in\beta+2 n^2}e^{i n(\phi-\phi')},
\label{Dreg-2}
\eea
where
\bea
\Psi_{R,L}(\phi',Q;n)&=&1-{i Q\over \beta}\cos\phi \nonumber \\
&\mp &{Q^2\over 4\beta}\sin 2\phi \pm {2 n Q\over \beta}\sin\phi.
\nonumber \\
\label{PsiRL-n-2}
\eea
The results (\ref{Dsing-2order})-(\ref{PsiRL-n-2}) allow
us to calculate the kernel $B_{xy}(\omega,\bq)$ in
the first non-vanishing order in $\beta^{-1}$,
see Sec.~\ref{IIIg}.

\section{Propagator for anisotropic systems}
\label{A5}
\renewcommand{\theequation}{E.\arabic{equation}}
\setcounter{equation}{0}

In this Appendix, we assume that the collision 
integral ${\hat C}$ induces a transport anisotropy, i.e. that
the scattering cross-section $W(\phi,\phi')$ is {\it not}
a function of $\phi-\phi'$. The propagator
${\cal D}(\omega,\bq;\phi,\phi')$ satisfies the
equation 
\bea
&&\left[-i\omega + i q v_F \cos(\phi-\phi_q)+
\omega_c{\partial\over\partial\phi} 
+{\hat C}\right]
{\cal D}(\omega,\bq;\phi,\phi') \nonumber \\
&&= 2\pi\delta(\phi-\phi'), 
\label{LB-anis} 
\eea
where
\be
[{\hat C}\Psi](\phi)=
\nu\int{d\phi'\over 2\pi}[\Psi(\phi)-\Psi(\phi')]W(\phi,\phi').
\label{coll-int-anis}
\ee
We first consider the diffusive limit
and concentrate on the leading
contribution ${\cal D}^{\rm s}$ governed by the diffusion mode.

This requires finding a lowest eigenvalue $\Lambda_0$
of the operator in the l.h.s. of (\ref{LB-anis})
and the corresponding left and right
eigenfunctions.
Treating the term $i q v_F \cos(\phi-\phi')$
perturbatively as in Appendix \ref{A5}, we find
\be
\Psi_{R,L}(\phi)=
1-i q v_F
\left(\pm\omega_c{\partial\over\partial\phi} +{\hat C}\right)^{-1}
\!\!\cos(\phi-\phi_q)
\label{PsiRL-anis}
\ee
and
\be
\Lambda_0=D_{\alpha\beta}q_\alpha q_\beta-i\omega,
\label{eigen-anis}
\ee
with the diffusion tensor
\be
D_{\alpha\beta}=v_F^2
\left\langle n_\alpha 
\left(\pm\omega_c{\partial\over\partial\phi} +{\hat C}\right)^{-1} 
n_\beta\right\rangle.
\label{diff-tens-anis}
\ee
We thus get the result (\ref{dmode-anis}) for the singular contribution 
${\cal D}^{\rm s}$, with  $\Psi_{R,L}$ given by (\ref{PsiRL-anis}).

In a strong magnetic field ($\omega_c\tau\gg 1$), 
we can go beyond the diffusive limit.
Proceeding as for an isotropic system, we perform
the transformation (\ref{sin-trans}).
Treating the $q$-dependent terms in the obtained
equation for $d(\omega,\bq;\phi,\phi')$ as a perturbation
and keeping the singular contribution only,
we come to the result (\ref{anis-ball-diffuson}),
where $\chi(\phi)$ can be represented symbolically as
\be
\chi(\phi)=1+\frac{i q v_F}{\omega_c^2}
\left(\partial\over\partial\phi\right)^{-1}{\hat C}\sin(\phi-\phi_q).
\label{chi-anis-C}
\ee
According to (\ref{DD})-(\ref{DnDnD}), we only
need to calculate averages of the type
$\langle{\cal D}\rangle$ and $\langle n_\alpha{\cal D}\rangle$,
so that it is sufficient to keep
the zero and first harmonics in $\phi$ in 
Eq.~(\ref{anis-ball-diffuson}).
Using 
\be
\label{t_xy}
\langle n_{\alpha}{\hat C}n_{\beta}\rangle=
\left(
\begin{array}{cc} \tau_x^{-1}      &   \\
              & \tau_y^{-1}
\end{array}
\right),
\ee
we then reduce (\ref{chi-anis-C}) to the form
(\ref{chi-anis}).

\vspace{-0.3cm}  



\begin{thebibliography}{1}
\bibitem{2pi} Since characteristic times corresponding to the 
temperature $T$ and the cyclotron frequency
$\omega_c$ are $t\sim(2\pi T)^{-1}$ (see the end of Sec.~\ref{IIIb})
and $t\sim 2\pi/\omega_c$, respectively,
a more accurate form of this condition is
$\omega_c\ll (2\pi)^2 T$.
\end{thebibliography}
\end{document}